\documentclass[journal]{IEEEtran}

\usepackage{color}
\usepackage{bm}
\usepackage{bbm}
\usepackage{amsmath}
\usepackage{amssymb}
\usepackage{amsthm}
\usepackage{graphicx}
\usepackage{mathrsfs}
\usepackage{empheq}	
\usepackage[mathcal]{euscript}
\usepackage{framed}
\usepackage{multirow}
\usepackage{cite}
\usepackage{enumitem} 
\graphicspath{{./figs/}}
\newtheorem{theorem}{Theorem}
\newtheorem{lemma}{Lemma}
\newtheorem{corollary}{Corollary}

\newtheorem{assumption}{Assumption}

\newtheorem{property}{Property}

\renewcommand{\P}{\mathbb{P}}
\newcommand{\E}{\mathbb{E}}
\newcommand{\VAR}{{\sf VAR}}
\newcommand{\beq}{\begin{equation}}
\newcommand{\eeq}{\end{equation}}
\newcommand{\beqa}{\begin{IEEEeqnarray}{rCl}}
\newcommand{\eeqa}{\end{IEEEeqnarray}}

\newcommand{\thetatrue}{\theta_0}
\newcommand{\T}{^\top}

\newcommand{\dfz}{\triangleq}

\newcommand{\chf}{\varphi_{\tilde{s}}}
\newcommand{\chfz}{\varphi_{\tilde{z}}}
\newcommand{\mnet}{{\sf m}_{\mathrm{ave}}}
\newcommand{\cnet}{{\sf c}_{\mathrm{ave}}}
\newcommand{\Cnet}{{\sf C}_{\mathrm{ave}}}
\newcommand{\omeganet}{\Lambda_{\mathrm{ave}}}
\newcommand{\xnet}{\bm{x}_{\mathrm{ave},i}}
\DeclareMathOperator*{\argmax}{arg\,max}

\begin{document}

\title{Adaptive Social Learning}
\author{Virginia~Bordignon,
	Vincenzo~Matta,
	and~Ali~H.~Sayed%
	\thanks{
		V. Bordignon and A.~H.~Sayed are with the \'Ecole Polytechnique F\'ed\'erale de Lausanne EPFL, School of Engineering, CH-1015 Lausanne, Switzerland (e-mails: \{virginia.bordignon, ali.sayed\}@epfl.ch).
		
V. Matta is with the Department of Information and Electrical Engineering and Applied Mathematics (DIEM), University of Salerno, via Giovanni Paolo II, I-84084, Fisciano (SA), Italy, and also with the National Inter-University Consortium for Telecommunications (CNIT), Italy (e-mail: vmatta@unisa.it).

		This work was supported in part by grant 205121-184999 from the Swiss National Science Foundation (SNSF). 
		
	An early version with partial results from this work was presented at EUSIPCO 2020~\cite{bordignon2020adaptation}.
	}
}


\maketitle

\begin{abstract}
This work proposes a novel strategy for social learning by introducing the critical feature of {\em adaptation}. 
In social learning, several distributed agents update continually their belief about a phenomenon of interest through: $i)$ direct observation of streaming data that they gather {\em locally}; and $ii)$ diffusion of their beliefs through {\em local cooperation} with their neighbors. 
Traditional social learning implementations are known to learn well the underlying hypothesis (which means that the belief of every individual agent peaks at the true hypothesis), achieving steady improvement in the learning accuracy under stationary conditions. 
However, these algorithms do not perform well under nonstationary conditions commonly encountered in online learning, exhibiting a significant inertia to track drifts in the streaming data.
In order to address this gap, we propose an Adaptive Social Learning (ASL) strategy, which relies on a small step-size parameter to tune the adaptation degree. 
First, we provide a detailed characterization of the {\em learning performance} by means of a steady-state analysis. 
Focusing on the small step-size regime, we establish that the ASL strategy achieves consistent learning under standard global identifiability assumptions. We derive reliable Gaussian approximations for the probability of error (i.e., of choosing a wrong hypothesis) at each individual agent. 
We carry out a large deviations analysis revealing the universal behavior of adaptive social learning: the error probabilities decrease exponentially fast with the inverse of the step-size, and we characterize the resulting exponential learning rate. 
Second, we characterize the {\em adaptation performance} by means of a detailed transient analysis, which allows us to obtain useful analytical formulas relating the {\em adaptation time} to the step-size. 
The revealed dependence of the adaptation time and the error probabilities on the step-size highlights the fundamental trade-off between adaptation and learning emerging in adaptive social learning.
\end{abstract}


\section{Introduction and Motivation}
\IEEEPARstart{S}{ocial} learning is a collective process whereby some agents form their opinions about a phenomenon of interest through the local exchange of information~\cite{ChamleyBook, Jad, Jad2, Shahrampour2015, Molavi2018, PoorSPmag2013, Krishnamurthy2014, ScaglioneSPmag2013, ScaglioneACM2013, AcemogluGEB2010, Golub2010}. 
More formally, given a set of $H$ hypotheses, $\Theta=\{1,2,\ldots,H\}$, there is one true state of nature $\theta_0\in\Theta$.
Each agent $k$, at time $i$, collects streaming data $\bm{\xi}_{k,i}$ (bold notation is used for random objects) drawn from a distribution that depends on the underlying hypothesis $\theta_0$. 
By exchanging local information with its neighbors, each agent assigns a belief $\bm{\mu}_{k,i}(\theta)$ to each hypothesis $\theta\in\Theta$, with the belief vector $\bm{\mu}_{k,i}=[\bm{\mu}_{k,i}(1),\bm{\mu}_{k,i}(2),\ldots,\bm{\mu}_{k,i}(H)]\T$ being a probability vector. 
Proper social learning occurs when the highest credibility is assigned to the true hypothesis, i.e., when the belief $\bm{\mu}_{k,i}(\theta)$ is maximized at $\theta=\theta_0$.

Several social learning strategies have been proposed in the literature. 
As a common feature, all of them exhibit the desirable property that, as time goes to infinity, the belief function converges to $1$ at $\theta_0$. In other words, if the amount of streaming data is sufficiently large, maximum credibility is assigned to the correct hypothesis whereas minimum (i.e., zero) credibility is assigned to the wrong hypotheses~\cite{AcemogluOzdaglar2011,Jadbabaie2013,Zhao,Salami,NedicTAC2017,Javidi,MattaSantosSayedICASSP2019,MattaBordignonSantosSayed2019}. 
Moreover, for most social learning implementations, convergence to the true hypothesis is exponentially fast.

However, such remarkably good convergence properties have a subtle consequence that has been overlooked so far in the literature. This is because the exponentially increasing accuracy in learning the true hypothesis makes all agents stubborn! 
We illustrate this phenomenon through a simple example. 

\begin{figure}[t]
	\centering
	\includegraphics[width=3in]{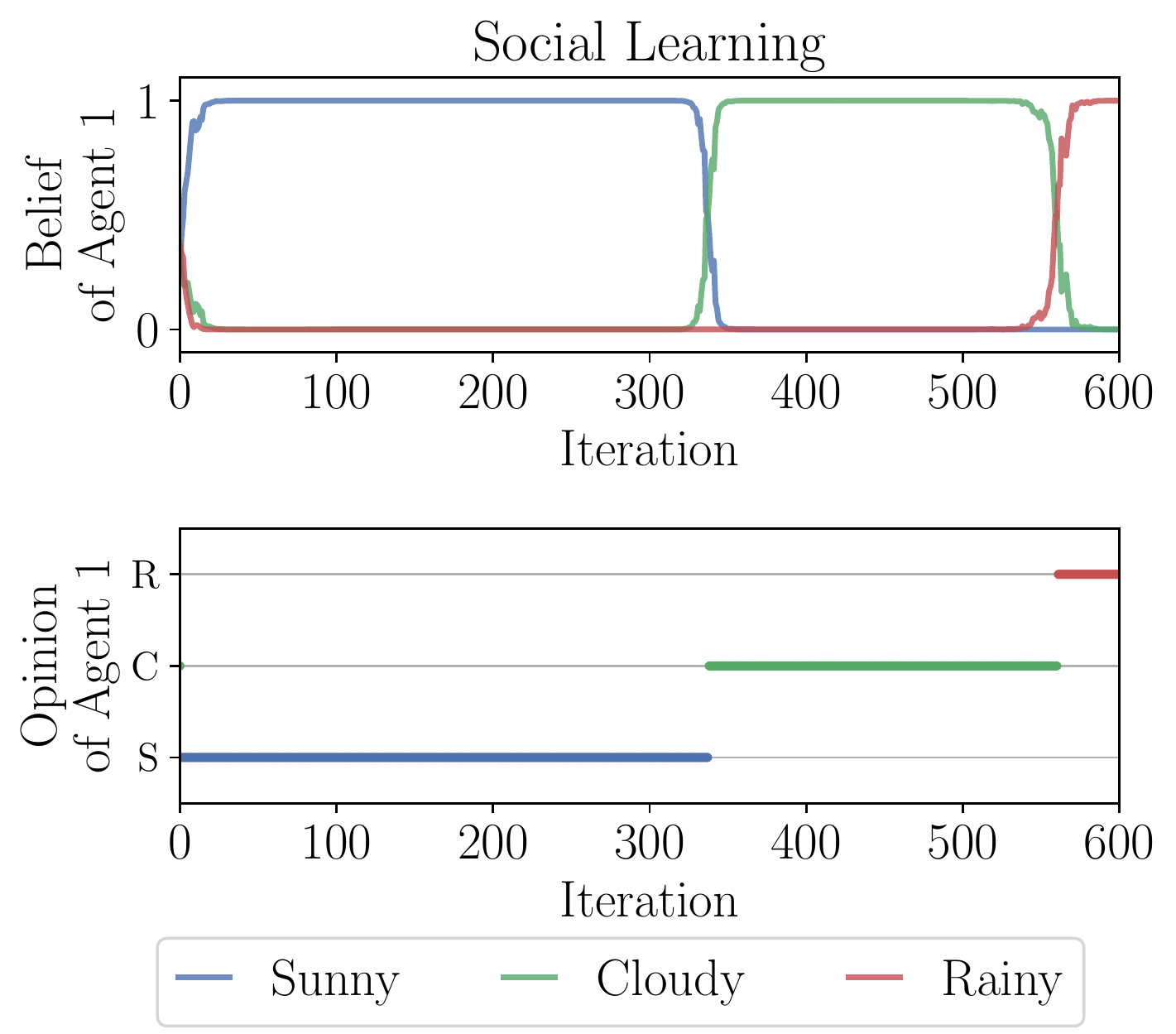}
	\caption{
		{\em Traditional} social learning strategy. 
		{\em Top panel}. Belief evolution of agent $1$, with a state of nature drifting at time $i=200$, from ``sunny'' to ``rainy''. {\em Bottom panel}. The instantaneous decision taken by agent $1$ by choosing the hypothesis that maximizes the current belief. We see that traditional social learning is not able to adapt to the new state of nature}.
	\label{fig:examplesl}
\end{figure}

Consider a weather forecast problem solved by an online social learning algorithm.
Assume that the agents are collecting data that drive them to believe that ``{\em tomorrow will be sunny}''. 
After some time, however, assume the streaming dataset available for the decision evolves in response to changes in weather conditions with the most recent evidences suggesting markedly that ``{\em tomorrow will be rainy}''. 
The traditional (existing) social learning algorithms discourage agents from changing their ``mind'' and it will be virtually impossible for the agents to adapt to the new situation and revise their earlier conclusion. 
This effect is clearly visible in the example of Fig.~\ref{fig:examplesl}. 
In this example we considered a network of $10$ agents that collect data originating from one of three possible hypotheses, namely, ``{\em sunny}'', ``{\em cloudy}'', ``{\em rainy}''. 
The data are initially consistent with the hypothesis ``{\em sunny}''. 
We observe from the blue curve in the top plot of Fig.~\ref{fig:examplesl} that the belief of agent $1$ for the hypothesis ``{\em sunny}'' approaches the value one and, therefore, this agent is able to arrive at the correct determination about the state of nature. However, in our simulation, the state of nature is made to change to ``{\em rainy}'' at instant $i=200$ (not shown in the figure). It is observed that the beliefs of agent $1$ start changing only around $i=350$ and the agent first transitions to believing that it is ``{\em cloudy}'' (the green curve) before switching to believing that it is ``{\em rainy}" many iterations later around $i=550$. This example shows that, under traditional social learning schemes, agents are not able to recover sufficiently fast to adapt their beliefs and track changes in the state of nature. 
The outcome of the social learning algorithm (we display in the figure the belief of agent $1$, with similar behavior being observed for other agents) shows clearly that the agents learn well until instant $i=200$, since they give almost full credibility to the hypothesis according to which the data are drawn, but react far slower afterward when the state of nature changes.
As a matter of fact, the traditional social learning algorithm has a delayed reaction to the change, only perceiving that something has changed at instant $i\approx 350$, but {\em still not detecting the true state}, because the agent gives maximum credibility to the wrong intermediate hypothesis ``{\em cloudy}''. After a prohibitive number of iterations, at $i\approx 550$, agents manage to overcome their stubbornness and opt for the correct hypothesis ``{\em rainy}''.

\begin{figure}[t]
	\centering
	\includegraphics[width=3in]{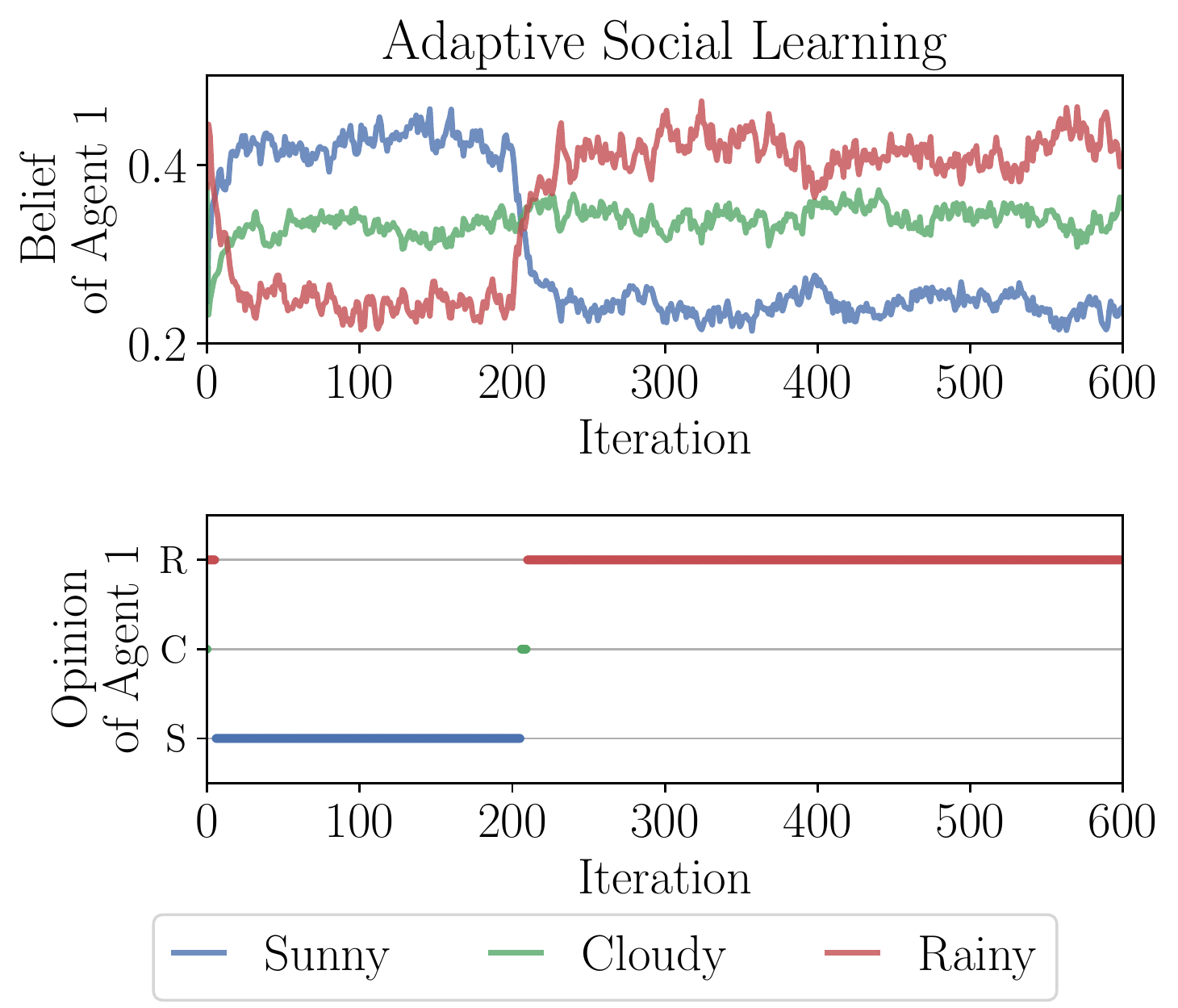}
	\caption{
	{\em Adaptive} social learning strategy proposed in this work. {\em Top panel}. Belief evolution of agent $1$, with a state of nature drifting at time $i=200$, from ``sunny'' to ``rainy''. {\em Bottom panel}. The instantaneous decision taken by agent $1$ by choosing the hypothesis that maximizes the current belief. 
		We see that the proposed strategy is able to adapt to the new state of nature.
	}	
	\label{fig:exampleasl}
\end{figure}

This behavior can be problematic for an online algorithm continuously fed by streaming data since, in many practical scenarios, the system operating conditions (e.g., the underlying state of nature as in the introductory example, or the network topology, the quality of data, the statistical models,...) are reasonably expected to undergo some changes over time.
For this reason, a good learning algorithm must be able to {\em adapt} to drifts in the streaming information collected by the agents.
This work proposes an Adaptive Social Learning (ASL) strategy to fill this gap. 
One instance of such strategy is shown in Fig.~\ref{fig:exampleasl} with reference to the same example from Fig.~\ref{fig:examplesl}. We see that the ASL algorithm reacts much faster (almost instantly) and is able to track the target change at instant $i\approx 200$, {\em exhibiting an adaptation capacity that is remarkably higher than that of the classic social learning algorithm}.

There are at least two advantages in devising the ASL algorithm.
The first one is related to a modeling perspective. 
As already indicated, the existing social learning strategies are not able to endow agents with adaptation abilities whereas the proposed ASL model will be able to do so. The second implication is related to a designing perspective. 
Social learning algorithms are useful not only in modeling opinion formation over social networks. They are also useful in designing man-made engineered systems (such as robotic swarms) tasked to solve decision problems collectively. Endowing such systems with adaptation abilities is critical for many applications. 

The main contributions of this work are as follows. 
First, we introduce a novel social learning strategy that enables adaptation. 
Then, we provide an accurate analytical characterization of this strategy. 
In particular, by exploiting recent advances in the field of distributed detection over adaptive networks --- see~\cite{MattaSayedCoopGraphSP2018} for an overview --- we furnish a detailed characterization of the social learning performance at each individual agent, in terms of $i)$ convergence of the system at the steady-state (Theorem~\ref{theor:steady}); $ii)$ achievability of consistent learning (Theorem~\ref{theor:weaklaw}); $iii)$ a Gaussian approximation for the learning performance (Theorem~\ref{theor:CLT}); $iv)$ the error exponents for the learning error probabilities (Theorem~\ref{theor:LD}); and $v)$ the transient evolution for the instantaneous error probabilities. As the analysis will show, the ASL model allows the user to design the adaptation time, at the expense of losing learning accuracy, i.e., agents no longer achieve full confidence around the true hypothesis. Instead, agents maintain some skepticism regarding the true hypothesis, as illustrated in the belief curves of Fig.~\ref{fig:exampleasl}.

\section{Background and Problem Formulation}
The agents of the network collect streaming observations (or data) about a phenomenon of interest. 
Agent $k=1,2,\ldots,N$, at time epoch $i=1,2,\ldots$, collects a ``private'' observation $\bm{\xi}_{k,i}$ belonging to a certain space $\mathcal{X}_k$. The qualification ``private'' comes from the assumption that the raw observations cannot be shared among agents in order to, for example, minimize communication costs or preserve secrecy.
The dependence of the space $\mathcal{X}_k$ upon $k$ allows for a possible heterogeneity in the types of data at the different agents. 
The data will be assumed statistically independent over time, i.e., over the index $i$, whereas they can be dependent across agents.\footnote{Some of the forthcoming results (Theorems~\ref{theor:CLT} and~\ref{theor:LD}) will be proved under the additional assumption of independence across agents.}

In social learning, since the inter-agent dependence is usually not known to the agents, the focus is on {\em marginal} distributions, i.e., on the distribution pertaining to any {\em individual} agent. 
Specifically, it is assumed that the distribution of $\bm{\xi}_{k,i}$ belongs to a set of $H$ admissible models that are identified by a discrete parameter (or hypothesis) $\theta\in\Theta=\{1,2,\ldots,H\}$. 
The likelihood of agent $k$ evaluated at $\theta$ is denoted by:
\beq
L_k(\xi|\theta),~~\xi\in\mathcal{X}_k.
\eeq
The presence of subscript $k$ highlights that the likelihoods are allowed to vary across the agents.
In our treatment, $L_k(\xi|\theta)$ (regarded as a function of $\xi$) can be either a probability density or mass function, depending on whether $\bm{\xi}_{k,i}$ is continuous or discrete, respectively.
Moreover, in order to avoid trivialities we assume the following regularity condition on Kullback-Leibler (KL) divergences~\cite{CT}. 
\begin{assumption}[{\bf Finiteness of KL divergences}]
	\label{assum:integrable}
	For each $k=1,2,\ldots,N$ and each pair of distinct hypotheses $\theta$ and $\theta'$, the Kullback-Leibler divergence between $L_k(\xi|\theta)$ and $L_k(\xi|\theta')$ is finite.
	~\hfill$\square$
\end{assumption}

In social learning implementations, the two main objects of the learning process are: an intermediate belief $\bm{\psi}_{k,i}(\theta)$, which each agent $k$ shares at time $i$ with its neighbors; and the belief $\bm{\mu}_{k,i}(\theta)$, which agent $k$ obtains at time $i$ by combining the intermediate beliefs received from its neighbors. 
For the algorithm initialization, we assume the following standard condition.
\begin{assumption}[{\bf Positive initial beliefs}]
	\label{assum:initbel}
	All agents start with a strictly positive belief for all hypotheses, i.e., $\bm{\mu}_{k,0}(\theta)>0$ for each agent $k$ and all $\theta\in \Theta$.~\hfill$\square$ 
\end{assumption}

In order to capture the essence of our {\em adaptive} social learning strategy, it is useful to introduce first some background on {\em traditional} social learning. We refer in particular to the social learning strategy presented in~\cite{NedicTAC2017,Javidi,MattaSantosSayedICASSP2019,MattaBordignonSantosSayed2019}, which is a two-step algorithm that iterates over time as follows. 

In the first step, each agent $k$ constructs an {\em intermediate} belief vector $\bm{\psi}_{k,i}$ by incorporating the current observation $\bm{\xi}_{k,i}$ into the belief of the preceding time epoch, $\bm{\mu}_{k,i-1}$, through the following {\em Bayesian update}:
\beq
\bm{\psi}_{k,i}(\theta)=\displaystyle{
	\frac{\bm{\mu}_{k,i-1}(\theta)L_k(\bm{\xi}_{k,i}|\theta)}
	{\sum_{\theta'\in\Theta}\bm{\mu}_{k,i-1}(\theta')L_k(\bm{\xi}_{k,i}|\theta')}
},
\label{eq:ASLintermold}
\eeq
where the denominator is a normalization factor that makes $\bm{\psi}_{k,i}$ a probability vector.

In the second step, each agent $k$ aggregates into its own current belief $\bm{\mu}_{k,i}$ the intermediate beliefs received from its neighbors by combining linearly the {\em logarithm} of the received intermediate beliefs, and then using exponentiation and normalization to get back an admissible probability vector. Specifically, each agent $k$ at time $i$ applies the following combination rule:
\beq
\bm{\mu}_{k,i}(\theta)=\displaystyle{
	\frac{\exp\Big\{\sum_{\ell\in\mathcal{N}_k}a_{\ell k}\log\bm{\psi}_{\ell,i}(\theta)\Big\}}
	{\sum_{\theta'\in\Theta}
		\exp\Big\{\sum_{\ell\in\mathcal{N}_k}a_{\ell k}\log\bm{\psi}_{\ell,i}(\theta')\Big\}
	}
},
\label{eq:ASLfinal}
\eeq
using a collection of convex combination weights:
\beq
0<a_{\ell k}<1,~~\sum_{\ell=1}^N a_{\ell k}=1,~~a_{\ell k}=0 \textnormal{ for }\ell\notin\mathcal{N}_k,
\eeq
where $\mathcal{N}_k$ denotes the neighborhood of agent $k$, with $k$ itself being included. 
The combination weights can be conveniently arranged into the {\em nonnegative and left-stochastic} combination matrix $A=[a_{\ell k}]$. 

In the forthcoming treatment, we assume that the network is strongly connected (i.e., for any two nodes $\ell$ and $k$, there exists always a path with nonzero weights linking them in both directions, and at least one node in the network has a self-loop, i.e., $a_{kk}>0$ for at least one agent $k$)~\cite{Sayed}. 
Under these assumptions, the (nonnegative and left-stochastic) matrix $A$ is {\em primitive}, implying, in view of the Perron-Frobenius theorem, that the Perron eigenvector $\pi$ associated with the matrix $A$ has all strictly positive entries~\cite{Sayed}[Lemma~F.4, p.~775]:
\beq
A\pi=\pi,~~\sum_{\ell=1}^N\pi_{\ell}=1,~~\pi_{\ell}>0 \textnormal{ for all }\ell=1,2,\ldots,N,
\eeq
and that the columns of the matrix powers $A^m$ converge, as $m\rightarrow\infty$, to the Perron eigenvector at an exponential rate governed by the second largest-magnitude eigenvalue of $A$, as stated in the following property~\cite{Johnson-Horn}[Th.~8.5.1, p.~516].
\begin{property}[{\bf Convergence of matrix powers}]
	\label{prop:Perron}
	Let $\beta_2$ be the second largest-magnitude eigenvalue of $A$. Then, for any positive $\beta$ such that $|\beta_2|<\beta<1$, there exists a positive constant $\kappa$ (depending only on $A$ and $\beta$), such that, for all $\ell,k=1,2,\ldots,N$, and for all $m=1,2,\ldots$, we have that:
	\beq
	\Big|
	[A^m]_{\ell k} - \pi_{\ell}
	\Big|
	\leq
	\kappa \beta^m.
	\label{eq:Perronbound}
	\eeq
	~\hfill$\square$
\end{property}

In traditional social learning, a stationary setting is assumed where the data collected by the agents are generated from one particular model (the {\em true} hypothesis) and the goal of social learning is to let the agents learn this hypothesis from the data. 
It has been shown that, under the aforementioned assumptions, the algorithm described by \eqref{eq:ASLintermold}--\eqref{eq:ASLfinal} leads each agent to learn the true hypothesis almost surely as $i\rightarrow\infty$~\cite{NedicTAC2017,Javidi,MattaSantosSayedICASSP2019,MattaBordignonSantosSayed2019}.

In the adaptive context, the statistical conditions governing the data can change over time. 
While effective learning must still be guaranteed under stationary conditions, it is also critical to guarantee that the learning algorithms are able to react fast to drifting conditions. 
Under these changing environments, traditional social learning algorithms do not perform well. 
The fundamental goal is therefore to devise a social learning algorithm that allows agents to promptly react to these drifts and start learning the ``new'' model.

\subsection{Adaptive Social Learning}
An \emph{adaptive} algorithm can be broadly defined as one that is tailored for environments in which the characteristics of the collected data might drift over time. The adaptive algorithm should be able to react promptly in view of these changes and deliver proper inference performance in a reasonable reaction time. 

While certainly intuitive, the above description of adaptation is only qualitative. 
In this work, we will present a rigorous analysis of the adaptation/learning trade-off in social learning. 
To this aim, it is necessary to identify formally the concepts of adaptation and learning, and the technical framework that will be used to characterize these concepts. 

{\em - Learning}. 
In the context of social learning, ``learning" means ``guessing the right hypothesis".
In order to quantify the learning performance, we specialize the standard prescriptions of adaptation theory to the social learning context. 
Given that the data are steadily generated according to a certain true likelihood model, what is the probability that an agent guesses the true state of nature?
In the theory of adaptation, this analysis is commonly referred to as {\em steady-state} analysis~\cite{Sayed}. 

{\em - Adaptation}. 
Assume that the system has been in operation for an arbitrary time. 
During this time, several phenomena can have occurred, i.e., variations of the true hypothesis, variations in the statistical conditions (i.e., malfunctioning of the system giving rise to distributions different from the nominal ones), missing observations, and so on. Due to the recursive nature of the social learning algorithms, at a given time $i_0$ all these variations are simply summarized in a certain initial belief vector $\mu_{i_0}$. From $i_0+1$ onward, assume that the system becomes stable and the data are steadily generated according to a given likelihood model.
Accordingly, the adaptation ability will be quantified by measuring how long it takes ({\em adaptation time}), given an arbitrary initial belief $\mu_{i_0}$, for an agent to enter the steady-state regime and reach a prescribed probability of guessing the true hypothesis.  
In the theory of adaptation, this analysis is commonly referred to as {\em transient} analysis~\cite{Sayed}.

\section{ASL Strategy}
Examining \eqref{eq:ASLintermold}, we see that the classic Bayesian update incorporates the new information into the past belief by giving {\em equal} weight to both $\bm{\mu}_{k,i-1}$ and the likelihood of the new data $L_k(\bm{\xi}_{k,i}|\theta)$. 
In order to promote adaptation, it is necessary to increase the relative credit given to the new data with respect to the belief accumulated over time by learning from past data. 
To this end, we turn the update step in \eqref{eq:ASLintermold} into the following {\em adaptive} form: 
\beq
\bm{\psi}_{k,i}(\theta)=\displaystyle{
	\frac{\bm{\mu}^{1-\delta}_{k,i-1}(\theta)L^{\delta}_k(\bm{\xi}_{k,i}|\theta)}
	{\sum_{\theta'\in\Theta}\bm{\mu}^{1-\delta}_{k,i-1}(\theta')L^{\delta}_k(\bm{\xi}_{k,i}|\theta')}
},
\label{eq:ASLinterm}
\eeq
where $0<\delta<1$ is a design parameter employed by each agent to modulate the relative weights assigned to the past and new information. 
In particular, relatively large values for $\delta$ give more importance to the new data, whereas small values for $\delta$ give more importance to the past beliefs. 
In this way, as we will show later in Sec.~\ref{sec:transient}, the step-size parameter $\delta$ infuses the social learning algorithm with an adaptation mechanism.

\subsection{ASL -- Stochastic Gradient Interpretation}
While the role of the parameter $\delta$ in favoring adaptation might be intuitive, the rationale behind the particular update rule in \eqref{eq:ASLinterm} might not be. 
In order to clarify this important issue, let us focus on the logarithmic belief ratio between two hypotheses $\theta$ and $\theta'$:
\beq
\log\frac{\bm{\mu}_{k,i}(\theta)}{\bm{\mu}_{k,i}(\theta')}.
\eeq
This quantity is the essential building block in the process of opinion formation, since an agent would opt for the hypothesis that maximizes the belief. In other words, agent $k$ at time $i$ would opt for opinion $\theta$ if its log-belief ratio $\log\frac{\bm{\mu}_{k,i}(\theta)}{\bm{\mu}_{k,i}(\theta')}$ is positive for all $\theta'\neq\theta$.
Combining \eqref{eq:ASLinterm} with \eqref{eq:ASLfinal} and developing the associated recursion we can obtain the time-evolution of the log-belief ratio as:
\beqa
\lefteqn{\log\frac{\bm{\mu}_{k,i}(\theta)}{\bm{\mu}_{k,i}(\theta')}=
	(1-\delta)\sum_{\ell\in\mathcal{N}_k}a_{\ell k}\log\frac{\bm{\mu}_{\ell,i-1}(\theta)}{\bm{\mu}_{\ell,i-1}(\theta')}}\nonumber\\
&+&\delta \sum_{\ell\in\mathcal{N}_k}a_{\ell k}\log\frac{L_{\ell}(\bm{\xi}_{\ell,i}|\theta)}{L_{\ell}(\bm{\xi}_{\ell,i}|\theta')}\nonumber\\
&=&
(1-\delta)^i\sum_{\ell=1}^N[A^i]_{\ell k}\log\frac{\bm{\mu}_{\ell,0}(\theta)}{\bm{\mu}_{\ell,0}(\theta')}\nonumber\\
&+&\delta \sum_{m=0}^{i-1}\sum_{\ell=1}^N (1-\delta)^m [A^{m+1}]_{\ell k}\log\frac{L_{\ell}(\bm{\xi}_{\ell,i-m}|\theta)}{L_{\ell}(\bm{\xi}_{\ell,i-m}|\theta')}.
\label{eq:ASLrationale1}
\eeqa
It is straightforward to prove that the time-evolution of the log-belief ratio in \eqref{eq:ASLrationale1} has the form of a {\em distributed stochastic gradient} algorithm with step-size $\delta$ and with quadratic cost function --- see~\cite{MattaSayedCoopGraphSP2018,Sayed}. In recent studies, it has been shown how this form of distributed adaptive inference can be useful for distributed binary hypothesis testing~\cite{MattaSayedCoopGraphSP2018,MattaBracaMaranoSayedTIT2016,MattaBracaMaranoSayedTSIPN2016}. 
These studies motivate our choice for \eqref{eq:ASLinterm}, whose goodness for social learning will be rigorously established in the forthcoming analysis.

\subsection{ASL -- Bayesian Update Interpretation}
Examining \eqref{eq:ASLinterm}, we see that the ASL strategy implements a convex combination of probability functions at the exponent, by discounting both the past belief and the new likelihood through the weights $1-\delta$ and $\delta$, respectively. However, the update \eqref{eq:ASLinterm} cannot be considered a {\em Bayesian} update because the likelihood exponentiated to $\delta$ does not integrate to one (w.r.t. $\xi$). 
Notably, the same problem does not occur for the discounted belief, since it can be normalized, both at the numerator and denominator of \eqref{eq:ASLinterm}, by dividing by $\sum_{\theta'\in\Theta}\bm{\mu}^{1-\delta}_{k,i-1}(\theta')$. Exploiting the latter property, we now show how it is possible to modify \eqref{eq:ASLinterm} to get an {\em adaptive Bayesian} update. 
First, agent $k$ takes the past belief $\bm{\mu}_{k,i-1}(\theta)$ and builds a new belief as:
\beq
\widetilde{\bm{\mu}}_{k,i-1}(\theta)=
\displaystyle{
	\frac
	{\bm{\mu}^{1-\delta}_{k,i-1}(\theta)}
	{\sum_{\theta'\in\Theta}\bm{\mu}^{1-\delta}_{k,i-1}(\theta')}
}.
\label{eq:flattenedbel}
\eeq
Second, agent $k$ implements the Bayesian update rule in \eqref{eq:ASLintermold} by replacing $\bm{\mu}_{k,i-1}(\theta)$ with $\widetilde{\bm{\mu}}_{k,i-1}(\theta)$, which yields:
\beq
\bm{\psi}_{k,i}(\theta)=\frac{\widetilde{\bm{\mu}}_{k,i-1}(\theta)L_k(\bm{\xi}_{k,i}|\theta)}{\sum_{\theta'\in\Theta}\widetilde{\bm{\mu}}_{k,i-1}(\theta')L_k(\bm{\xi}_{k,i}|\theta')}.
\label{eq:altbayes}
\eeq
Notably, the exponentiation and normalization in \eqref{eq:flattenedbel} has the physical meaning of {\em flattening} the belief vector, i.e., of making it more uniform across $\theta$. In this way, if an agent had a particularly peaked belief around a certain hypothesis, perhaps due to a bias accumulated over time, flattening the belief helps to give more credit to new data.  
Observe from \eqref{eq:altbayes} that the limiting choice $\delta=0$ (i.e., no adaptation) gives back the classic Bayesian update in \eqref{eq:ASLintermold}. 
In contrast, the update in \eqref{eq:ASLinterm} cannot be reduced to \eqref{eq:ASLintermold} for any selection of $\delta\in(0,1)$.
This notwithstanding, we now show that the ASL strategies \eqref{eq:ASLinterm} and \eqref{eq:altbayes} are in fact equivalent. 
To this aim, we can develop the recursion obtained by combining \eqref{eq:ASLfinal}, \eqref{eq:flattenedbel}, and \eqref{eq:altbayes} and get:
\beqa
\lefteqn{\log\frac{\bm{\mu}_{k,i}(\theta)}{\bm{\mu}_{k,i}(\theta')}=
	(1-\delta)\sum_{\ell\in\mathcal{N}_k}a_{\ell k}\log\frac{\bm{\mu}_{\ell,i-1}(\theta)}{\bm{\mu}_{\ell,i-1}(\theta')}}\nonumber\\
&+&\sum_{\ell\in\mathcal{N}_k}a_{\ell k}\log\frac{L_{\ell}(\bm{\xi}_{\ell,i}|\theta)}{L_{\ell}(\bm{\xi}_{\ell,i}|\theta')}\nonumber\\
&=&
(1-\delta)^i\sum_{\ell=1}^N[A^i]_{\ell k}\log\frac{\bm{\mu}_{\ell,0}(\theta)}{\bm{\mu}_{\ell,0}(\theta')}\nonumber\\
&+&\sum_{m=0}^{i-1}\sum_{\ell=1}^N (1-\delta)^m [A^{m+1}]_{\ell k}\log\frac{L_{\ell}(\bm{\xi}_{\ell,i-m}|\theta)}{L_{\ell}(\bm{\xi}_{\ell,i-m}|\theta')}.
\label{eq:ASLrationale2}
\eeqa
Let us now compare \eqref{eq:ASLrationale1} against \eqref{eq:ASLrationale2}. We see that in both equations there is a term that dies out exponentially fast with time, and which is due to the initialization term $\log\frac{\bm{\mu}_{\ell,0}(\theta)}{\bm{\mu}_{\ell,0}(\theta')}$. We remark that this initialization term is determined by the beliefs set by the agents in the {\em absence of data}. Thus, it is zero if the agents set a uniform, non-informative prior, or it can be non-uniform if the agents have unbalanced prior convictions. 
The relevant term that determines the algorithms' evolution over time is given by the double summations appearing in \eqref{eq:ASLrationale1} and \eqref{eq:ASLrationale2}. Comparing these summations, we see that they differ only by a scaling factor $\delta$. 
As a result, we conclude that the time-evolution of the log-belief ratios for the two ASL strategies is equivalent. For example, the opinion that maximizes the belief function would be the same under both strategies, implying the same error probability. In fact, proportionality of the log-belief ratios implies that the belief function of one strategy is simply an exponentiated (and normalized) version of the belief function of the other strategy. This does not mean that the beliefs of the two strategies would take on the same values. In particular, our results will show that, as $\delta\rightarrow 0$, the steady-state log-belief ratios are stable under \eqref{eq:ASLinterm}, which immediately implies that they diverge (i.e., achieving a belief close to $1$ at the true hypothesis) under \eqref{eq:altbayes}. While immaterial from a technical perspective, these differences might matter from a {\em behavioral} perspective~\cite{Molavi2018}, namely, to understand which update strategy reflects better the way of reasoning that an individual agent uses in social learning environments. For the sake of clarity, in the presentation of our technical results we opt for sticking to the update rule in \eqref{eq:ASLinterm}, since it automatically stabilizes the log-belief ratio without necessity of additional scaling factors.

The adaptation properties of the ASL strategy are enabled by a learning mechanism that is fundamentally different from that of classic social learning. To see why, let us assume that the true hypothesis remains stable for a sufficiently long time interval. 
Different from what happens in classic social learning --- e.g., in \eqref{eq:ASLintermold} --- in the ASL strategy the belief will {\em not} converge as time $i$ increases. 
In contrast, the belief will {\em vary} indefinitely, preserving a random behavior also in the steady state. 
The learning performance will then be assessed by examining the statistical behavior of the beliefs in steady state. 
We will provide an accurate characterization of such statistical behavior in the regime of small step-sizes, i.e., by performing an asymptotic analysis as $\delta\rightarrow 0$. Under this regime, we will show that the probability of guessing the right hypothesis approaches $1$ for sufficiently small step-sizes. We will furthermore characterize the transient performance by obtaining closed-form relationships that reveal how the adaptation time grows with smaller step-sizes. The overall analysis will highlight well the adaptation/learning trade-off: small (resp., large) values of $\delta$ mean less (resp., more) adaptation and higher (resp., lower) learning accuracy. 

\section{Statistical Descriptors of the Learning Performance}
Assume that the algorithm has been running until a certain time $i_0$, with the evolution of the system up to $i_0$ being summarized in the ``initial'' belief vectors $\bm{\mu}_{k,i_0}$.
Starting from $i_0$, the ASL algorithm behavior will exhibit two important phases: a {\em transient} phase where, given the (possibly wrong) initial belief, each agent must suddenly adapt in order to depart from $\bm{\mu}_{k,i_0}$ and start learning the correct hypothesis; and a {\em steady-state} phase where, given sufficient time to learn ($i\rightarrow\infty$), each agent must achieve high confidence in learning the correct hypothesis. 
According to the theory of adaptive inference, the performance of an adaptive learning strategy is characterized under the {\em steady-state} regime.

The following property is relevant for steady-state analysis. 
By examining the algorithm recursions \eqref{eq:ASLfinal}--\eqref{eq:ASLinterm}, it is straightforward to see that, in light of Assumptions~\ref{assum:integrable} and~\ref{assum:initbel}, the belief remains always nonzero at any $\theta$ during the algorithm evolution.\footnote{This property follows by induction once we observe that, starting from a belief that is nonzero at any $\theta$: $i)$ the intermediate belief remains nonzero at any $\theta$ because the likelihoods in the update step \eqref{eq:ASLinterm} cannot be zero (but for an ensemble of zero probability) otherwise Assumption~\ref{assum:integrable} would be violated; and $ii)$ the final belief in \eqref{eq:ASLfinal} remains nonzero at any $\theta$ since the combination weights are convex.} 
Now, assume that the algorithm has been running up to time $i_0$, and that from $i_0+1$ onward the system remains stationary for sufficiently long time, with the data being generated according to hypothesis $\theta_0$. 
In order to perform a steady-state analysis from $i_0+1$ onward, we need to consider $\bm{\mu}_{k,i_0}$ as initial state. Since we have observed that the beliefs are always nonzero, we can see that the initial belief vector $\bm{\mu}_{k,i_0}$ fulfills Assumption~\ref{assum:initbel}.

In summary, for the purpose of the steady-state analysis and without loss of generality, we will assume that the steady-state analysis starts at time $i_0=0$ and consider an initial belief vector $\bm{\mu}_{k,0}$ that fulfills Assumption~\ref{assum:initbel}. 
The true hypothesis $\theta_0$ is kept constant over time, yielding:
\beq
\bm{\xi}_{k,i}\sim L_k(\xi|\theta_0),~~k=1,2,\ldots,N, ~~i=1,2,\ldots
\eeq
Therefore, for the purpose of the steady-state analysis, we will always imply that expectations and probabilities are evaluated under the distributions $L_k(\xi|\theta_0)$. 
Note also that, under the steady-state regime, the data $\{\bm{\xi}_{k,i}\}$ are independent and identically distributed (i.i.d.) over time, i.e., over the index $i$. 
We will assume that they can have different distributions across the agents, i.e., across the index $k$. 
Statistical independence across the agents will be only used to prove some of the forthcoming results (Theorems~\ref{theor:CLT} and~\ref{theor:LD} further ahead).

\subsection{Log-Belief Ratios and Error Probabilities}
In order to characterize the learning performance, it is convenient to introduce the logarithm of the ratio between the belief evaluated at $\theta_0$ and the belief evaluated at a generic hypothesis $\theta\neq\theta_0$:
\beq
\bm{\lambda}^{(\delta)}_{k,i}(\theta)\triangleq \log\frac{\bm{\mu}_{k,i}(\theta_0)}{\bm{\mu}_{k,i}(\theta)},
\label{eq:logbeldef}
\eeq
which is well-defined since, as already remarked, the belief remains nonzero at any $\theta$ during the algorithm evolution.
Before continuing, it is important to make a notational remark. 
With the symbol $\bm{\lambda}^{(\delta)}_{k,i}(\theta)$ we denote a {\em random} (bold notation) function of: the agent index $k=1,2,\ldots,N$, the time index $i=0,1,\ldots$, the hypothesis $\theta\in\Theta\setminus{\theta_0}$, and the adaptation parameter $\delta$. 
When we omit the argument $\theta$ and write $\bm{\lambda}^{(\delta)}_{k,i}$, we will be referring to the $(H-1)\times 1$ vector of log-belief ratios, namely,
\beq
\bm{\lambda}^{(\delta)}_{k,i}=\left[\bm{\lambda}^{(\delta)}_{k,i}(\theta_1),\bm{\lambda}^{(\delta)}_{k,i}(\theta_2),\ldots,\bm{\lambda}^{(\delta)}_{k,i}(\theta_{H-1})\right]\T,
\label{eq:logbelvec}
\eeq 
where the elements in the set of wrong-hypotheses have been indexed as:
\beq
\Theta\setminus{\theta_0}=\{\theta_1,\theta_2,\ldots,\theta_{H-1}\}.
\label{eq:wrongset}
\eeq 
One natural way for the agents to choose a hypothesis is to select the hypothesis that maximizes the belief. 
Therefore, the error probability at each time $i$ can be expressed as
\beq
p^{(\delta)}_{k,i}=\P\left[\argmax_{\theta\in\Theta}\bm{\mu}_{k,i}(\theta)\neq\thetatrue\right].
\label{eq:perrfirstdef}
\eeq
It is useful to rewrite the error probability as a function of the log-belief ratios. 
To this end, observe that the event within brackets in \eqref{eq:perrfirstdef} corresponds to saying that the belief is not maximized at $\thetatrue$, which in turn corresponds to saying that the log-belief ratios in \eqref{eq:logbeldef} are less than or equal to zero for at least one $\theta\neq\thetatrue$. Therefore, the \emph{instantaneous} error probability can be equivalently rewritten as:
\beq
p^{(\delta)}_{k,i}=\P\left[\exists \theta\neq\thetatrue: \bm{\lambda}^{(\delta)}_{k,i}(\theta)\leq 0\right].
\label{eq:errprob}
\eeq
Finally, we introduce the {\em steady-state} error probability:
\beq
p_k^{(\delta)}\triangleq\lim_{i\rightarrow\infty} p_{k,i}^{(\delta)}.
\label{eq:steadyprob}
\eeq
There are two fundamental questions related to the concept of steady-state error probability. 
The first question regards its {\em existence}, which is in principle not guaranteed. Theorem~\ref{theor:steady} will provide an affirmative answer to this question by characterizing the steady-state behavior of the log-belief ratios. 
The second question regards the {\em evaluation} of the steady-state error probability. 
An exact evaluation is generally a formidable task. Therefore, to tackle this critical problem, we will perform an asymptotic analysis in the regime of small $\delta$, which will allow us to obtain reliable predictions of the steady-state performance.

\begin{figure}[t]
	\centering
	\includegraphics[width=.95\linewidth]{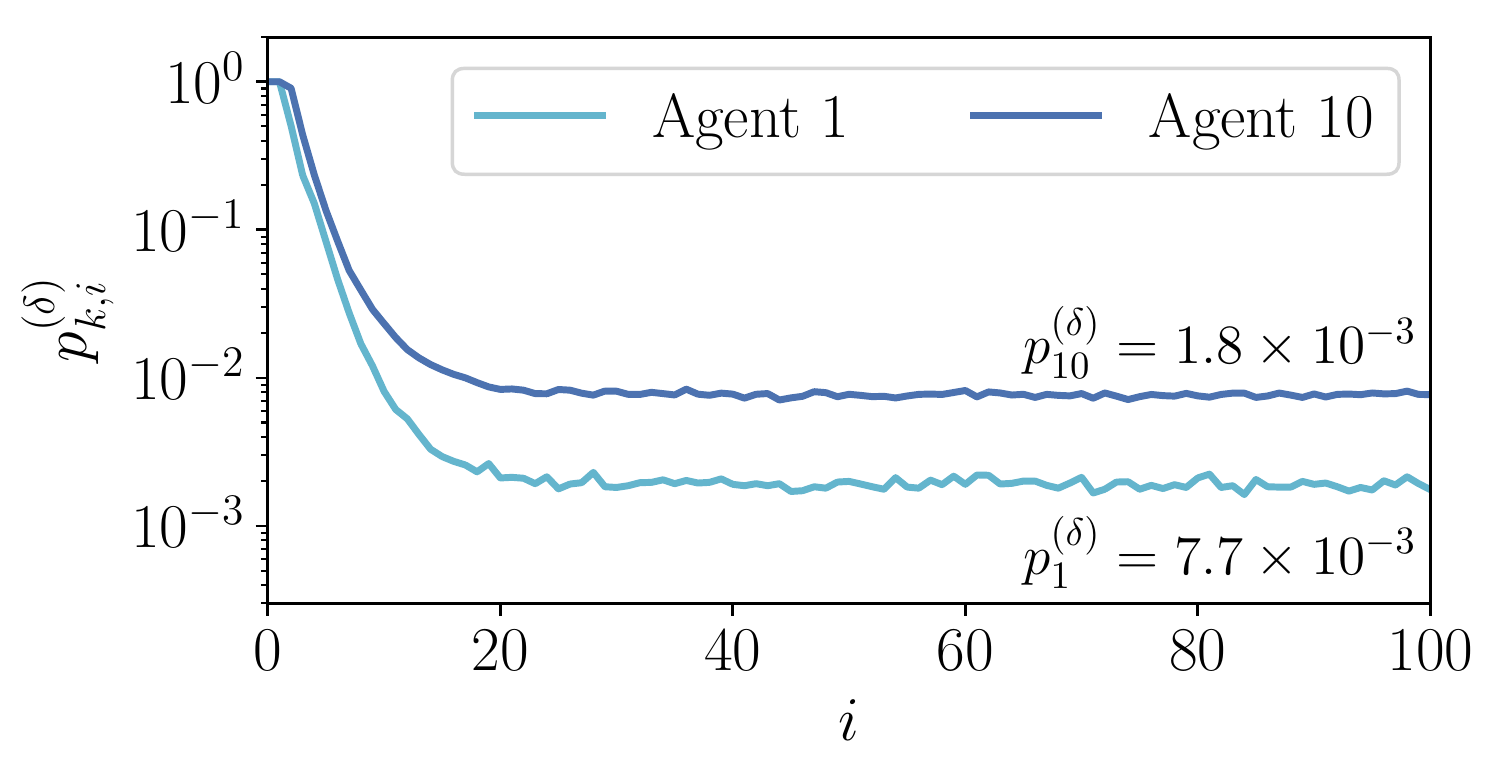}
	\caption{Evolution of the error probability of two agents in a network running the ASL algorithm.}
	\label{fig:pe_ex}
\end{figure}

In Fig.~\ref{fig:pe_ex} we show an example of evolution for the error probability of two agents in a network implementing the ASL strategy.\footnote{The details of the network topology as well as of the statistical learning problem are immaterial at this stage of the presentation.} All the probabilities are estimated empirically by Monte Carlo simulation. 
We see how the instantaneous error probability $p_{k,i}^{(\delta)}$ converges to a steady-state {\em nonzero} value $p_k^{(\delta)}$ as $i$ increases. It is useful to remark that this behavior is different from that of classic social learning, where, {\em under stationary conditions}, the error probability of each agent vanishes as time elapses. 
This is one instance of the adaptation/learning trade-off: non-adaptive strategies can increase their accuracy indefinitely under stationary conditions. However, astronomically low values of the error probabilities lead to a detrimental inertia in responding to possible changes.

\subsection{Log-Likelihood Ratios}
For $k=1,2,\ldots,N$, $i=0,1,\ldots$, and $\theta\neq\theta_0$, we introduce the log-likelihood ratio:
\beq
\bm{x}_{k,i}(\theta)\triangleq \log\frac{L_k(\bm{\xi}_{k,i}|\thetatrue)}{L_k(\bm{\xi}_{k,i}|\theta)},
\label{eq:xkidefin}
\eeq
and its expectation: 
\beq
d_k(\theta)\triangleq\E[\bm{x}_{k,i}(\theta)]<\infty,
\label{eq:delldef}
\eeq
namely, the KL divergence between $L_k(\xi|\thetatrue)$ and $L_k(\xi|\theta)$, which is finite in view of Assumption~\ref{assum:integrable}, implying that the log-likelihood ratios cannot diverge (but for an ensemble of realizations with zero probability). We recall that the expectation in \eqref{eq:delldef} is computed assuming that the random variable $\bm{\xi}_{k,i}$ is distributed according to model $L_k(\xi|\thetatrue)$. Since we focus on the steady state, this distribution is constant over time, which explains why $d_k(\theta)$ does not depend on $i$. 
Furthermore, since the true hypothesis $\theta_0$ is held fixed during the steady-state analysis, in order to avoid a heavier notation we are not emphasizing the dependence of the KL divergence $d_k(\theta)$ on $\theta_0$.

We continue by introducing an average variable that will play a role in the forthcoming results, namely, the {\em network average} of log-likelihood ratios, for all $\theta\neq\theta_0$:
\beq
\xnet(\theta)=\sum_{\ell=1}^N \pi_{\ell} \bm{x}_{\ell,i}(\theta).
\label{eq:avlik}
\eeq
The random variable $\xnet(\theta)$ appearing in \eqref{eq:avlik} is obtained by combining linearly the local log-likelihood ratios $\bm{x}_{\ell,i}(\theta)$. 
The combination weight assigned to the log-likelihood ratio of the $\ell$-th agent is given by the {\em limiting} combination weight, i.e., by the $\ell$-th entry, $\pi_{\ell}$, of the Perron eigenvector. 
We will see in the following that the asymptotic properties of the ASL strategy as $\delta\rightarrow 0$ are directly related to the statistical properties of the vector of average variables, $\xnet$.

\section{Steady-State Analysis}
As we have remarked in the introduction, different from the classic social learning setting, in the {\em adaptive} setting the belief will not converge as $i\rightarrow\infty$. In contrast, the belief of each agent will preserve a {\em random} behavior. This everlasting randomness is critical to ensure that the algorithm will adapt quickly to a change in the environment.
On the other hand, it makes the steady-state analysis more difficult, since the beliefs preserve a random character even when $i\rightarrow\infty$. 
In order to carry out a meaningful steady-state analysis, the fundamental preliminary step becomes then to establish whether such random fluctuations lead to {\em stable} random variables as $i\rightarrow\infty$. 
Theorem~\ref{theor:steady} further ahead ascertains that this is the case.  

Before stating the theorem, let us examine the evolution of the log-belief ratios. 

Exploiting \eqref{eq:ASLfinal} and \eqref{eq:ASLinterm}, we end up with the following recursion, for every $\theta\neq\theta_0$:
\begin{equation}
\bm{\lambda}^{(\delta)}_{k,i}(\theta)=\sum_{\ell\in\mathcal{N}_k} a_{\ell k}\left\{
(1-\delta)\bm{\lambda}^{(\delta)}_{\ell,i-1}(\theta) 
+ 
\delta \bm{x}_{\ell,i}(\theta)
\right\},
\label{eq:mainASLrec}
\end{equation}
which can be rewritten as the following two-step recursion:
\begin{IEEEeqnarray}{rCl}
	\bm{\nu}_{\ell,i}^{(\delta)}(\theta)&=&(1-\delta)\bm{\lambda}^{(\delta)}_{\ell,i-1}(\theta) 
	+ 
	\delta\,\bm{x}_{\ell,i}(\theta),\label{eq:buASLrec}\\\bm{\lambda}^{(\delta)}_{k,i}(\theta)&=&\sum_{\ell\in\mathcal{N}_k} a_{\ell k}\,\bm{\nu}_{\ell,i}^{(\delta)}(\theta).
	\label{eq:combASLrec}
\end{IEEEeqnarray}
The time-evolution of the log-belief ratios in \eqref{eq:buASLrec} and \eqref{eq:combASLrec} is in the form of a {\em diffusion} 
algorithm with constant {\em step-size} $\delta$ --- see, e.g.~\cite{Sayed}.
This is why we referred to $\delta$ as the step-size.

Developing the recursion in \eqref{eq:mainASLrec} and recalling that $A=[a_{\ell k}]$ is the combination matrix we can write, for all $\theta\neq\theta_0$:
\beqa
\bm{\lambda}^{(\delta)}_{k,i}(\theta)
&=&
\underbrace{(1-\delta)^i 
	\sum_{\ell=1}^N [A^i]_{\ell k} \bm{\lambda}_{\ell,0}(\theta)
}_{\textnormal{transient term}}
\nonumber\\
&+&\delta \sum_{m=0}^{i-1}\sum_{\ell=1}^N (1-\delta)^m [A^{m+1}]_{\ell k}\, \bm{x}_{\ell,i-m}(\theta).\nonumber\\
\label{eq:withtransient}
\eeqa
Since the transient term dies out as $i\rightarrow\infty$, in order to evaluate the steady-state behavior of $\bm{\lambda}_{k,i}(\theta)$, we can ignore it and focus on the second term:
\beq
\widehat{\bm{\lambda}}^{(\delta)}_{k,i}(\theta)=\delta \sum_{\ell=1}^N \sum_{m=0}^{i-1} (1-\delta)^m [A^{m+1}]_{\ell k}\, \bm{x}_{\ell,i-m}(\theta).
\label{eq:lambdarec}
\eeq

\subsection{Steady-State Log-Belief Ratios}
The goal of the steady-state analysis is to evaluate the performance (i.e., the error probability) for large $i$. 
For this evaluation to be meaningful, we must ascertain that the error probability in \eqref{eq:errprob} converges as $i\rightarrow\infty$. To this end, we will now establish that there exists a certain limiting random vector, $\widetilde{\bm{\lambda}}^{(\delta)}_k$, such that the {\em probability distribution} of the vector of log-belief ratios, $\widehat{\bm{\lambda}}^{(\delta)}_{k,i}$, converges, as $i\rightarrow\infty$, to the probability distribution of $\widetilde{\bm{\lambda}}^{(\delta)}_k$. This notion of convergence can be formally defined as follows.

We say that the sequence (over the index $i$) of random vectors $\widehat{\bm{\lambda}}^{(\delta)}_{k,i}$ converges {\em in distribution} or {\em weakly} as $i\rightarrow\infty$ if we can define a random vector $\widetilde{\bm{\lambda}}^{(\delta)}_k$ such that~\cite{shao}:
\beq
\lim_{i\rightarrow\infty}\P\left[\widehat{\bm{\lambda}}^{(\delta)}_{k,i}\in\mathcal{B}\right]=
\P\left[\widetilde{\bm{\lambda}}^{(\delta)}_k\in\mathcal{B}\right]
\label{eq:convdistdef0}
\eeq
for all measurable sets $\mathcal{B}$ whose boundary $\partial\mathcal{B}$ has zero probability under the limiting distribution, namely, for all measurable sets $\mathcal{B}$ fulfilling the condition:
\beq
\P\left[\widetilde{\bm{\lambda}}^{(\delta)}_k\in\partial\mathcal{B}\right]=0.
\eeq
In the following, weak convergence will be compactly denoted as:
\beq
\widehat{\bm{\lambda}}^{(\delta)}_{k,i}\stackrel{i\rightarrow\infty}{\rightsquigarrow} \widetilde{\bm{\lambda}}^{(\delta)}_k,
\label{eq:convdistdef1}
\eeq
and the vector $\widetilde{\bm{\lambda}}^{(\delta)}_k$ will be referred to as the {\em steady-state} log-belief vector, since it provides the statistical characterization of the log-belief vector $\widehat{\bm{\lambda}}^{(\delta)}_{k,i}$ as $i\rightarrow\infty$.

We are now ready to present the theorem that establishes the existence of steady-state log-belief ratios.
\begin{theorem}[{\bf Steady-state log-belief ratios}]
	\label{theor:steady}
	Let Assumptions~\ref{assum:integrable} and~\ref{assum:initbel} hold, and let 
	\beq
	\widetilde{\bm{\lambda}}_{k,i}^{(\delta)}(\theta)\dfz
	\delta \sum_{\ell=1}^N \sum_{m=0}^{i-1} (1-\delta)^m [A^{m+1}]_{\ell k}\, \bm{x}_{\ell,m+1}(\theta)
	\label{eq:lambdarecrevdef}
	\eeq
	be the random sum obtained from \eqref{eq:lambdarec} by taking the summands in reversed order. 
	
	First, we have that all the $N$ inner sums in \eqref{eq:lambdarecrevdef} are almost-surely absolutely convergent as $i\rightarrow\infty$, implying that $\widetilde{\bm{\lambda}}_{k,i}^{(\delta)}(\theta)$ converges almost surely to the random series:
	\beq
	\widetilde{\bm{\lambda}}_k^{(\delta)}(\theta)\dfz\delta\sum_{\ell=1}^N \sum_{m=0}^{\infty} (1-\delta)^m [A^{m+1}]_{\ell k}\, \bm{x}_{\ell,m+1}(\theta).
	\label{eq:convseries}
	\eeq
	Second, we have that the vector of log-belief ratios $\widehat{\bm{\lambda}}^{(\delta)}_{k,i}$ (with the original, i.e., non-reversed ordering of summation) converges in distribution to the vector $\widetilde{\bm{\lambda}}^{(\delta)}_k$, namely,
	\beq
	\widehat{\bm{\lambda}}^{(\delta)}_{k,i}\stackrel{i\rightarrow\infty}{\rightsquigarrow} \widetilde{\bm{\lambda}}^{(\delta)}_k.
	\label{eq:convdistheorem}
	\eeq
\end{theorem}
\begin{IEEEproof}
	See Appendix~\ref{app:Th1}.
\end{IEEEproof}

It is useful to make some comments on Theorem~\ref{theor:steady}.
First, finiteness of the expectation of $\bm{x}_{k,i}$ is sufficient (through Assumption~\ref{assum:integrable}) to guarantee the existence of a steady-state random variable. No assumption is made on higher-order moments.

\begin{figure}[t]
	\centering
	\includegraphics[width=.92\linewidth]{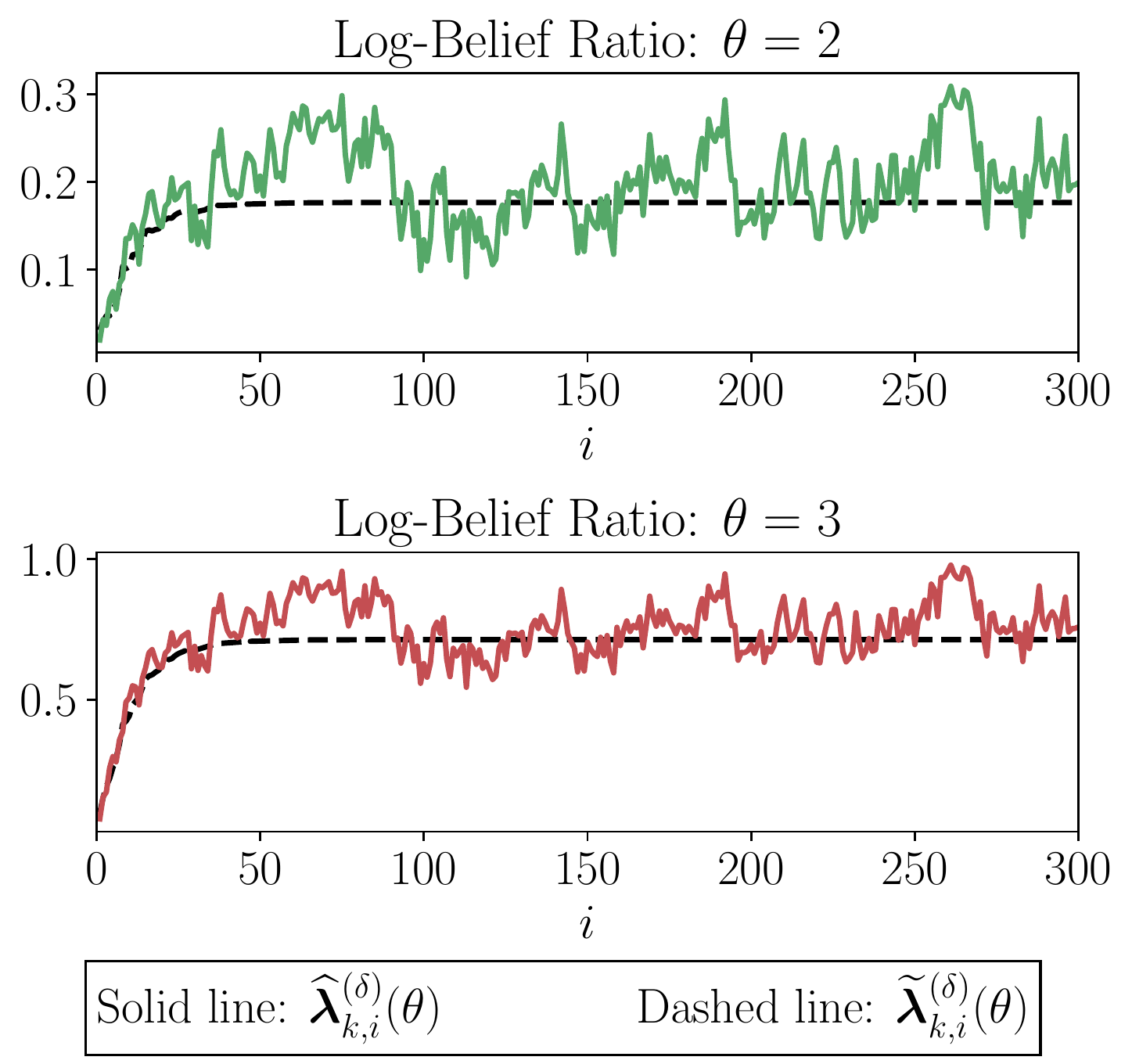}
	\caption{
		Comparison of the random sequences $\widehat{\bm{\lambda}}_{k,i}^{(\delta)}(\theta)$ and $\widetilde{\bm{\lambda}}_{k,i}^{(\delta)}(\theta)$ for $\delta=0.1$, for the Gaussian setting described in Sec.~\ref{sec:1} further ahead.}
	\label{fig:comprandsums}
\end{figure}

Second, it is important to notice that \eqref{eq:convseries} does not correspond to letting $i\rightarrow\infty$ in the summation in \eqref{eq:lambdarec}. In order to explain why, let us compare the random sums:
\beq
\widehat{\bm{\lambda}}^{(\delta)}_{k,i}(\theta)=
\delta \sum_{m=0}^{i-1}\sum_{\ell=1}^N (1-\delta)^m [A^{m+1}]_{\ell k}\, \bm{x}_{\ell,i-m}(\theta),
\label{eq:lambdarecompare}
\eeq
and
\beq
\widetilde{\bm{\lambda}}_{k,i}^{(\delta)}(\theta)=
\delta \sum_{m=0}^{i-1}\sum_{\ell=1}^N (1-\delta)^m [A^{m+1}]_{\ell k}\, \bm{x}_{\ell,m+1}(\theta).
\label{eq:lambdarecrevcompare}
\eeq
In Fig.~\ref{fig:comprandsums} we examine a sample path for these sums, and we can see that they exhibit different behavior.
The random sum in \eqref{eq:lambdarecompare}, displayed with solid line in Fig.~\ref{fig:comprandsums}, exhibits steadily random fluctuations as time elapses.
In contrast, the random sum in \eqref{eq:lambdarecrevcompare}, displayed with dashed line, converges as time elapses, specifically to the random value $\widetilde{\bm{\lambda}}^{(\delta)}_k(\theta)$ defined in \eqref{eq:convseries}.
Both behaviors are consistent with what we have already shown in Theorem~\ref{theor:steady}. 
These profoundly different behaviors depend on the different ordering of the summands in \eqref{eq:lambdarecompare} and \eqref{eq:lambdarecrevcompare}. 
In particular, in \eqref{eq:lambdarecrevcompare} the most recent term, $\bm{x}_{\ell,i}(\theta)$, takes the smallest weight $(1-\delta)^{i-1}$, which lets the remainder of the series vanish (almost surely). 
In contrast, in \eqref{eq:lambdarecompare} the most recent term, $\bm{x}_{\ell,i}(\theta)$ takes the highest weight $(1-\delta)^0=1$, thus keeping fluctuations (hence, adaptation) alive.

Even though the sums in \eqref{eq:lambdarecompare} and \eqref{eq:lambdarecrevcompare} exhibit a markedly different behavior in terms of their time-evolution (i.e., on the sample paths), one notable conclusion from Theorem~\ref{theor:steady} is that their {\em probability distributions}  converge to the same distribution, that is the distribution of the limiting variable $\widetilde{\bm{\lambda}}^{(\delta)}_k$. 
This equivalence can be explained as follows. With reference to the top panel in Fig.~\ref{fig:comprandsums}, consider a sufficiently large $i$ (say, $i=300$) and take the corresponding values of the dashed curve and the solid curve, namely, $\widehat{\bm{\lambda}}_{k,300}(2)$ and $\widetilde{\bm{\lambda}}_{k,300}(2)$. These values are different. 
However, if we now repeat the experiment in Fig.~\ref{fig:comprandsums} several times, the realizations of $\widehat{\bm{\lambda}}_{k,300}(2)$ across different experiments will be distributed in the same way as the realizations of $\widetilde{\bm{\lambda}}_{k,300}(2)$.

The existence of a limiting distribution for the log-belief vector $\widehat{\bm{\lambda}}^{(\delta)}_{k,i}$ makes the definition of a {\em steady-state} error probability meaningful, since from Eqs. \eqref{eq:errprob} and \eqref{eq:steadyprob} we see that the steady-state error probability can be computed as:\footnote{According to the definition of convergence in distribution, the result in \eqref{eq:steadyprobexplicit} holds provided that the limiting random variable $\widetilde{\bm{\lambda}}_k^{(\delta)}$ has no point mass at $0$. However, we rule out such pathological case that is in practice the exception rather than the rule.} 
\beq
p_{k}^{(\delta)}=\P\left[\exists \theta\neq\thetatrue: \widetilde{\bm{\lambda}}^{(\delta)}_k(\theta)\leq 0\right].
\label{eq:steadyprobexplicit}
\eeq
However, it should be noticed that Theorem~\ref{theor:steady} constitutes only a first, albeit fundamental step towards the characterization of the ASL performance, since it establishes only the existence of a steady-state error probability without providing any explicit characterization thereof. Such characterization is in general not available. 
In the next sections we tackle this challenging problem by focusing on an asymptotic characterization of $\widetilde{\bm{\lambda}}_k^{(\delta)}$ in the regime of small $\delta$.

\section{Small-$\delta$ Analysis}
\label{sec:smalldelta}
We have ascertained that it makes sense to define {\em steady-state} random variables characterizing the log-belief ratios. 
Then, the steady-state learning performance can be determined by examining the probability that these random variables fulfill certain conditions. For example, the steady-state probability that an agent learns the truth is the probability that the steady-state log-belief ratio of that agent is positive only at the true value $\theta_0$. 
However, in general the exact characterization of these steady-state variables is a formidable task. 
For this reason we will resort to an asymptotic analysis in the regime of small $\delta$. 
We will provide three types of asymptotic results. 
\begin{itemize}
	\item
		{\bf Sec.~\ref{sec:weaklaw}:
	Weak law of small step-sizes (Theorem~\ref{theor:weaklaw})}.  
	We will show that, for small $\delta$, the steady-state vector $\widetilde{\bm{\lambda}}^{(\delta)}_k$ concentrates around the weighted average of the agents' KL divergences defined in \eqref{eq:mnet}. This concentration property guarantees that, with high probability as $\delta\rightarrow 0$, the true hypothesis is chosen by each agent. 
	This result will require only finiteness of the first moments of the log-likelihood ratios, i.e., finiteness of the KL divergences.
	\item {\bf Sec.~\ref{sec:normal}:
	Asymptotic normality (Theorem~\ref{theor:CLT})}. We will obtain a Central Limit Theorem (CLT) that will provide a normal approximation, holding for small $\delta$, for the error probabilities of each individual agent. 
	This result will be proved assuming independence across agents and will require finiteness of the variance of the log-likelihood ratios. 
	We remark that previous results of asymptotic normality for adaptive distributed detection assumed finiteness of higher-order moments~\cite{MattaBracaMaranoSayedTIT2016}. To the best of our knowledge, the result in Theorem~\ref{theor:CLT} (which is based on part $5$ of Lemma~\ref{lem:mainlemma}) is the first result that assumes the minimal requirement of finiteness of second moments. 
	\item {\bf Sec.~\ref{sec:largedev}:
	Large deviations analysis (Theorem~\ref{theor:LD})}. We will characterize the exponential rate of decay of the error probabilities as $\delta\rightarrow 0$. This result will be proved assuming independence across agents and will require the existence of the moment generating function of the log-likelihood ratios.
\end{itemize}
Notably, the above three steps reflect perfectly a classic path in asymptotic statistics.  
However, in order to avoid misunderstandings, it is necessary to clarify one fundamental difference between the small-$\delta$ analysis and classic results. 
In order to illustrate this difference let us refer, for example, to the CLT result. 
In the traditional setting of asymptotic statistics, one examines the asymptotic behavior of sums of random variables when the number of terms of the sum goes to infinity. 
In contrast, the CLT proved in this work does {\em not} affirm that the sums involved in \eqref{eq:lambdarec} converge to a Gaussian as $i\rightarrow\infty$.
As a matter of fact, we have shown in Theorem~\ref{theor:steady} that the sums in \eqref{eq:lambdarec} converge to certain random variables, but these variables are {\em not Gaussian}, in general. 
The CLT that we prove deals instead with the behavior, as $\delta$ goes to zero, of the steady-state random vector $\widetilde{\bm{\lambda}}^{(\delta)}_k$. The same distinction applies to the other two types of asymptotic results, namely, the weak law and the large deviations analysis.
For this reason, as explained in~\cite{MattaSayedCoopGraphSP2018}, the correct way to deal with the asymptotic regime of small step-sizes in the adaptation context is made of two steps:
\begin{itemize} 
	\item
	First introduce a proper steady-state vector $\widetilde{\bm{\lambda}}^{(\delta)}_k$, which already embodies the effect of combining an infinite number of summands. This steady-state vector will be non-degenerate (i.e., no weak law as $i\rightarrow\infty$), will be non-Gaussian (i.e., no CLT as $i\rightarrow\infty$), and will be non-vanishing (i.e., no large deviations as $i\rightarrow\infty$). 
	\item
	Then, characterize the asymptotic behavior of the steady-state random vector $\widetilde{\bm{\lambda}}^{(\delta)}_k$ as $\delta$ goes to zero.
\end{itemize}
It is worth noticing that, in the adaptation literature, the critical role of the first step is usually not emphasized. 
This is because the adaptation literature mostly focuses on {\em estimation} problems, where one usually quantifies the performance by evaluating convergence of the {\em moments}~\cite{Sayed}. In contrast, when dealing with {\em decision} problems (as in our case), the performance is quantified through {\em probabilities}, namely, the probabilities of making a wrong (or correct) decision.
In order to evaluate probabilities at the steady state, it is critical to obtain first a representation of the steady-state random variables~\cite{MattaSayedCoopGraphSP2018}.

\subsection{Consistent Social Learning}\label{sec:weaklaw}
We will establish that the ASL strategy achieves consistent social learning under the following standard assumption of global identifiability.
\begin{assumption}[{\bf Global identifiability}]
	\label{assum:globo}
	For each wrong hypothesis $\theta\neq\thetatrue$, there is at least one agent that has strictly positive KL divergence.~\hfill$\square$
\end{assumption}
Let us provide some intuition behind Assumption~\ref{assum:globo}. 
Consider agent $k$ and hypothesis $\theta\neq\thetatrue$. Now, if the likelihoods $L_k(\xi|\theta)$ and $L_k(\xi|\thetatrue)$ are equal, $\theta$ is not distinguishable from $\thetatrue$ at agent $k$, i.e., the classification problem is {\em locally non-identifiable}. 
Clearly, if there exists a hypothesis $\theta$ that is indistinguishable from $\thetatrue$ at {\em all} agents, there is no hope for the system to classify correctly, because the agents will be necessarily uncertain between $\theta$ and $\thetatrue$. 
Therefore, a minimal requirement for global identifiability is that, for each $\theta\neq\thetatrue$, there exists at least one agent for which model $L_k(\xi|\theta)$ is distinct from $L_k(\xi|\thetatrue)$. This is exactly what Assumption~\ref{assum:globo} requires. 
It is also useful to highlight that Assumption~\ref{assum:globo} does not imply in any manner that agent $k$ would be able to classify locally. In fact, saying that agent $k$ is able to distinguish $\theta$ from $\thetatrue$ does not mean that it can distinguish $\thetatrue$ from the remaining hypotheses $\theta'\notin\{\theta,\thetatrue\}$.

We are now ready to state the theorem that establishes achievability of consistent learning. 
To this end, it is useful to introduce the expectation of the average log-likelihood ratio in \eqref{eq:avlik}:
\beq
\mnet(\theta)\triangleq \E[\xnet(\theta)]=\sum_{\ell=1}^N \pi_{\ell} d_{\ell}(\theta),
\label{eq:mnet}
\eeq
which does not depend on $i$ owing to the identical distribution over time implied by the steady-state analysis.

\begin{theorem}[{\bf Consistency of ASL}]
	\label{theor:weaklaw}
	Under Assumptions~\ref{assum:integrable} and~\ref{assum:initbel}, we have the following convergence:
	\beq
	\widetilde{\bm{\lambda}}^{(\delta)}_k\stackrel{\delta\rightarrow 0}{\longrightarrow}\mnet~~\textnormal{ in probability}.
	\label{eq:wlawintermediate}
	\eeq
	Since under Assumption~\ref{assum:globo} all entries of $\mnet$ are strictly positive, Eq. \eqref{eq:wlawintermediate} implies that each agent learns correctly the true hypothesis as $\delta\rightarrow 0$, namely, for all $\theta\neq\theta_0$ we have that the steady-state error probability of all agents $k=1,2,\ldots,N$ converges to zero as $\delta$ approaches zero:
	\beq
	\lim_{\delta\rightarrow 0}p_k^{(\delta)}=0.
	\eeq
\end{theorem}
\begin{IEEEproof}
	See Appendix~\ref{app:Th2}.
\end{IEEEproof}

The result of Theorem~\ref{theor:weaklaw} relies on the weak law of small step-sizes proved in Lemma~\ref{lem:mainlemma}, part $3$. Technically, this law requires finiteness of only the first moments $d_{\ell}(\theta)$, which is guaranteed by Assumption~\ref{assum:integrable}. 
Moreover, the result of Theorem~\ref{theor:weaklaw} requires that $\mnet(\theta)>0$ for all $\theta\neq\theta_0$. 
Since the entries of the Perron eigenvector are all strictly positive, we see that $\mnet(\theta)$ is strictly greater than zero for every $\theta$ if, for every $\theta$, there exists at least one agent $\ell$ for which the KL divergence $d_{\ell}(\theta)$ is strictly positive. In other words, in order to achieve consistent learning, it is sufficient that at least one of the first moments (i.e., the KL divergence) is nonzero, which is guaranteed by Assumption~\ref{assum:globo}.

Therefore, we see that Assumption~\ref{assum:globo} provides one important motivation for agents' cooperation in social learning. 
In fact, we assume that the learning problem can be non-identifiable (i.e., can be singular) {\em locally}, meaning that an individual agent can have one or more hypotheses $\theta\neq\theta_0$ that are indistinguishable from the true one (zero KL divergence). 
If this happens, an individual agent is not able to learn properly. 
On the other hand, under a {\em global} identifiability condition, the network is able (as shown in Theorem~\ref{theor:weaklaw}) to identify the true hypothesis by fusing the information coming from distinct agents.

We have shown that the ASL strategy allows correct learning of the true hypothesis for sufficiently small step-sizes. 
In other words, we have established that the error probability vanishes as $\delta\rightarrow 0$. 
On the other hand, we have not established {\em how} it vanishes. 
There are at least two good reasons to examine the way this probability converges to zero.
The first reason is to get manageable formulas for the evaluation of the social learning performance. 
The second reason is to characterize the fundamental scaling laws of the system. 
We will see that the ASL strategy is characterized by an exponential law, since the error probability of each individual agent decays exponentially fast as a function of the inverse step-size $1/\delta$.

\subsection{Normal Approximation for Small $\delta$}\label{sec:normal}
We will now prove a central limit theorem for the steady-state random vector $\widetilde{\bm{\lambda}}^{(\delta)}_k$. 
To this end, we will assume finiteness of second-order moments for the log-likelihoods. We furthermore assume statistical independence across the agents. 

In order to state the CLT, it is convenient to define some useful quantities. First, we introduce the covariance between the log-likelihood ratios at $\theta$ and $\theta'$, that is:
\beq
\rho_{\ell}(\theta,\theta')=\E\left[
\Big(
\bm{x}_{\ell,i}(\theta)
-
d_{\ell}(\theta)
\Big)
\Big(
\bm{x}_{\ell,i}(\theta')
-
d_{\ell}(\theta')
\Big)
\right].
\eeq
Then we introduce the covariance between the average variables $\xnet(\theta)$ and $\xnet(\theta')$ which, exploiting independence across agents, can be evaluated as:
\beq
\cnet(\theta,\theta')\triangleq
\sum_{\ell=1}^N \pi^2_{\ell} \rho_{\ell}(\theta,\theta').
\label{eq:cnet}
\eeq
Next, it is necessary to examine the behavior of the first two moments of the log-belief ratios. 
In view of Lemma~\ref{lem:mainlemma}, part $2$, it is possible to conclude that the expectation of the steady-state random vector $\widetilde{\bm{\lambda}}^{(\delta)}_k$ can be expressed as:
\beq
{\sf m}^{(\delta)}_k(\theta)\triangleq \E\left[\widetilde{\bm{\lambda}}^{(\delta)}_k(\theta)\right]
=\mnet(\theta)+O(\delta),
\label{eq:mean}
\eeq
where $O(\delta)$ is a quantity such that the ratio $O(\delta)/\delta$ remains bounded as $\delta\rightarrow 0$.
Likewise, using part $4$ of Lemma~\ref{lem:mainlemma}, we conclude that the covariance of the steady-state random vector $\widetilde{\bm{\lambda}}^{(\delta)}_k$ is: 
\beqa
\lefteqn{c^{(\delta)}_k(\theta,\theta')}\nonumber\\&\triangleq& \E\left[
\Big(\widetilde{\bm{\lambda}}^{(\delta)}_k(\theta)-{\sf m}^{(\delta)}_k(\theta)\Big)
\Big(\widetilde{\bm{\lambda}}^{(\delta)}_k(\theta')-{\sf m}^{(\delta)}_k(\theta')\Big)
\right]\nonumber\\
&=&
\frac{\cnet(\theta,\theta')}{2}\,\delta
+O(\delta^2).
\label{eq:covar}
\eeqa
Equations \eqref{eq:mean} and \eqref{eq:covar} can be rewritten in vector and matrix form, respectively as:
\beq
{\sf m}^{(\delta)}_k=\mnet+O(\delta),\quad
{\sf C}^{(\delta)}_k=\frac{\Cnet}{2}\,\delta+O(\delta^2),
\label{eq:meancovarmat}
\eeq
where ${\sf C}^{(\delta)}_k=[c^{(\delta)}_k(\theta,\theta')]$ and $\Cnet=[\cnet(\theta,\theta')]$ are the matrices that collect the individual covariances. 
We see from \eqref{eq:meancovarmat} that, as $\delta\rightarrow 0$, there is a leading term that does not depend on the agent index $k$ (whose impact is implicitly included in the higher order corrections, i.e., the $O(\cdot)$ terms). 

The first relation in \eqref{eq:meancovarmat} reveals that the expectation vector of the steady-state log-belief ratios, ${\sf m}^{(\delta)}_k$, approximates, for small $\delta$, the expectation vector of the {\em average} log-likelihood ratios, $\mnet$.
In comparison, the second relation in \eqref{eq:meancovarmat} reveals that the covariance matrix of the steady-state log-belief ratios, ${\sf C}^{(\delta)}_k$, goes to zero as $\Cnet \,\delta/2$, where $\Cnet$ is the covariance matrix of the {\em average} log-likelihood ratios, namely,
\beq
\lim_{\delta\rightarrow 0}\frac{2{\sf C}^{(\delta)}_k}{\delta}=\Cnet.
\eeq
We are now ready to state our central limit theorem.

\begin{theorem}[{\bf Asymptotic normality}]
	\label{theor:CLT}
	Assume that the data $\{\bm{\xi}_{k,i}\}$ are independent across the agents (recall that they are always assumed i.i.d. over time), and that the log-likelihood ratios have finite variance. 
	Then, under Assumptions~\ref{assum:integrable},~\ref{assum:initbel} and~\ref{assum:globo}, the following convergence holds:
	\beq
	\frac{\widetilde{\bm{\lambda}}_k^{(\delta)} - \mnet}{\sqrt{\delta}}
	\stackrel{\delta\rightarrow 0}{\rightsquigarrow} \mathscr{G}\left(0,\frac{\Cnet}{2}\right),
	\label{eq:CLTstatement2}
	\eeq
	where the symbol $\rightsquigarrow$ denotes convergence in distribution, and $\mathscr{G}(0,C)$ is a zero-mean multivariate Gaussian with covariance matrix equal to $C$.
\end{theorem}
\begin{IEEEproof}
	See Appendix~\ref{app:Th3}.
\end{IEEEproof}


Theorem~\ref{theor:CLT} entails the following approximation, holding for $\delta\approx 0$:
\beq
\widetilde{\bm{\lambda}}^{(\delta)}_k\approx\mathscr{G}\left(\mnet, \frac{\Cnet}{2}\,\delta\right).
\label{eq:CLTfirstapp}
\eeq
We see that such approximation does {\em not} depend on the agent index $k$. 
As shown in~\cite{MattaSayedCoopGraphSP2018}, in order to capture differences in performance across the agents, it is possible to replace the limiting expectation vector $\mnet$ and the limiting covariance matrix $\Cnet\,\delta/2$ with their exact counterparts, i.e., with the series appearing in \eqref{eq:mean} and \eqref{eq:covar}, yielding the refined approximation:
\beq
\widetilde{\bm{\lambda}}^{(\delta)}_k\approx\mathscr{G}\left({\sf m}^{(\delta)}_k, {\sf C}^{(\delta)}_k\right).
\label{eq:CLTsecondapp}
\eeq
The approximations in \eqref{eq:CLTfirstapp} and \eqref{eq:CLTsecondapp} will be tested in the section devoted to numerical experiments.

\subsection{Large Deviations for Small $\delta$}\label{sec:largedev}
In this section we focus on another relevant type of asymptotic analysis, namely, a {\em large deviations} analysis~\cite{DemboZeitouni,DenHollander}. 
The application of large deviations to adaptive networks was used in~\cite{MattaSayedCoopGraphSP2018,MattaBracaMaranoSayedTIT2016,MattaBracaMaranoSayedTSIPN2016}.

The basic aim of the LD analysis is to estimate the exponential decay rate of the probabilities associated to certain {\em rare} events. 
In our setting, the rare event is the probability that an agent opts for the wrong hypothesis. 
We will show that, at the steady state, this type of event becomes in fact rare as $\delta$ approaches zero.

More formally, the LD analysis will furnish the following type of representation for the steady-state error probability~\cite{DemboZeitouni,DenHollander}:
\beq
p^{(\delta)}_k\stackrel{{\bm \cdot}}{=}
e^{-\Phi/\delta},
\label{eq:LDef}
\eeq
where the notation $\stackrel{{\bm \cdot}}{=}$ means equality to the leading exponential order (as $\delta\rightarrow 0$) or, more explicitly:
\beq
\lim_{\delta\rightarrow 0} \delta\log p^{(\delta)}_k=-\Phi,
\label{eq:LDprinc}
\eeq
for a certain value $\Phi$ that is called the {\em error exponent}. 
Notably, in the exponent $\Phi$ we did not put any dependence on the agent index $k$. This is because, as shown in Theorem~\ref{theor:LD} further ahead, {\em all agents will exhibit the same error exponent}.

On the other hand, it should be remarked that the equality at the leading exponential order in \eqref{eq:LDef} does not imply in any way that we can approximate the probability of error as $e^{-\Phi/\delta}$, namely,  
	\beq
	p^{(\delta)}_k\not\approx e^{-\Phi/\delta}.
	\eeq
This is because any LD analysis neglects sub-exponential corrections. For example, it is immediate to check that the probabilities $e^{-\Phi/\delta}$ and $100\,e^{-\Phi/\delta}$ have the same LD exponent (equal to $\Phi$), but the second probability is two orders of magnitude larger. 
These sub-exponential corrections embody higher-order differences in the error probabilities (see, e.g., Fig.~\ref{fig:pe_ex}) that can arise across the agents due to different factors, for example, due to differences between very ``central'' agents with a high number of neighbors as opposed to ``peripheral'' agents with few neighbors. 
In order to compensate for sub-exponential corrections, a refined LD framework exists, usually referred to as ``exact asymptotics'', which has been applied to binary adaptive detection in~\cite{MattaSayedCoopGraphSP2018,MattaBracaMaranoSayedTSIPN2016}.

In summary, the aim of a large deviations analysis is to evaluate the asymptotic decay rate of the error probabilities, which is a meaningful and significant index of the inferential performance. 
Since the error exponent is a compact statistical descriptor of the learning performance, it can be useful to compare different systems (e.g., ASL strategies with different network graphs) and/or to optimize some system parameters (e.g., the network graph) to achieve the fastest learning rate. 

Before stating the main result about the LD analysis, it is necessary to introduce the Logarithmic Moment Generating Function (LMGF), a.k.a. cumulant generating function, of the log-likelihood ratios:
\beq
\Lambda_k(t;\theta)=\log\E\left[
e^{t \,\bm{x}_{k,i}(\theta)}
\right].
\label{eq:LMGF}
\eeq
We recall that, in the steady-state regime, the expectation is computed under the true model $L_k(\xi|\thetatrue)$, which does not change over time, and this explains why $\Lambda_k(t;\theta)$ does not depend on $i$.
It is also useful to introduce the LMGF of the average variable $\xnet(\theta)$ which, under the assumption that the data are independent across the agents, is:
\beq
\omeganet(t;\theta)=\log\E\left[
e^{t \,\xnet(\theta)}
\right]=\sum_{\ell=1}^N \Lambda_{\ell}(\pi_{\ell} t;\theta).
\label{eq:LMGFav}
\eeq

\begin{theorem}[{\bf Error exponents}]
	\label{theor:LD}
	Assume that the data $\{\bm{\xi}_{k,i}\}$ are independent across the agents (recall that they are always assumed i.i.d. over time), and that the logarithmic moment generating function of $\bm{x}_{k,i}(\theta)$ exists everywhere, namely, for all $k=1,2,\ldots,N$ and $\theta\neq\theta_0$: 
	\beq
	\Lambda_k(t;\theta)<+\infty ~~\forall t\in\mathbb{R}.
	\eeq
	Let
	\beq
	\phi(t;\theta)=\int_0^{t} \frac{\omeganet(\tau;\theta)}{\tau} d\tau.
	\label{eq:phittheta}
	\eeq
	Then, under Assumptions~\ref{assum:integrable},~\ref{assum:initbel} and~\ref{assum:globo} we have the following two results holding for every agent $k=1,2,\ldots,N$. 
	First, we have that:
	\beq
	\P\left[\widetilde{\bm{\lambda}}^{(\delta)}_k(\theta)\leq 0\right]
	\stackrel{{\bm \cdot}}{=}
	e^{-\Phi(\theta)/\delta}, ~~~ \Phi(\theta)=-\inf_{t\in\mathbb{R}}\phi(t;\theta).
	\label{eq:theorpreclaim}
	\eeq
	Second, the error probability is dominated by the worst-case (i.e., smaller) exponent:
	\beq
	p^{(\delta)}_k\stackrel{{\bm \cdot}}{=}
	e^{-\Phi/\delta}, ~~~ \Phi=\min_{\theta\neq\theta_0}\Phi(\theta).
	\label{eq:theorLDclaim}
	\eeq
\end{theorem}
\begin{IEEEproof}
	See Appendix~\ref{app:Th4}
\end{IEEEproof}

The main message conveyed by Theorem~\ref{theor:LD} is that the steady-state error probability of each individual agent converges to zero as $\delta\rightarrow 0$, exponentially fast as a function of $1/\delta$. 
This exponential law provides a {\em universal} law for adaptive social learning, which reflects the universal scaling law of distributed adaptive detection --- see~\cite{MattaSayedCoopGraphSP2018}.
The exponent $\Phi$ governing such an exponential decay is computed from the logarithmic moment generating function of the average log-likelihood, where the weights of this average are the limiting weights, i.e., the entries of the Perron eigenvector. 

The need for cooperation has been already motivated in relation to social learning problems that are locally non-identifiable. 
Theorem~\ref{theor:LD} implies another potential benefit of cooperation, namely, that {\em cooperation improves the learning accuracy}. 
We will illustrate this aspect through one example. 
Assume the most favorable case where all agents could learn the true hypothesis individually. 
Consider then a doubly-stochastic combination matrix, yielding a Perron eigenvector with uniform entries $\pi_\ell=1/N$ for all $\ell=1,2,\ldots,N$. 
Exploiting \eqref{eq:theorLDclaim}, we can easily see that in this particular case the error exponent of the network is given by:
\beq
\Phi=N\Phi_{\mathrm{ind}},
\label{eq:Ntimes}
\eeq 
where $\Phi_{\mathrm{ind}}$ is the error exponent of an individual agent. 
According to \eqref{eq:Ntimes}, we see that the network error exponent is $N$ times larger than the individual error exponent, which in turn implies an $N$-fold exponential improvement in the learning accuracy. Intuitively, a network of $N$ agents observes $N$ times as much data as a single agent at each time instant. The strong-connectivity of the network allows for the data to fully propagate across agents and yields the aforementioned learning performance improvement. 

\section{Transient Analysis}
\label{sec:transient}
\subsection{Qualitative Description of the Transient Phase}
\label{sec:qualtrans}
Preliminarily, we deem it is useful to provide a qualitative overview of the transient behavior of adaptive social learning in comparison to traditional social learning. 
To this end, we consider initially a simple example consisting of a single-agent (indices $k$ and $\ell$ dropped) binary ($\Theta =\{1,\,2\}$) problem, with symmetric KL divergences:
\beq 
\E_1 \left[
\log \frac{L(\bm{\xi}_{i}|1)}{L(\bm{\xi}_{i}|2)}
\right]
=-\E_2 \left[
\log \frac{L(\bm{\xi}_{i}|1)}{L(\bm{\xi}_{i}|2)}
\right]\dfz x>0,
\label{eq:symmKLdivdef}
\eeq 
where $\E_\theta$ denotes expectation under the distribution $L(\xi|\theta)$.  
We assume that at time $i=1$, the true underlying hypothesis is $\theta_0=1$, and the situation remains stationary until a certain time ${\sf T}_1$, after which data start being generated according to $\theta_0=2$, and that is why a transient analysis is necessary to see how the learning algorithm is able to track this drift.

In order to examine how the learning process progresses over time, it is sufficient to consider the time-evolution of the log-belief ratio: 
\beq
\bm{r}_i\triangleq\log \frac{\bm{\mu}_{i}(1)}{\bm{\mu}_{i}(2)},
\eeq 
whose positive (resp., negative) values will let the agent opt for $\theta=1$ (resp., $\theta=2$).  
Specializing \eqref{eq:ASLintermold} and \eqref{eq:ASLfinal} to the single-agent binary setting, traditional social learning evolves according to the recursion (we add a superscript to distinguish traditional from adaptive social learning):
\beq
\bm{r}_{i}^{{\sf SL}}=\bm{r}_{i-1}^{{\sf SL}}+\log \frac{L(\bm{\xi}_{i}|1)}{L(\bm{\xi}_{i}|2)},
\qquad\bm{r}_{0}^{{\sf SL}}=0.
\label{eq:recursionTSL}
\eeq
Likewise, replacing \eqref{eq:ASLintermold} with \eqref{eq:ASLinterm}, the adaptive social learning strategy in this single-agent binary case evolves according to the recursion:
\begin{equation}
\bm{r}_{i}^{{\sf ASL}}=(1-\delta)\bm{r}_{i-1}^{{\sf ASL}}+\delta\log \frac{L(\bm{\xi}_{i}|1)}{L(\bm{\xi}_{i}|2)},
\qquad\bm{r}_{0}^{{\sf ASL}}=0.
\label{eq:recursion}
\end{equation}
For the sake of concreteness, in both \eqref{eq:recursionTSL} and \eqref{eq:recursion} we assume flat initial priors (i.e., $\bm{r}_{0}^{{\sf SL}}=\bm{r}_{0}^{{\sf ASL}}=0$).

\begin{figure}[t]
	\centering
	\includegraphics[width=.9\linewidth]{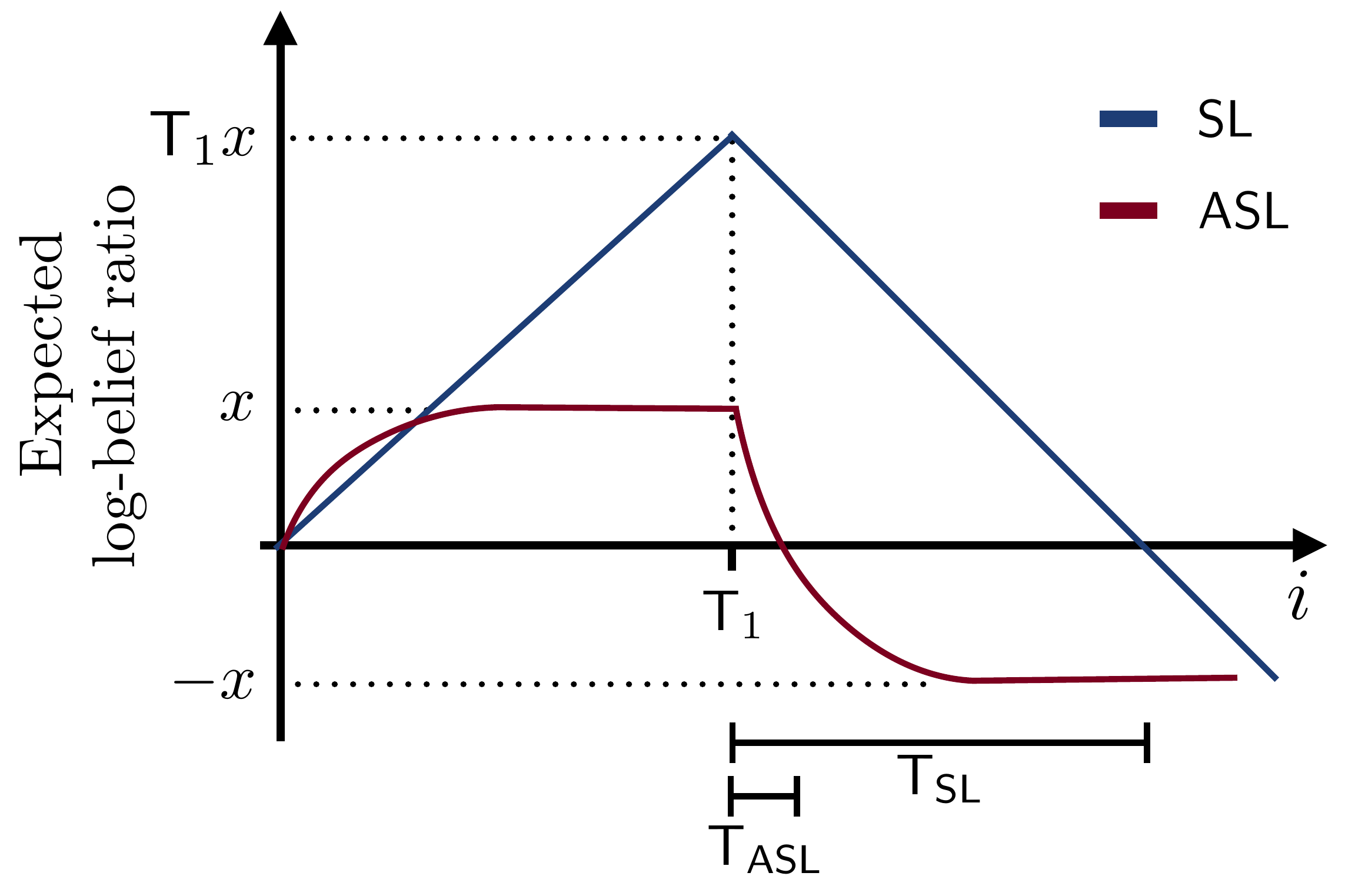}
	\caption{Diagram of the time evolution of the log-belief ratio in expectation for the traditional social learning strategy (in blue) and for the ASL strategy (in violet) within the single-agent case.}\label{fig:evol}
\end{figure}

In order to get a flavor of the main trade-offs involved in the transient behavior, let us focus on the time-evolution of the {\em expected values}. 
Taking expectations in \eqref{eq:recursionTSL}, at time ${\sf T}_1$ we have:
\begin{equation}
\E\left[\bm{r}^{{\sf SL}}_{{\sf T}_1}\right]={\sf T}_1 x,
\label{eq:slexp}
\end{equation}
where $x$ is the symmetric KL divergence introduced in \eqref{eq:symmKLdivdef}. Equation \eqref{eq:slexp} shows that the expected value of the log-belief ratio grows linearly with the stationarity interval ${\sf T}_1$.
This linear growth is a reflection of the increasing knowledge acquired by the agent as it aggregates new information represented by the log-likelihood ratio $\log \frac{L(\bm{\xi}_{i}|1)}{L(\bm{\xi}_{i}|2)}$. In a virtual asymptotic regime, this knowledge becomes a certainty, i.e., as ${\sf T}_1\longrightarrow+\infty$, $\bm{r}^{{\sf SL}}_{{\sf T}_1}\rightarrow+ \infty$, which implies that if hypothesis $1$ remains in force indefinitely, the belief of the agent regarding this hypothesis achieves full confidence. 
Unfortunately, this increasing confidence comes at a cost in terms of an elephant memory that makes the algorithm slow in adaptation. 
Indeed, since from time ${\sf T}_1+1$ the true hypothesis is $\theta_0=2$, from \eqref{eq:recursionTSL} and \eqref{eq:slexp} we have that:
\beq
\E\left[\bm{r}^{{\sf SL}}_{i}\right]=
\E\left[\bm{r}^{{\sf SL}}_{{\sf T}_1}\right] 
-i x=({\sf T}_1 - i)x.
\label{eq:TSLtransienteq}
\eeq
Now, the adaptation time can be roughly identified by considering the time necessary to overcome the initial bias towards hypothesis $1$ once the true hypothesis switches from $1$ to $2$. In terms of our qualitative mean-value analysis, this is the time necessary for the expected log-belief ratio to change from positive to negative, which, in view of \eqref{eq:TSLtransienteq} implies that the adaptation time for the traditional social learning strategy is on the order of:
\beq
{\sf T}_{{\sf SL}}={\sf T}_{1}.
\label{eq:timeTSLtransientqual}
\eeq
This behavior is clearly not admissible for an adaptive algorithm, since it implies that the time necessary to recover from a wrong opinion is proportional to the stationarity interval where this opinion was actually true! This behavior is illustrated in Fig.~\ref{fig:evol}.

Let us switch to the adaptive strategy. Developing the recursion until time ${\sf T}_1$, from \eqref{eq:recursion} we get, respectively:
\begin{equation}
\E\left[ \bm{r}^{{\sf ASL}}_{{\sf T}_1}\right]=
\delta\sum_{m=0}^{i-1}(1-\delta)^m x=\left(1-(1-\delta)^{{\sf T}_1}\right)x\approx x,
\label{eq:aslexp}
\end{equation}
where the approximation is motivated from assuming a sufficiently large ${\sf T}_1$. 
Considering then that from time ${\sf T}_1+1$ onward the true hypothesis is $\theta_0=2$, Eqs. \eqref{eq:recursion} and \eqref{eq:aslexp} yield, for any $i>{\sf T}_1$:
\beqa
\E \left[ \bm{r}^{{\sf ASL}}_{i}\right]&=&
(1-\delta)^i \E\left[ \bm{r}^{{\sf ASL}}_{{\sf T}_1}\right]
-\delta\sum_{m=0}^{i-1}(1-\delta)^m x\nonumber\\
&\approx&
-\left(1-2(1-\delta)^{i}\right)x.
\label{eq:ASLtransientqual}
\eeqa
Now, equating \eqref{eq:ASLtransientqual} to zero to evaluate the adaptation time, we obtain:
\beq
{\sf T}_{\sf ASL}=\frac{\log 2}{\log(1-\delta)^{-1}}\approx \frac{\log 2}{\delta}.
\label{eq:timeASLtransientqual}
\eeq
A visual comparison of the enhanced adaptation provided by the ASL strategy is exemplified in Fig.~\ref{fig:evol}.

Comparing \eqref{eq:timeASLtransientqual} against \eqref{eq:timeTSLtransientqual}, we see that, in contrast to the undesirable behavior exhibited by traditional social learning, the adaptive formulation exhibits a controlled initial bias. 
This is because, after a relatively long stationarity interval ${\sf T}_1$, the expected log-belief is concentrated around a fixed value $x$, and the adaptation time will then increase roughly as $1/\delta$. 
In a nutshell, while the reaction capacity of traditional social learning is not controlled by design and is severely affected by the duration of previous stationarity intervals, in adaptive social learning the adaptation time is not affected by previous stationarity intervals, and the effective memory is controlled through the step-size.
This enhanced adaptivity comes at the price of learning accuracy. In fact, as we have established in the previous sections, the steady-state error probability does not converge to zero as time elapses, but converges to some stable value. However, this value vanishes exponentially fast as a function of $1/\delta$, highlighting the fundamental trade-off of adaptive social learning: the smaller the step-size $\delta$, the smaller the error probability and the slower the adaptation. 

In the theory of adaptation and learning, the transient analysis is typically performed by characterizing the evolution of suitable higher-order moments, such as second or fourth order moments of the pertinent statistics~\cite{Sayed}. 
However, this analysis is more appropriate for estimation/regression problems where the focus of the transient analysis is to ascertain how long it takes for the pertinent system state to attain a prescribed neighborhood of the expected value. 
In our social learning setting, it is more appropriate to identify an adaptation time in terms of {\em error probabilities}. 
As established in Theorem~\ref{theor:LD}, the behavior of these probabilities is governed by the logarithmic moment generating function of the observations which, as the name itself suggests, incorporates dependence upon {\em all moments}. 
Accordingly, a meaningful way to perform the transient analysis is to examine the time-evolution of logarithmic moment generating functions, rather than individual moments. This characterization constitutes the core of Theorem~\ref{theor:instanterrprobfin}, which is introduced in the next section.

\subsection{Quantitative Description of the Transient Phase}
In this section, we provide a rigorous analysis to support the qualitative description of the transient behavior, seen in Sec.~\ref{sec:qualtrans}. 
We assume that the ASL strategy has been in operation for a certain arbitrary time $i_0$. 
All the knowledge accumulated by the agents until this time is summarized in the belief vector $\bm{\mu}_{i_0}$. 
We remark that the evolution of the statistical models from $i=0$ to $i=i_0$ is left completely arbitrary, that is, the system could have experienced several drifts in the statistical conditions, including change of the underlying hypotheses, data generated according to models that do not match the assumed likelihoods, and so on. 
From the ASL algorithm viewpoint, all these effects are summarized in the belief vector $\bm{\mu}_{i_0}$ that acts as initial state at time $i_0$. 
In order to perform the transient analysis, we assume that from $i_0+1$ onward, the true hypothesis is steadily equal to $\theta_0$, and will establish how much time is necessary to stay sufficiently close to the steady-state learning performance starting from a given (arbitrary) realization $\mu_{i_0}$. 
As done before, to simplify the notation we set $i_0=0$ and the initial state becomes $\mu_{0}$. 

As noticed at the end of the previous section, in a social learning problem the adaptation time should be properly related to the time-evolution of the error probability, and particularly to the time necessary for the {\em instantaneous} error probability to approach the {\em steady-state} error probability. 
Accordingly, in the next theorem we start by providing an upper bound on the {\em instantaneous} error probability introduced in \eqref{eq:errprob}.

\begin{theorem}[{\bf Bounds on the instantaneous error probability}]
	\label{theor:instanterrprobfin}
	The claim of the theorem holds under the same assumptions of Theorem~\ref{theor:LD}. 
	Let $\kappa$ and $\beta$ be the constants defined in Property~\ref{prop:Perron}, and let $t^{\star}_{\theta}<0$ be the unique solution to the equation:
	\beq
	\frac{\Lambda_{\mathrm{ave}}(t^{\star}_{\theta};\theta)}{t^{\star}_{\theta}}=0.
	\label{eq:stateqapp}
	\eeq
	Let 
	\beq
	\lambda_{\mathrm{ave},0}(\theta)=\sum_{\ell=1}^N \pi_{\ell} \lambda_{\ell,0}(\theta)
	\eeq
	be the network average of the initial log-belief ratios $\lambda_{\ell,0}(\theta)$, and let, for all $\theta\neq \theta_0$:
	\beqa
	{\sf K}_1(\theta)&\dfz&|t^{\star}_{\theta}|\,\Big[\mnet(\theta)-\lambda_{\mathrm{ave},0}(\theta)\Big],\label{eq:C1theta}\\
	{\sf K}_2(\theta)&\dfz&\kappa |t^{\star}_{\theta}|\,\sum_{\ell=1}^N |\lambda_{\ell,0}(\theta)|.\label{eq:C2theta}
	\eeqa
	Then, the instantaneous error probability $p_{k,i}^{(\delta)}$ is upper bounded as:
	\beqa
	p^{(\delta)}_{k,i}&\leq& \sum_{\theta\neq\theta_0}
	e^{\frac{1}{\delta}\left[
		-\Phi(\theta)+{\sf K}_1(\theta)(1-\delta)^i+{\sf K}_2(\theta) (1-\delta)^i \beta^i+\mathcal{O}(\delta)
		\right]},\nonumber\\
	\label{eq:insterrprobmainbound}
	\eeqa
	where the notation $\mathcal{O}(\delta)$ signifies that the ratio $\mathcal{O}(\delta)/\delta$ stays bounded as $\delta\rightarrow 0$.
\end{theorem}

\begin{IEEEproof}
	See Appendix~\ref{app:theor5proof}.
\end{IEEEproof}
Theorem~\ref{theor:instanterrprobfin} reveals the main behavior of the transient error probability. 
Examining the error exponent of the upper bound in \eqref{eq:insterrprobmainbound} we see, up to higher-order small-$\delta$ corrections embodied in the term $\mathcal{O}(\delta)$, the emergence of three terms: the {\em steady-state} error exponent $\Phi(\theta)$ already identified in Theorem~\ref{theor:LD}, and two other terms that characterize the {\em transient} behavior. 
The first transient term decays as $(1-\delta)^i$, and is thus influenced solely by the step-size. 
The second transient term, $(1-\delta)^i \beta^i$, decays faster and is influenced also by the parameter $\beta$. This parameter, according to Property~\ref{prop:Perron}, is determined by the second largest-magnitude eigenvalue of $A$, and accordingly determines the mixing properties of $A$ (i.e., the convergence rate of $[A^{i}]_{\ell k}$ to the Perron eigenvector entry $\pi_{\ell}$). 
Therefore, the second transient term, with rate $(1-\delta)^i\beta^i$, determines a transient phenomenon that is related to the convergence of the matrix-powers to a ``centralized'' solution with combination weights $\pi_{\ell}$. 
In comparison, the first term, with rate $(1-\delta)^i$, determines a transient phenomenon ruled by the step-size only.

In summary, Theorem~\ref{theor:instanterrprobfin} provides an upper bound on the instantaneous error probability that converges, as $i\rightarrow\infty$, to a sum of exponential terms with steady-state error exponent $\Phi=\min_{\theta\neq\theta_0}\Phi(\theta)$.
Accordingly, we identify as a meaningful definition for the {\em adaptation time} the critical time instant after which the error probability decays with an error exponent $(1-\epsilon)\Phi$, for some small $\epsilon$. This is made precise in the following corollary.

\begin{corollary}[{\bf Adaptation time}]
	\label{cor:corollaryinsterr}
	Under the same notation and assumptions of Theorem~\ref{theor:instanterrprobfin}, let
	\beqa
	{\sf K}_1&\dfz&\max_{\theta\neq\theta_0} {\sf K}_1(\theta)=
	\max_{\theta\neq\theta_0}\left\{|t^{\star}_{\theta}|\,\Big[\mnet(\theta)-\lambda_{\mathrm{ave},0}(\theta)\Big]\right\},\nonumber\\
	{\sf K}_2&\dfz&\max_{\theta\neq\theta_0} {\sf K}_2(\theta)=
	\kappa\,\max_{\theta\neq\theta_0}\left\{|t^{\star}_{\theta}|\,\sum_{\ell=1}^N |\lambda_{\ell,0}(\theta)|\right\}.
	\eeqa
	Then, the upper bound:
	\beq
	p^{(\delta)}_{k,i}\leq e^{-\frac{1}{\delta}[(1-\epsilon)\Phi+\mathcal{O}(\delta)]}
	\label{eq:whatwemeanbyadaptime}
	\eeq
	holds for all $i>{\sf T}_{\sf ASL}$, where ${\sf T}_{\sf ASL}$ is given by the following rules:
	\begin{itemize}
		\item[i)] (Favorable case, all initial states are good). 
		
		\vspace*{10pt}
		\noindent If $\lambda_{\mathrm{ave},0}(\theta)\geq \mnet(\theta)$ for all $\theta\neq\theta_0$:
		\beq
		{\sf T}_{\sf ASL}=\frac{1}{\log\beta^{-1}}\log\frac{{\sf K}_2}{\epsilon\,\Phi},\qquad\epsilon<\frac{{\sf K}_2}{\Phi}.
		\label{eq:adaptime1}
		\eeq
		\item[ii)] (Unfavorable case, at least one initial state is bad). 
		\vspace*{10pt}
		
		\noindent
		If $\lambda_{\mathrm{ave},0}(\theta)<\mnet(\theta)$ for at least one $\theta\neq\theta_0$:
		\beq
		{\sf T}_{\sf ASL}=\frac{1}{\log(1-\delta)^{-1}}\log\frac{{\sf K}_1}{\epsilon\,\Phi},\qquad \epsilon<\frac{{\sf K}_1}{\Phi}.
		\label{eq:adaptime2}
		\eeq
	\end{itemize}
\end{corollary}
\begin{IEEEproof}
	See Appendix~\ref{app:corolla}.
\end{IEEEproof}

Let us now examine the main parameters and phenomena affecting the adaptation time ${\sf T}_{\sf ASL}$. 

{\em -- Memory}.
The memory coming from the past algorithm evolution is summarized in the starting belief vector $\mu_{0}$, which in turn determines the average log-belief $\lambda_{\mathrm{ave},0}(\theta)$.

First of all, we notice that an average initial state $\lambda_{\mathrm{ave},0}(\theta)$ greater than $\mnet(\theta)$ creates already a (favorable) bias toward the true hypothesis. 
Accordingly, when $\lambda_{\mathrm{ave},0}(\theta)\geq\mnet(\theta)$ the transient term ${\sf K}_1(\theta)(1-\delta)^i$ reduces the error probability since ${\sf K}_1(\theta)<0$. 
In this case, the dominant transient term is $(1-\delta)^i\beta^i$, and the corresponding adaptation time in \eqref{eq:adaptime1} is essentially determined by the mixing parameter $\beta$, i.e., by how fast the combination weights converge to the Perron eigenvector. 
Under this regime, the adaptation time {\em does not depend critically on the step-size}. 

In comparison, the case where $\lambda_{\mathrm{ave},0}(\theta)<\mnet(\theta)$ is the unfavorable case where we are, as $\lambda_{\mathrm{ave},0}(\theta)$ decreases, progressively far from the steady-state. 
Under this regime, for small $\delta$ the dominant transient term is ${\sf K}_1(\theta)(1-\delta)^i$, and the adaptation time {\em scales with the step-size as $1/\log(1-\delta)^{-1}\approx1/\delta$}.

One particularly interesting case is when the average initial state is negative. This happens, for example, when the initial state comes from a previous learning cycle where the agent converged to a certain hypothesis that has then changed at the beginning of the subsequent learning cycle. 
In line with intuition, the adaptation time \eqref{eq:adaptime2} increases with increasing size of the wrong starting conditions. Moreover, this dependence upon the past states is only logarithmic, which reveals that the past algorithm evolution has not a dramatic impact on the adaptation time. 

{\em -- KL Divergences and Error Exponent}. By ignoring the initial state, Eq. \eqref{eq:adaptime2} becomes:
\beq
{\sf T}_{\sf ASL}=\frac{1}{\log(1-\delta)^{-1}}\log\frac{\max_{\theta\neq\theta_0}\left[|t^{\star}_{\theta}|\mnet(\theta)\right]}
{\epsilon\,\Phi}.
\label{eq:simplifiedadaptime2}
\eeq
From Property P2) in Lemma~\ref{lem:uspropLMGFapp} (see Appendix~\ref{app:theor5proof}), we know that:
\beq
\Phi(\theta)\leq |t^{\star}_{\theta}| \mnet(\theta),
\eeq
which shows that the ratio $\max_{\theta\neq\theta_0}\left[|t^{\star}_{\theta}|\mnet(\theta)\right]/\Phi$ appearing in \eqref{eq:simplifiedadaptime2} is greater than $1$. 
Even if declaring a general behavior for this ratio for all statistical models is not obvious, we see that the numerator and the denominator are not independent. 
For example, having an ``easier'' detection problem where the KL divergences (numerator) increase typically corresponds to an increase of the error exponent (denominator) as well.
However, in all cases the dependence on these parameters is not critical, since it is logarithmic.

{\em -- Parameter $t^{\star}_{\theta}$}. 
First of all, to evaluate and interpret the bound on the adaptation time it is useful to remark that the term $|t^{\star}_{\theta}|$ is comprised between $1/\pi_{\max}$ and $1/\pi_{\min}$ --- see property P3) in Lemma~\ref{lem:uspropLMGFapp}.
Apparently, these bounds introduce a dependence on the network parameters (i.e., on the Perron eigenvector). 
However, we should be careful here, and recall that the network error exponent $\Phi$ depends on the whole network as well. 
In order to get insights on this dependence, let us ignore the initial state and consider the case where all likelihoods are equal across agents and the combination matrix is doubly stochastic (yielding a uniform Perron eigenvector). Under these assumptions, from property P3) in Lemma~\ref{lem:uspropLMGFapp} we get $t^{\star}_{\theta}=-N$, and using \eqref{eq:Ntimes} we obtain:
\beq
{\sf T}_{\sf ASL}=\frac{1}{\log(1-\delta)^{-1}}\log\frac{\max_{\theta\neq\theta_0}\left[\mnet(\theta)-\lambda_{\mathrm{ave},0}(\theta)\right]}{\epsilon\,\Phi_{\mathrm{ind}}},
\eeq
which shows how the network size appearing in the parameter $t^{\star}_{\theta}=-N$ is perfectly compensated by the network size embodied in the network exponent $\Phi=N\Phi_{\mathrm{ind}}$. 
Accordingly, we expect that the network parameters have a reduced impact on the transient time in \eqref{eq:adaptime2}, while, as observed before, the effect of the network is embodied in the parameter $\beta$ controlling the higher-order transient term $(1-\delta)^i\beta^i$ in \eqref{eq:insterrprobmainbound}, which is neglected in the small-$\delta$ regime.

{\em -- Parameter $\epsilon$}. 
The smaller $\epsilon$ is, the closer the error exponent to the steady-state exponent $\Phi$ will be. Remarkably, the dependence is logarithmic in $1/\epsilon$, which means that this parameter is not critical.

{\em -- Step-Size}. 
Finally, in the (more interesting) case where the initial state is not good, see \eqref{eq:adaptime2}, the adaptation time scales as $1/\delta$. We remark that this behavior matches well the qualitative analysis of Sec.~\ref{sec:qualtrans}. 

The bottom line of Corollary~\ref{cor:corollaryinsterr} is that the adaptive capabilities of the ASL strategy are enhanced by a larger value of $\delta$, by yielding a reduced adaptation time. A larger $\delta$ however is not always desirable, since it can reduce the accuracy in the decision-making process (as seen in Theorem~\ref{theor:LD}, the steady-state probability of error is increased for larger $\delta$). Both phenomena represent the trade-off adaptation vs. learning present in the ASL strategy and should be taken into account when designing $\delta$. 
Such trade-off can be better summarized by combining Theorem~\ref{theor:LD} and Corollary~\ref{cor:corollaryinsterr}, which shows that the error probability decays exponentially fast with the adaptation time, roughly as:
\beq
p_k^{(\delta)}\approx \exp\left\{-\frac{\Phi}{\log[{\sf K}_1\times(\epsilon\,\Phi)^{-1}]}\,{\sf T}_{\sf ASL}\right\}.
\eeq

{\em -- Stability over Successive Learning Cycles}.
The characterization of the transient stage provided by Theorem~\ref{theor:instanterrprobfin} and the related corollary is valid under an arbitrary choice of the starting state $\lambda_{k,0}$. 
However, as we have commented in the previous section, if we start from a wrong state the level of this state affects adversely the adaptation time. 
Therefore, some fundamental questions arise. 
Assume that the time axis is divided into successive intervals (learning cycles) wherein the system evolves under stationary conditions. 
Then, the belief accumulated at the end of a learning cycle can be wrong in relation to the subsequent learning cycle. 
How ``wrong'' are the initial beliefs at the beginning of a learning cycle as the algorithm progresses? Do these initial states compromise the learning capability of the algorithm over successive cycles?
These fundamental questions can be answered by combined steady-state and transient analyses. 
In fact, from the steady-state analysis carried out in the previous sections, we learned that the steady-state log-belief ratios fluctuate in a small neighborhood (of size $\sim\sqrt{\delta}$) of the expected values of the pertinent KL divergences. 
This means that at the end of each cycle the ASL strategy converges to some {\em stable} state, i.e., a state that does not diverge as the step-size $\delta$ becomes small. As a result, the initial states of each learning cycle would evolve in a stable manner and, hence, do not compromise the learning performance of the algorithm, provided that the adaptation time is smaller than the duration of the learning cycles. These aspects will be more quantitatively illustrated in Sec.~\ref{sec:learningcycles}, with reference to specific illustrative examples.

\section{Illustrative Examples}
\label{sec:1}
\begin{figure}[t]
	\centering
	\includegraphics[width=.88\linewidth]{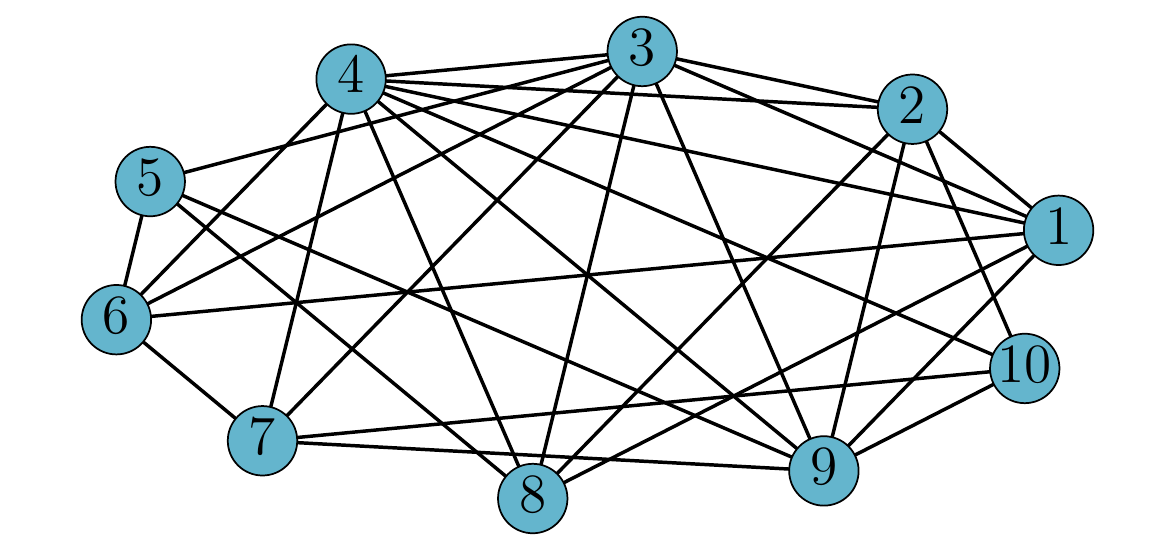}
	\caption{Strongly-connected network topology with $N=10$ agents.}
	\label{fig:network}
\end{figure}

We consider the strongly-connected network of $N=10$ agents displayed in Fig.~\ref{fig:network}. 
We assume that all agents have a self-loop (not displayed in the figure). 
Besides, the combination matrix is designed using an averaging rule, resulting in a left-stochastic matrix~\cite{Sayed}.

\begin{figure}[t]
	\centering
	\includegraphics[width=.85\linewidth]{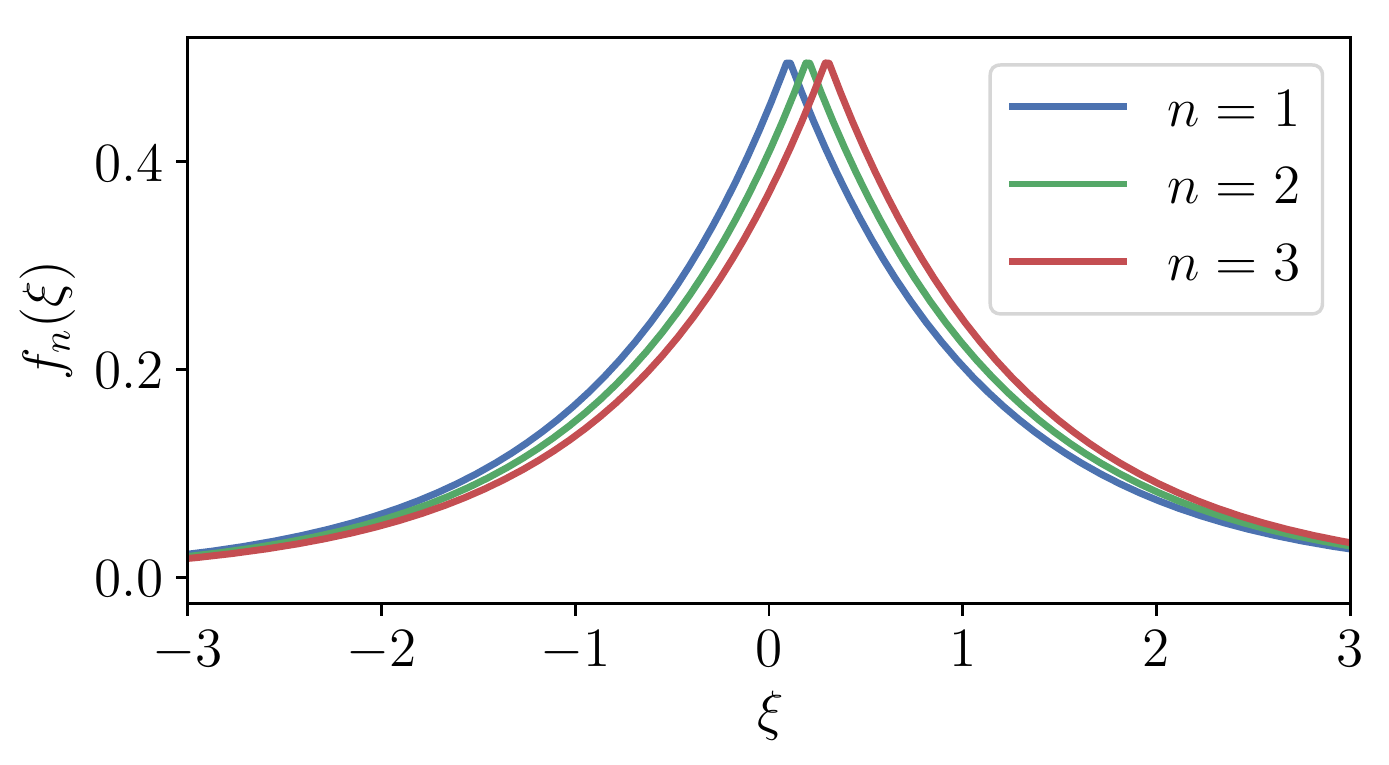}
	\caption{Family of Laplace likelihood functions.}
	\label{fig:gaussian}
\end{figure}

The network is faced with the following statistical learning problem.
We consider a family of Laplace likelihood functions with scale parameter $1$, seen in Fig.~\ref{fig:gaussian}. 
Formally, we are given three Laplace densities:
\beq
f_n(\xi)=\frac{1}{2}\exp\left\{-|\xi-0.1 n|\right\},
\label{eq:Laplacepdf}
\eeq
for $n\in\{1,2,3\}$.
The likelihoods of the data collected by the agents are chosen from among these Laplace densities.  

To make things more interesting, we assume that the inference problem is {\em not locally identifiable}. The setup for each agent's family of likelihood functions can be seen in Table~\ref{tab:id}.
\begin{table}[ht]
	\def\arraystretch{1.3}%
	\caption{Identifiability setup for the network in Fig.~\ref{fig:network}.}
	\begin{center}
		\begin{tabular}{|p{0.1\linewidth}|p{0.1\linewidth}|p{0.1\linewidth}|p{0.1\linewidth}|}
			\hline
			\multirow{2}{*}{\textbf{Agent} $k$}&\multicolumn{3}{|c|}{\textbf{Likelihood Function}: $L_k(\xi|\theta)$} \\
			\cline{2-4} 
			&$\theta=1$ & $\theta=2$& $\theta=3$\\
			\hline
			$1-3$& $f_1(\xi)$& $f_1(\xi)$& $f_3(\xi)$ \\
			\hline
			$4-6$& $f_1(\xi)$& $f_3(\xi)$& $f_3(\xi)$ \\
			\hline
			$7-10$& $f_1(\xi)$& $f_2(\xi)$& $f_1(\xi)$ \\
			\hline
		\end{tabular}
		\label{tab:id}
	\end{center}
\end{table}

In summary, the data $\{\bm{\xi}_{k,i}\}$ are i.i.d. (across time and agents) Laplace random variables, with expectations that depend both on the agent $k$ and the hypothesis $\theta$. Accordingly, we will use the notation $e_k(\theta)$ to denote the expectation of $\bm{\xi}_{k,i}$, computed under likelihood $L_k(\xi|\theta)$. 
For example, using Table~\ref{tab:id}, we see that:
\beq
e_1(1)=0.1,~~e_4(3)=0.3,~~e_7(2)=0.2.
\eeq
We are now ready to delve into a detailed illustration of the numerical experiments. 
In particular, in this section we will test how the empirical performance matches the steady-state performance as characterized in Theorems~\ref{theor:steady}--\ref{theor:LD}. In order to examine the steady-state behavior {\em empirically}, we need that the ASL algorithm run for a sufficiently long period of time. In line with the prescriptions from Sec.~\ref{sec:transient}, the duration of this this period is chosen as at least one order of magnitude larger than the inverse of the step-size, $1/\delta$.

\subsection{Consistency}
\begin{figure}[t]
	\centering
	\includegraphics[width=.95\linewidth]{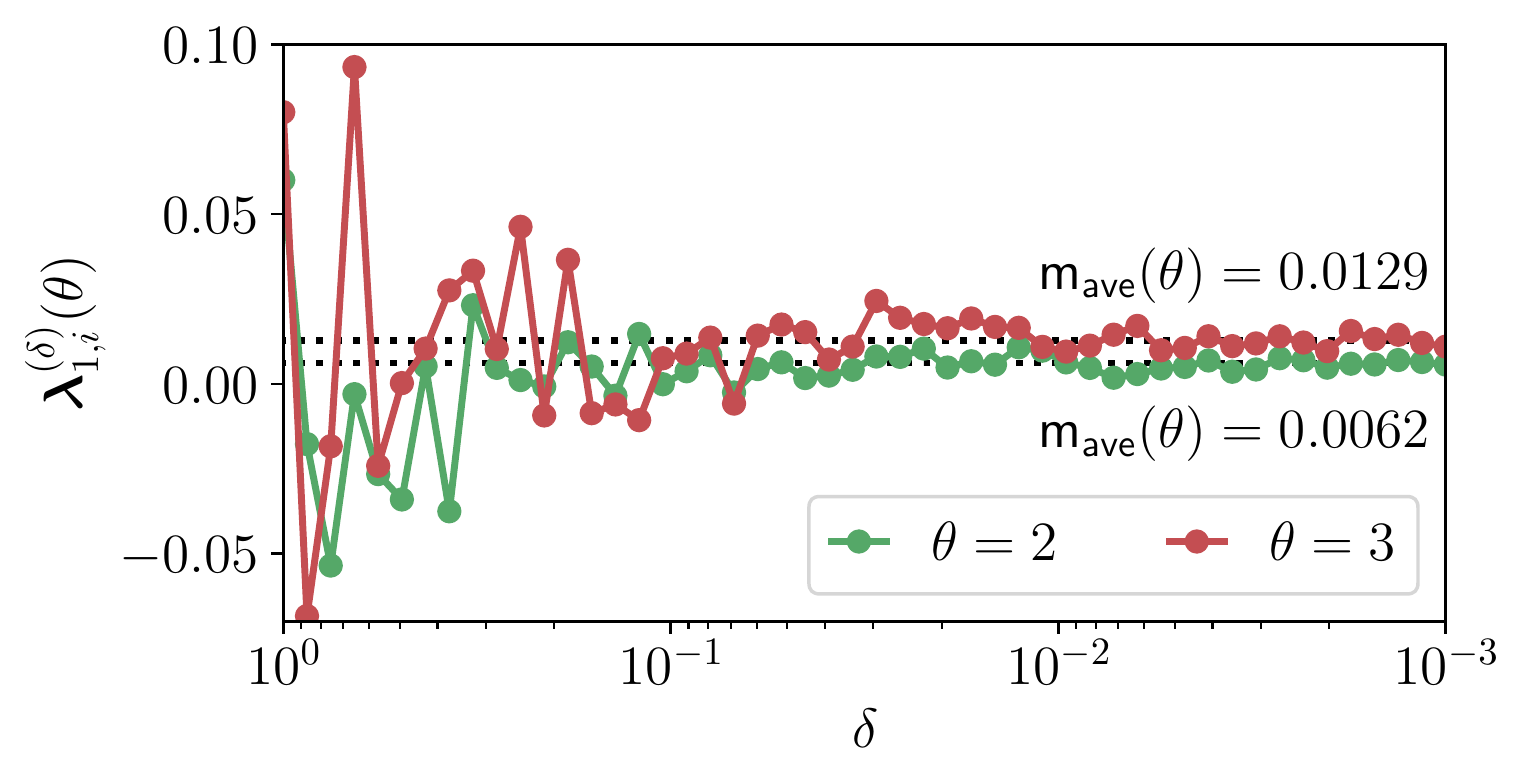}
	\caption{Consistency of the ASL strategy (Theorem~\ref{theor:weaklaw}). According to the weak-law of small step-sizes, the steady-state log-belief ratios for agent $1$ concentrate around the predicted expectation values in $\mnet$ as $\delta$ approaches zero.}
	\label{fig:theo1}
\end{figure}

We consider that all agents are running the ASL algorithm for a fixed $\thetatrue=1$ over $8000$ time samples (after which we consider that they achieved the steady state). From Theorem~\ref{theor:weaklaw}, we saw that as $\delta$ approaches zero, all agents $k$ are able to consistently learn --- see \eqref{eq:wlawintermediate}. 
In order to show this effect, for each value of $\delta$ (50 sample points in the interval $\delta\in[0.001,1]$ are taken), we consider a different realization of the observations. In Fig.~\ref{fig:theo1}, for agent $1$ and $\theta=2, 3$, we show how the log-belief ratios $\bm{\lambda}_1^{(\delta)}(\theta)$ behave for decreasing values of $\delta$. 
We see the weak-law of small step-sizes arising, since the limiting log-belief ratios tend to concentrate around ${\sf m}_{\sf ave}$. 

\subsection{Asymptotic Normality}
\begin{figure}[t]
	\centering
	\includegraphics[width=.95\linewidth]{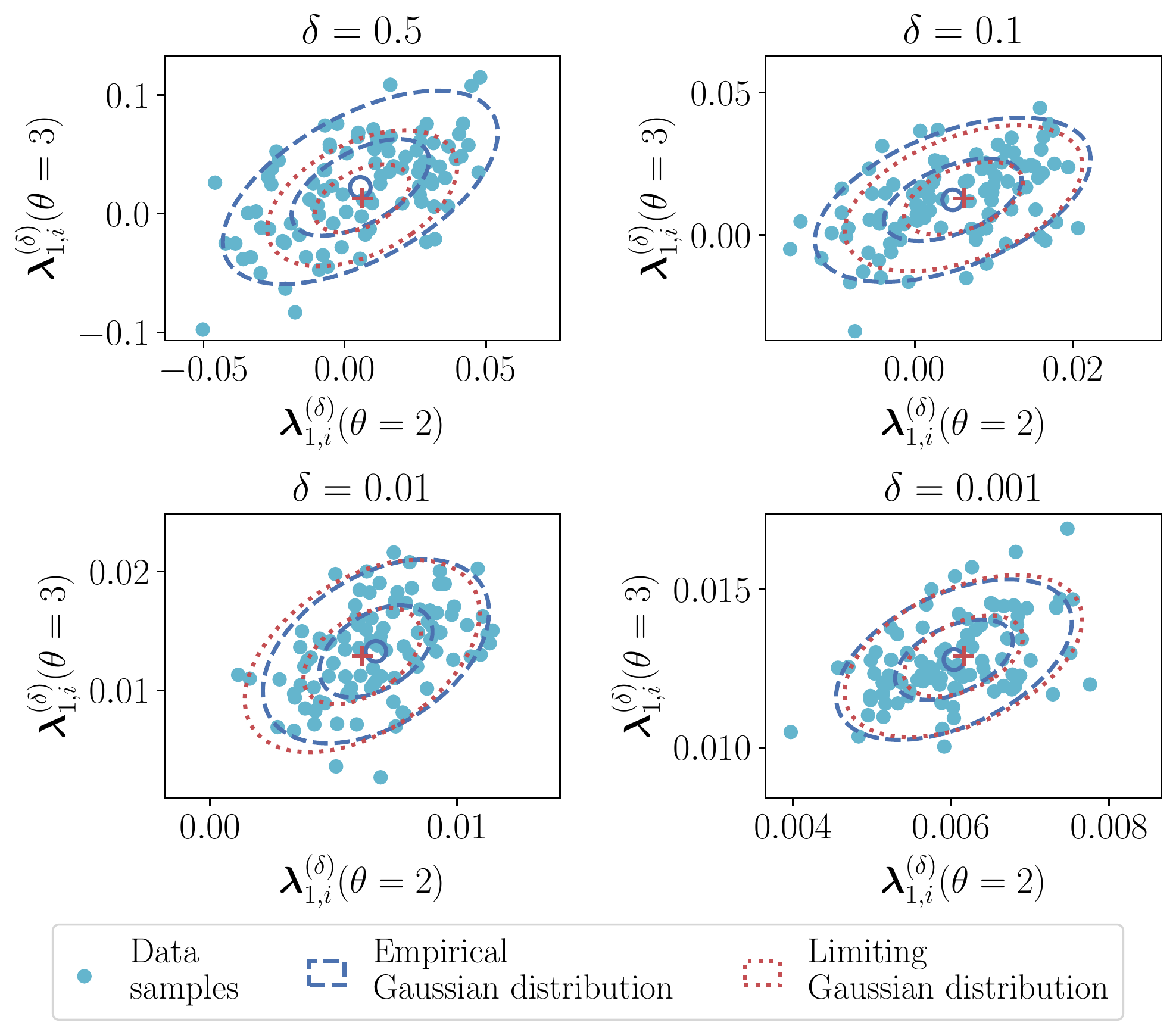}
	\caption{Distribution of data samples at steady state compared with the limiting and empirical Gaussian distributions.}
	\label{fig:theo2}
\end{figure}

We consider $10000$ time samples, where again all agents are collecting data under a true hypothesis $\thetatrue=1$. 
From Theorem~\ref{theor:CLT}, we saw that in steady state we can approximate the log-belief ratios distribution by a multivariate Gaussian pdf, see Eqs. \eqref{eq:CLTfirstapp} and \eqref{eq:CLTsecondapp}. 
In Fig.~\ref{fig:theo2}, we assume that the ASL algorithm has reached the steady state at $i=10000$, and display the log-likelihood ratios corresponding to instant $i=10000$. The experiment is repeated over $100$ Monte Carlo runs, such that we obtain $100$ realizations of the steady-state variable $\bm{\lambda}_k^{(\delta)}$. 
Moreover, we consider $4$ values of $\delta$. 

In dashed blue lines we see the ellipses representing the confidence intervals relative to one and two standard deviations computed for the empirical Gaussian approximation seen in \eqref{eq:CLTsecondapp}: the smaller ellipse encompasses approximately $68\%$ of the samples whereas the larger ellipse encompasses $95\%$. In red dotted lines, we see the corresponding ellipses for the limiting theoretical Gaussian approximation seen in \eqref{eq:CLTfirstapp}, with the red cross indicating the limiting theoretical expectation ${\sf m}_{\sf ave}$. Note how as $\delta$ decreases, the ellipses tend to be smaller, which is in accordance with the scaling of the covariance matrices by $\delta$ in \eqref{eq:CLTfirstapp} and \eqref{eq:CLTsecondapp}, and the distributions tend to overlap, which is in accordance with the behavior predicted by Theorem~\ref{theor:CLT}.

\subsection{Error Exponents}
We start by evaluating the theoretical exponents for the Laplace example at hand. To this aim, we need to compute first the logarithmic moment generating function of the log-likelihood ratios $\bm{x}_{k,i}(\theta)$ in \eqref{eq:xkidefin}. Since the data follow a Laplace distribution, the log-likelihood ratio is:
\begin{equation}
\bm{x}_{k,i}(\theta)=|\bm{\xi}_{k,i}-e_{k}(\theta)|-|\bm{\xi}_{k,i}-e_{k}(\thetatrue)|.
\end{equation}
Before we proceed to characterize the random variable $\bm{x}_{k,i}(\theta)$, let us define the auxiliary quantity:
\beq
\Delta_{k,\theta} \triangleq e_k(\theta)-e_k(\thetatrue).
\eeq
We also introduce the centered variable $\widetilde{\bm{\xi}}_{k,i}=\bm{\xi}_{k,i}-e_k(\theta_0)$, and therefore we can write:
\begin{equation}
\bm{x}_{k,i}(\theta)=|\widetilde{\bm{\xi}}_{k,i}-\Delta_{k,\theta}|-|\widetilde{\bm{\xi}}_{k,i}|.
\end{equation}
For the case in which $\Delta_{k,\theta}>0$, the random variable $\bm{x}_{k,i}(\theta)$ depends on the random variable $\widetilde{\bm{\xi}}_{k,i}$ in the following manner:
\begin{equation}
\bm{x}_{k,i}(\theta)=\begin{cases}
-\Delta_{k,\theta}, & \text{ if } \widetilde{\bm{\xi}}_{k,i}>\Delta_{k,\theta},\\
\Delta_{k,\theta}-2\widetilde{\bm{\xi}}_{k,i}, & \text{ if } \widetilde{\bm{\xi}}_{k,i}\in\left[0,\Delta_{k,\theta}\right],\\
\Delta_{k,\theta}, & \text{ if } \widetilde{\bm{\xi}}_{k,i}<0.
\end{cases}
\end{equation}
We can then express the cumulative distribution function of $\bm{x}_{k,i}(\theta)$ as
\begin{IEEEeqnarray}{rCl}
	\IEEEeqnarraymulticol{3}{l}{\mathbb{P}[\bm{x}_{k,i}(\theta)\leq x]}\IEEEnonumber\\&=&\begin{cases}
		0,&\text{ if }x<-\Delta_{k,\theta},\\
		\mathbb{P}\left[\widetilde{\bm{\xi}}_{k,i}\geq \frac{\Delta_{k,\theta}-x}{2} \right], &\text{ if }x\in\left[-\Delta_{k,\theta},\Delta_{k,\theta}\right],\\
		1,&\text{ if }x>\Delta_{k,\theta},\label{eq:cdf}
	\end{cases}
\end{IEEEeqnarray}
where $\mathbb{P}[\mathcal{A}]$ is the probability of event $\mathcal{A}$, computed from the distribution of $\widetilde{\bm{\xi}}_{k,i}$. Note that its probability density function is given by $L_k(\xi+e_k(\thetatrue)|\thetatrue)$, which is a Laplace distribution with zero mean and scale parameter 1.

From the cumulative distribution function in \eqref{eq:cdf}, we can derive the density function of $\bm{x}_{k,i}(\theta)$ as:
\begin{IEEEeqnarray}{rCl}
	p(x)&=&\mathbb{P}\left[\widetilde{\bm{\xi}}_{k,i}>\Delta_{k,\theta}\right]\delta(x+\Delta_{k,\theta})\IEEEnonumber\\&+&\mathbb{P}\left[\widetilde{\bm{\xi}}_{k,i}<0\right]\delta(x-\Delta_{k,\theta})\IEEEnonumber\\&+&\frac{1}{2}L_k\left(\frac{\Delta_{k,\theta}-x}{2}+e_{k}(\thetatrue)\Big| \thetatrue\right)\mathrm{rect}\left(\frac{x}{2\Delta_{k,\theta}}\right),\IEEEnonumber\\
	\quad&=&\frac{1}{2}\exp\left[-\Delta_{k,\theta}\right]\delta(x+\Delta_{k,\theta})+\frac{1}{2}\delta(x-\Delta_{k,\theta})\IEEEnonumber\\&+&\frac{1}{4}\exp\left[-\frac{(\Delta_{k,\theta}-x)}{2}\right]\mathrm{rect}\left(\frac{x}{2\Delta_{k,\theta}}\right),\label{eq:lmgfpdf}
\end{IEEEeqnarray}
where $\mathrm{rect}(\cdot)$ is the rectangle function, i.e., it is equal to $1$ in the interval $]-\frac{1}{2},\frac{1}{2}[$ and $0$ elsewhere. Also we should distinguish the notation $\delta(x)$, which represents the Dirac delta-function, from the notation $\delta$, which refers to the step-size parameter. 

\begin{figure}[t]
	\centering
	\includegraphics[width=.95\linewidth]{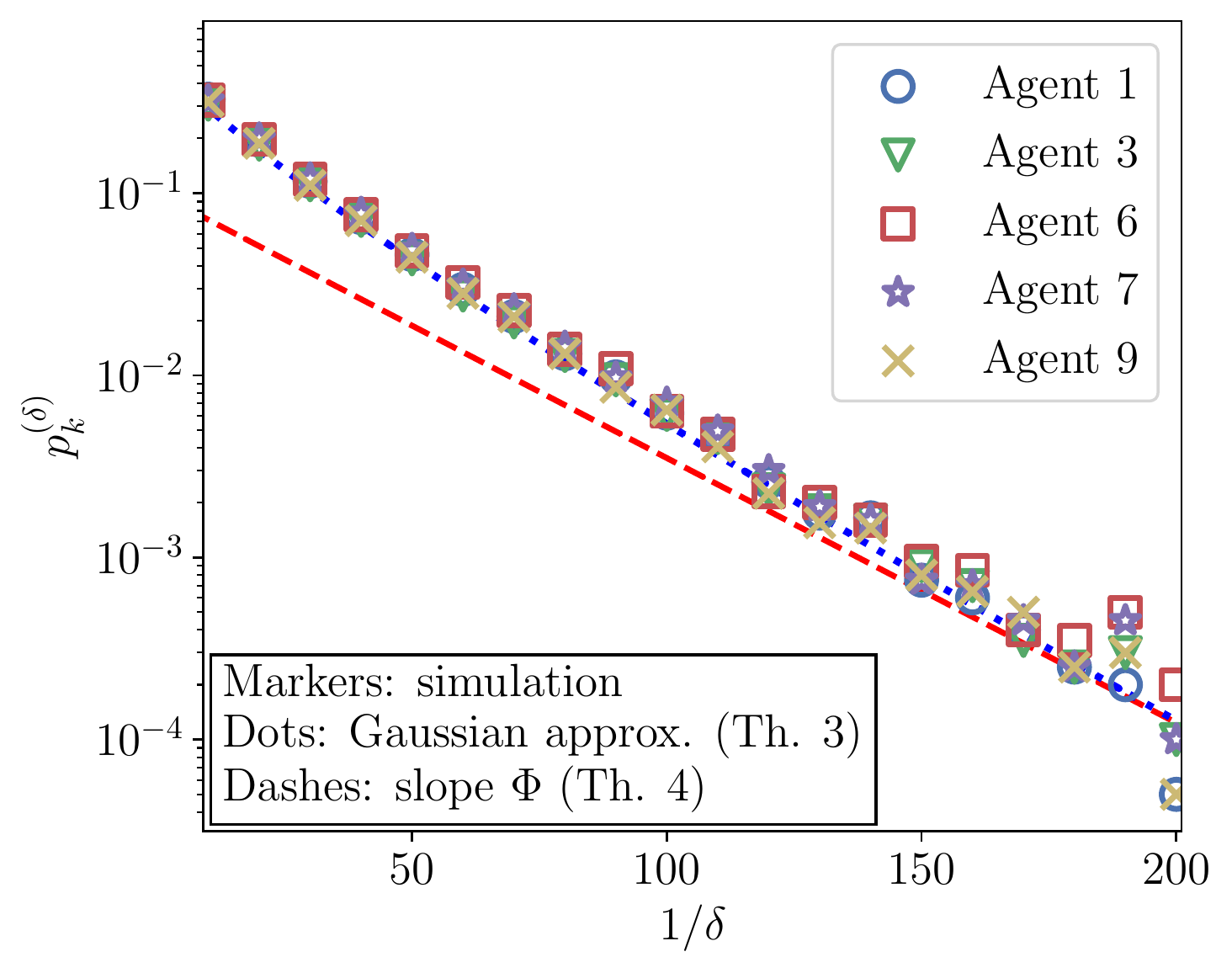}
	\caption{Steady-state error probability. 
		Markers refer to the empirical probability curves estimated via Monte Carlo simulation. 
		The dotted line refers to the theoretical error probability in \eqref{eq:errprob} computed using the Gaussian approximation in \eqref{eq:CLTfirstapp}. 
		The slope of the probability curves is compared against the slope $\Phi$ (i.e., the error exponent) predicted by Theorem~\ref{theor:LD}, and shown with dashed line.
	}
	\label{fig:theo4}
\end{figure}

The LMGF of variable $\bm{x}_{k,i}(\theta)$, whose expression was seen in \eqref{eq:LMGF}, can be explicitly computed using \eqref{eq:lmgfpdf}:
\begin{IEEEeqnarray}{rCl}
	\Lambda_k(t;\theta)&=&\log\left( \int_{\mathbb{R}}e^{tx}p(x)dx \right)\IEEEnonumber\\
	&=&\log\left[\frac{1}{2}\exp(-\Delta_{k,\theta}(t+1) )+\frac{1}{2}\exp(\Delta_{k,\theta}t)\right.\IEEEnonumber\\&+&\left.\frac{1}{2}\exp\left(-\frac{\Delta_{k,\theta}}{2}\right)\frac{\sinh(\Delta_{k,\theta}(t+1/2))}{t+1/2}\right].
\end{IEEEeqnarray}
If similar steps are followed for the case $\Delta_{k,\theta}<0$, we would find the following expression for the LMGF:
\begin{IEEEeqnarray}{rCl}
	\Lambda_k(t;\theta)&=&\log\left[\frac{1}{2}\exp(\Delta_{k,\theta}(t+1) )+\frac{1}{2}\exp(-\Delta_{k,\theta} t)\right.\IEEEnonumber\\&&\left.-\frac{1}{2}\exp\left(\frac{\Delta_{k,\theta}}{2}\right)\frac{\sinh(\Delta_{k,\theta}(t+1/2))}{t+1/2}\right].
\end{IEEEeqnarray}
Assuming that the true state is $\thetatrue=1$, we can then evaluate numerically $\Phi(\theta)$ by employing the expressions in Theorem \ref{theor:LD}, for $\theta=2$ and $\theta=3$, from which we obtain $\Phi(2)=0.03348$ and $\Phi(3)=0.05051$. Finally, the error probability dominant exponent is given by:
\beq
\Phi=\min_{\theta\in\{2,3\}}\Phi(\theta)=0.03348
\eeq 
Now we illustrate the details of the numerical experiments. 
We consider that the true state of nature is set as $\thetatrue=1$, and we let all agents execute the ASL algorithm for $3000$ iterations and for $20$ values of $\delta$ in the interval $[1/150,1/10]$. We run $20000$ Monte Carlo experiments and we compute the steady-state empirical probability of error for each agent and each value of $\delta$.
In Fig.~\ref{fig:theo4}, the empirical probability curves of agents $1,3,6,7,9$ are compared against the theoretical error probability in \eqref{eq:errprob} computed using the Gaussian approximation in \eqref{eq:CLTfirstapp}. The slope of these curves is compared against the slope $\Phi$ (i.e., the error exponent) predicted by Theorem~\ref{theor:LD}.

\section{Evolution over Successive Learning Cycles}
\label{sec:learningcycles}
In this section we would like to focus on a specific nonstationary setting to illustrate in more detail the role of adaptation. 
We consider the time axis can be divided into successive {\em random} intervals ({\em learning cycles}) wherein the system conditions remain stationary. We do not focus here on situations where the system parameters can vary smoothly at each time instant following some ``trajectory'', as happens, e.g., in tracking applications. 
While from the analysis of similar algorithms we can expect that the ASL strategy possesses some inherent tracking ability, the study of this scenario is left for future work.

\begin{figure}[t]
	\centering
	\includegraphics[width=\linewidth]{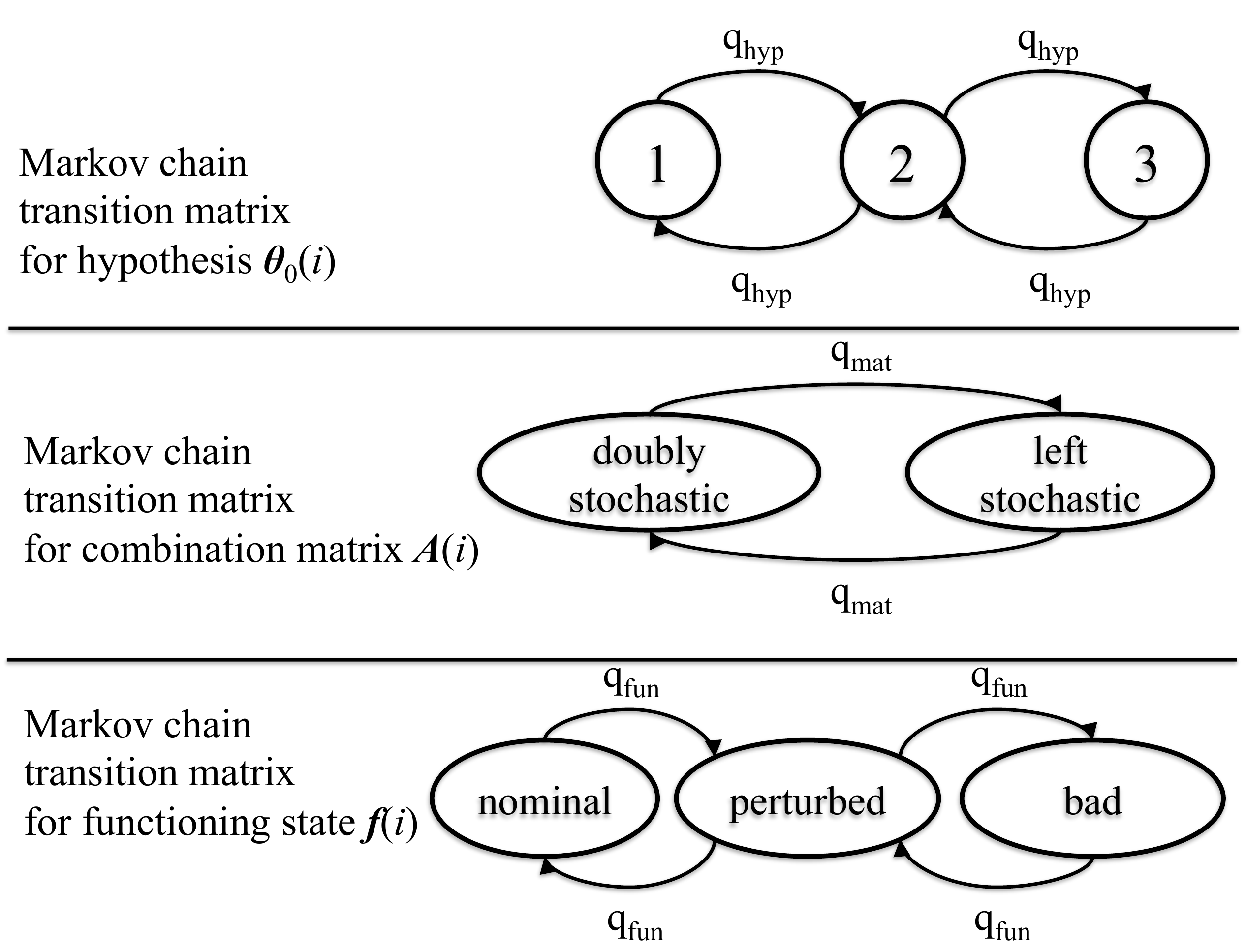}
	\caption{Transition matrices of the Markov chains corresponding to the sources of nonstationarity illustrated in Sec.~\ref{sec:learningcycles}. } 
	\label{fig:MCtmat}
\end{figure}

We examine an environment where there are three different sources of nonstationarity, which will be modeled as (mutually independent) homogeneous Markov chains, as now specified:

\begin{itemize}
	\item
	The true hypothesis can change over time.
	For $i=1,2,\ldots$, the true state of nature at time $i$, denoted by $\bm{\theta}_0(i)$, follows a Markov process with possible states in $\Theta=\{1,2,3\}$ and with transition probabilities described by the finite-state diagram in Fig.~\ref{fig:MCtmat} (where only transition probabilities are displayed, with the complementary probabilities of remaining in a state being omitted).  
	\item
	The combination policy can change over time. 
	We assume that the agents employ two possible combination matrices, one doubly-stochastic, the other left-stochastic. 
	For $i=1,2,\ldots$, the combination matrix in force at time $i$, denoted by $\bm{A}(i)$, follows a Markov process with transition matrix represented by the corresponding finite-state diagram in Fig.~\ref{fig:MCtmat}.
	\item
	The system can be in one of three possible functioning states, namely, nominal, perturbed, and bad. 
	For $i=1,2,\ldots$, the operating state at time $i$ is denoted by $\bm{f}(i)$. 
	Under state $\bm{f}(i)=\textnormal{nominal}$, the data are generated according to the true likelihood corresponding to hypothesis $\theta_0(i)$. 
	Under state $\bm{f}(i)=\textnormal{perturbed}$, some noise is added to perturb the true data model (while the agents still rely on the nominal likelihood to run their ASL strategy). State $\bm{f}(i)=\textnormal{bad}$ corresponds to a failure of the system, where a large amount of noise is added to the data so as to impair the learning process.
	The transition matrix of the functioning process is encoded in the pertinent finite-state diagram in Fig.~\ref{fig:MCtmat}. 
\end{itemize}

Let us evaluate the average duration of a learning cycle. In order to be conservative, we focus on the worst case, i.e., on the shorter average duration, which is obtained when the system is in the most unstable case (i.e., the state where transitions are more frequent).
Examining Fig.~\ref{fig:MCtmat}, the most unstable state is obtained when: $i)$ the hypothesis in force is $\bm{\theta}_0(i)=2$, since from such intermediate state the Markov chain can move leftward or rightward, while from the other states it cannot; $ii)$ the combination policy is either left stochastic or doubly stochastic; and $iii)$ the system works under a perturbed state of functioning, for the same reasons as in point $i)$. 
Now, given that the overall system is in the joint state $\{\bm{\theta}_0(i)=2,\bm{A}(i)=\textnormal{left stochastic},\bm{f}(i)=\textnormal{perturbed}\}$, the probability that the system remains stable for a single step is equal to:
\beq
q^{\star}=(1-2\,q_{\mathrm{hyp}})(1-q_{\mathrm{mat}})(1-2\,q_{\mathrm{fun}}).
\eeq
Likewise, the probability that the system remains stable for a certain number of steps is ruled by a geometric distribution of parameter $q^{\star}$, yielding the following average duration for the worst-case learning cycle:
\beq
{\sf T}_{\sf LC}=\frac{q^{\star}}{1-q^{\star}}.
\label{eq:avTLC}
\eeq
In order to model a nonstationary environment where the system parameters remain stable during the learning cycles, we take inspiration from the Gilbert-Elliott model typically employed to model random bursts of errors over communication channels~\cite{Gilbert,Elliott}. According to the Gilbert-Elliott model, the transition probabilities between states of the chain are kept small so as to ensure that the chain remains in the same state for some contiguous time samples (i.e., we have ``bursts'' where the same state is repeatedly observed). 

For what concerns the nominal likelihood models, we use the following family of Laplace likelihood functions, for $n\in\{1,2,3\}$:
\begin{equation}
f_n(\xi)=\frac{1}{2}\exp\left\{-|\xi - n| \right\},
\end{equation}
under the same identifiability setup as in Table~\ref{tab:id}.
The network topology is the same as in Fig.~\ref{fig:network}, on top of which we build two possible combination matrices: a left-stochastic matrix obtained through a uniform-averaging combination policy, and a doubly-stochastic matrix obtained through a Laplacian combination policy~\cite{Sayed}.
Under this setting, we evaluate the adaptation time exploiting \eqref{eq:adaptime2}. 
Regarding the initial states appearing in \eqref{eq:adaptime2}, we assume that in a given learning cycle the system comes from a previous learning cycle where the agents converged to a hypothesis different from that in force during the current learning cycle. 
Then we consider the worst-case initial state, and further the worst-case over all possible $\theta$ and $\theta_0$. 
With these conservative choices, the time necessary to stay at $3$ dB from the exponent $\Phi$ is equal to:
\beq
{\sf T}_{\sf ASL}\approx \frac{2.7286}{\delta}.
\label{eq:adaptimeeval}
\eeq
We now examine two settings that correspond to (relatively) short and long learning cycles, respectively.

\begin{figure}[t]
	\centering
	\includegraphics[width=\linewidth]{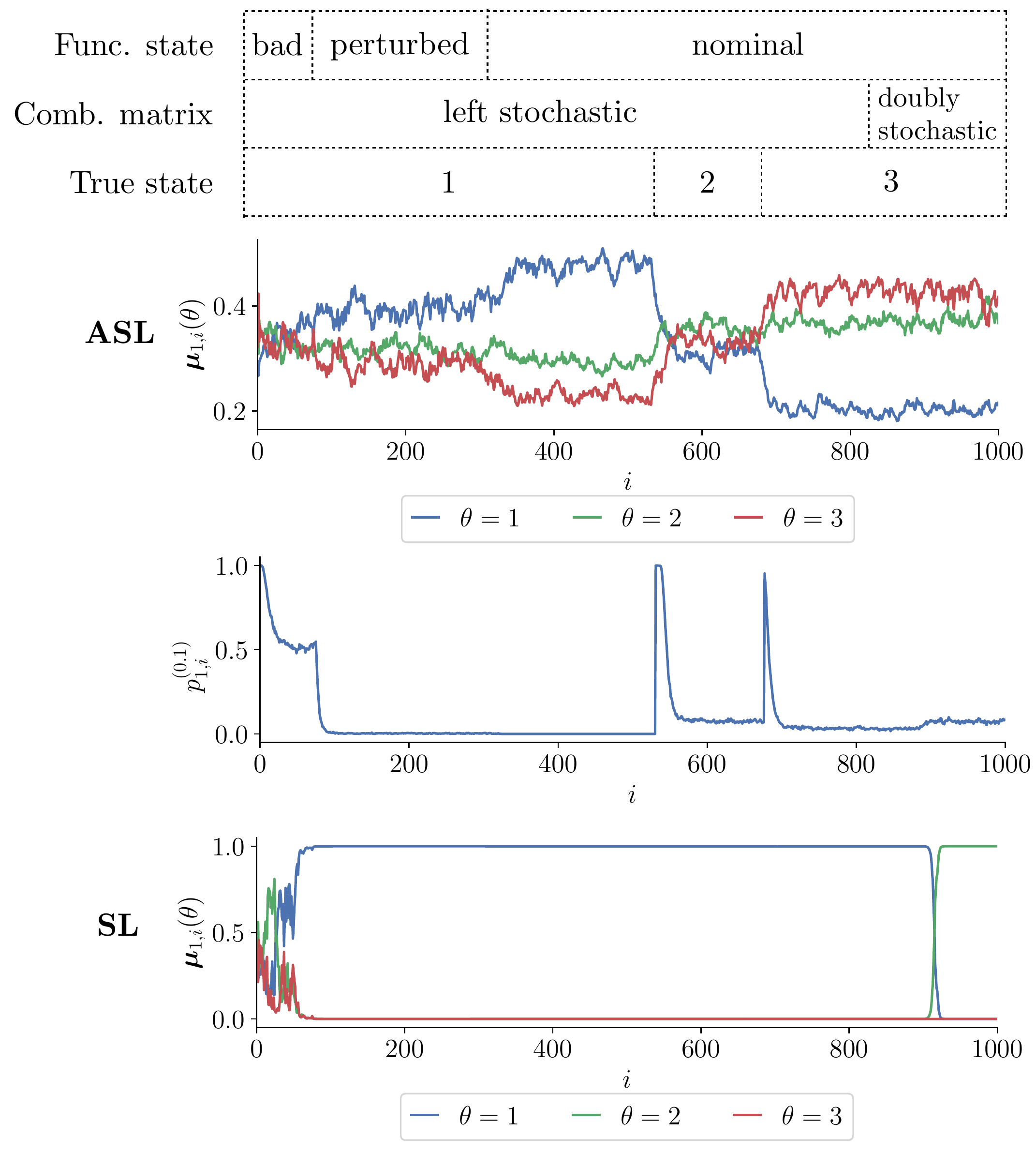}
	\caption{Evolution of the learning strategies over successive learning cycles, with step-size $\delta=0.1$ and average learning-cycle duration ${\sf T}_{\sf LC}\approx 100$. 
			{\em First (top) row}. Observed transitions for the three sources of nonstationarity illustrated in the main text, namely, state of functioning, combination matrix, and hypothesis.
			{\em Second row}. Time-evolution of the belief at agent $1$ for the {\em adaptive} social learning strategy. 
			{\em Third row}. Time-evolution of the error probability at agent $1$ for the {\em adaptive} social learning strategy.
			{\em Fourth row}. Time-evolution of the belief at agent $1$ for the {\em traditional} social learning strategy. } 
	\label{fig:bel1}
\end{figure}
\begin{figure}[t]
	\centering
	\includegraphics[width=\linewidth]{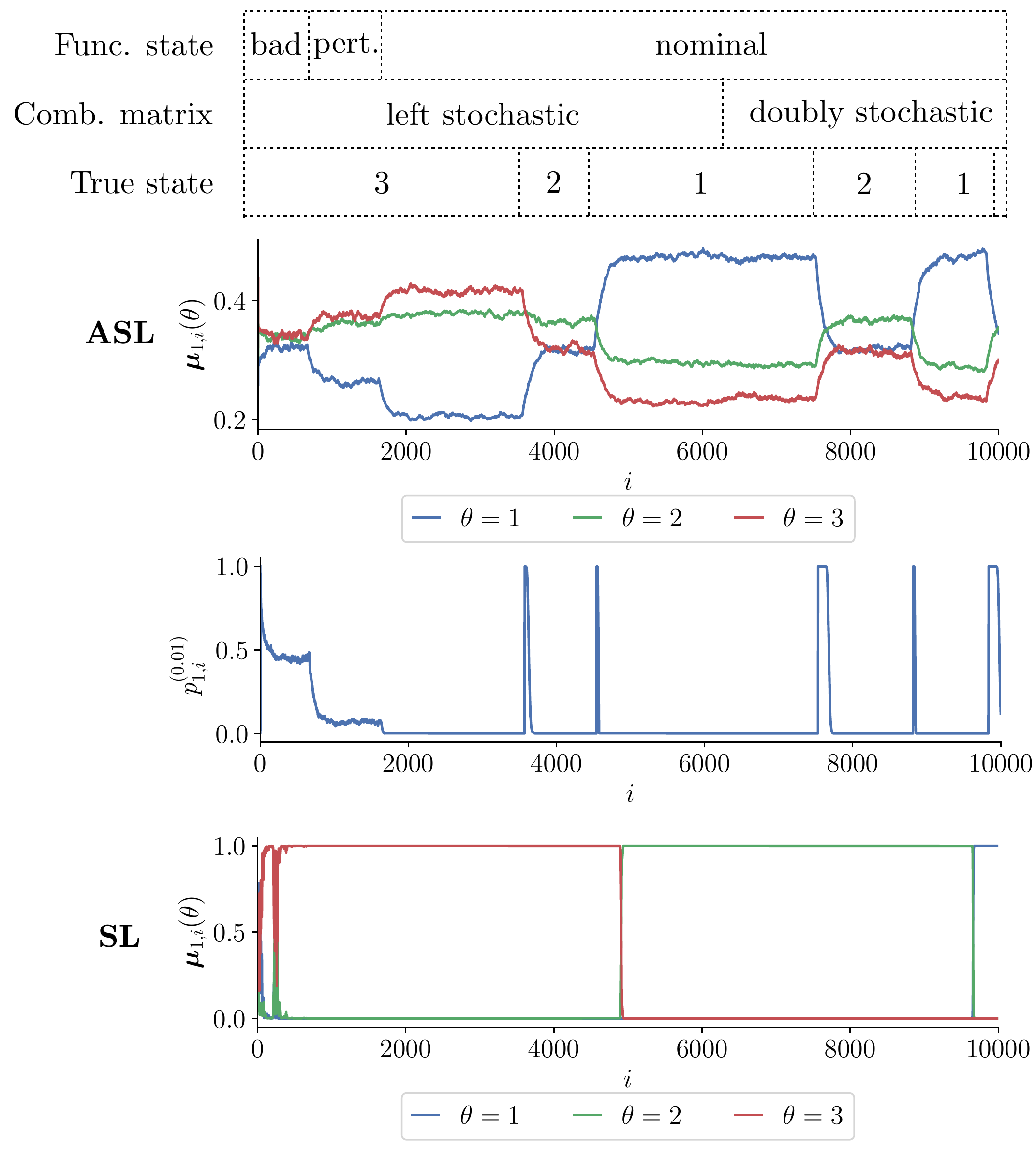}
	\caption{Evolution of the learning strategies over successive learning cycles, with step-size $\delta=0.01$ and average learning-cycle duration ${\sf T}_{\sf LC}\approx 1000$. 
		{\em First (top) row}. Observed transitions for the three sources of nonstationarity illustrated in the main text, namely, state of functioning, combination matrix, and hypothesis.
		{\em Second row}. Time-evolution of the belief at agent $1$ for the {\em adaptive} social learning strategy. 
		{\em Third row}. Time-evolution of the error probability at agent $1$ for the {\em adaptive} social learning strategy.
		{\em Fourth row}. Time-evolution of the belief at agent $1$ for the {\em traditional} social learning strategy. } 
	\label{fig:bel2}
\end{figure}

{\em -- ``Short'' Learning Cycles}. 
First of all, we consider that malfunctioning events and variations of the combination matrix are rare as compared to changes in the hypothesis. In particular, we set:
\beq
q_{\mathrm{hyp}}=5\times10^{-3},~~q_{\mathrm{mat}}=10^{-3},~~q_{\mathrm{fun}}=10^{-3}.
\eeq
Exploiting \eqref{eq:avTLC}, the average duration of a learning cycle can be approximated as ${\sf T}_{\sf LC}\approx 76$.
If we equate the value found for ${\sf T}_{\sf LC}$ to the adaptation time in \eqref{eq:adaptimeeval}, we get $\delta=0.035$. 
For proper learning, we need that the adaptation time is smaller than the average duration of a learning cycle to ensure convergence to the correct hypothesis. 
In the experiments shown in Fig.~\ref{fig:bel1} we made the choice:
\beq
\delta=0.1,
\eeq
which corresponds to an adaptation time not larger than one third of the average worst-case learning cycle.
In Fig.~\ref{fig:bel1}, we display: in the second row, the time-evolution of the beliefs at agent $1$ corresponding to one realization of the process; in the third row, the corresponding error probability; and in the fourth row, the time-evolution of the beliefs at agent $1$ for a traditional social learning strategy (same realization considered for the ASL strategy). 
During the considered time interval, several variations occurred, according to the nonstationary model described before. 

First, we see that, except for the learning cycle corresponding to a bad state of functioning, the ASL strategy is able to learn well in all learning cycles, after a relatively short transient at the beginning of each learning cycle. 
The ability of learning well is showed by the time-evolution of the beliefs (second row), which shows how the maximum belief corresponds to the true hypothesis, after relatively short adaptation intervals necessary to react in face of nonstationarities. 
More quantitatively, the ability of learning is showed by the time-evolution of the error probabilities (third row), where we see some peaks (error probability close to $1$) that clearly correspond to the changes, and that have a short duration dictated by the adaptation times. 
In sharp contrast, the traditional social learning strategy looses irremediably its learning ability yet after the first learning cycle.

Zooming in on Fig.~\ref{fig:bel1}, we see that nonstationarities in the hypotheses induce a perceivable change in the learning performance, whereas nonstationarities in the combination policy or in the state of functioning deserve a separate analysis.

For what concerns the combination policies, we see that the learning ability is preserved in face of a change, i.e., the system does {\em not} undergo an interval of poor performance. This behavior makes perfect sense, since from the theoretical analysis we know that the ASL strategy must consistently learn both with a left-stochastic or a doubly-stochastic matrix. 
What can be different is the learning performance, which depends on the Perron eigenvector. In this particular example, we have verified that, as opposed to the uniform Perron eigenvector corresponding to the doubly-stochastic matrix, the eigenvector of the left-stochastic matrix features higher weights corresponding to more informative agents (i.e., agents with higher KL divergences), which provides an explanation of the slightly increased performance observed in Fig.~\ref{fig:bel1}.

Regarding the state of functioning, we see that during the ``bad'' functioning state the data do not provide useful information, and the system undergoes an interval of failure (error probability $\approx 0.5$). Remarkably, the adaptivity of the ASL strategy allows the agents to recover from this failure state in the successive learning cycles. 
In particular, the agents are able to recover and learn well already during the ``perturbed'' state of functioning. 
Actually, this regime of operation where the data do not follow any of the nominal likelihoods is not covered by our steady-state analysis. Our results could be in principle extended by allowing arbitrary distributions for the true data. 
In this case, as an inherent property of the ASL recursion, what is certainly preserved is the convergence to a {\em stable limiting state that does not diverge as the step-size $\delta$ becomes small}. What is lost is the relation between the limiting statistics and the KL divergences. Accordingly, the possibility of consistent learning would depend on the particular behavior of the limiting random variables under the non-nominal distributions. In light of these observations, it is not unexpected that, for reasonable amounts of perturbation, the agents are still able to learn, as happens in the considered example. 
Intuitively, we expect that, passing from a perturbed to a nominal state, the performance improves. 
Visually, this effect would be more clearly appreciated in the subsequent example shown in Fig.~\ref{fig:bel2}.

In summary, we see that the starting values at the beginning of each learning cycle are stable, since they arise as steady-state limiting values from at the end of the previous learning cycle. As such, these starting values do not diverge as time elapses, guaranteeing proper learning over successive learning cycles. 
This is a critical property, since it reveals that the number of variations of the underlying statistical conditions occurring during the entire algorithm evolution does not impair successful learning of the ASL strategy. 
What really matters is that the duration of the learning cycle is sufficiently large to allow a (small) value of $\delta$ to enable accurate learning.

{\em -- ``Long'' Learning Cycles}. In Fig.~\ref{fig:bel2}, we consider the more favorable situation where the average duration of the learning cycle is increased by one order of magnitude, using the following transition probabilities for the pertinent Markov chains: 
\beq
q_{\mathrm{hyp}}=5\times 10^{-4},~~q_{\mathrm{mat}}=10^{-4},~~q_{\mathrm{fun}}=10^{-4}.
\eeq
Accordingly, we expect that the adaptation properties of the system will be preserved if we reduce the step-size by one order of magnitude, yielding:
\beq
\delta=0.01.
\eeq
Comparing Fig.~\ref{fig:bel2} against Fig.~\ref{fig:bel1}, we see that the general behavior is perfectly confirmed, and two notable effects emerge. 
First, the adaptation properties are preserved, i.e., the system is able to adapt to the changes sufficiently fast to guarantee a stable evolution over successive learning cycles. 
Second, the fluctuations around the limiting steady-state are reduced w.r.t. Fig.~\ref{fig:bel1}, yielding a smaller error probability, as it must be according to the theoretical analysis carried out in the previous sections since we are now using a smaller step-size $\delta=0.01$.

\section{Concluding Remarks}
Social learning is a relevant inferential paradigm lying at the core of many multi-agent systems. 
Under this paradigm, the agents are able to learn progressively an underlying state of nature by continually updating their beliefs based on the incoming streaming data and the beliefs exchanged with their neighbors. 

Several social learning implementations are currently available. 
However, these implementations are not open to deal with {\em nonstationary} data. For example, even if the agents learned well the true state, when this state changes at a certain instant, in the traditional social learning implementations the agents tend to be stubborn and keep on believing in the old state. 
In this work we proposed an Adaptive Social Learning (ASL) strategy that overcomes this issue and examined its performance and convergence guarantees in some great detail. 
The key insight is the introduction of an {\em adaptive update} depending on a step-size parameter $\delta$ that allows to tune the degree of adaptation. The introduction of the step-size $\delta$ allows the user to explore the trade-off between accuracy in decision making and adaptation speed.

A careful analysis of the system performance has been provided. 
In the steady-state phase, with focus on the small step-size regime, we have ascertained that the ASL strategy is able to learn consistently, and we have provided reliable performance characterization of the learning performance at each individual agent. In the transient phase, we have shown how the learning performance evolves over time and how the choice of the step-size affects the adaptation time.
	
Several useful extensions and generalizations are possible. 
One extension refers to removing the assumption that the agents' updates and data sharing are synchronous. An asynchronous version of the ASL strategy could be devised and examined exploiting the tools applied in the context of adaptive diffusion algorithms~\cite{sayedzhao}. 

Another interesting generalization concerns the analysis of the ASL performance in {\em tracking} applications where some system parameters vary ``smoothly'', as opposed to the setting considered in this work where the system evolves under stationary conditions during individual learning cycles.

Finally, the technical machinery used to prove our results can be exploited to characterize adaptive social learning under broader non-ideal settings, for example: when the true distribution is different from the nominal likelihoods assumed by the agents~\cite{NedicTAC2017,MattaBordignonSantosSayed2019}; when the learning task amounts to solving an optimization problem with multiple optimal solutions (i.e., there does not exist a unique truth from the agents' perspective)~\cite{NedicTAC2017}; or when the network is weakly connected, giving rise to a dichotomy between influential and influenced agents~\cite{MattaBordignonSantosSayed2019}.

\section*{Acknowledgment}
The authors wish to thank the anonymous Reviewers for the careful reading of the manuscript and for the valuable suggestions that helped improve the quality of the work.

\begin{appendices}

	\section{}
	\label{app:mainlemmapp}
	In the following, the symbols $\mathcal{S}^{o}$ and $\overline{\mathcal{S}}$ denote the interior and the closure of set $\mathcal{S}$, respectively. 
	
	\begin{lemma}[{\bf Asymptotic properties of random series useful for adaptation}]
		\label{lem:mainlemma}
		For $m=0,1,\ldots$, let $\{\bm{z}_m\}$ be a sequence of i.i.d. integrable random variables with:
		\beq
		{\sf m}_z\triangleq\E\Big[\bm{z}_m\Big],\qquad
		{\sf m}^{\mathrm{abs}}_z\triangleq\E\Big[|\bm{z}_m|\Big]<\infty.
		\label{eq:absint}
		\eeq
		Let also $0<\delta<1$, and consider the following partial sums:
		\beq
		\bm{s}_i(\delta)=\delta
		\sum_{m=0}^{i}(1-\delta)^m \alpha_m \bm{z}_m,
		\label{eq:randomseries}
		\eeq
		where $0<\alpha_m\leq 1$, with $\alpha_m$ converging to some value $\alpha>0$ and obeying the following upper bound for all $m$:
		\beq
		|\alpha_m - \alpha| \leq \kappa \beta^m,
		\label{eq:exprate}
		\eeq
		for some constant $\kappa>0$ and for some $0<\beta<1$.
		Then, we have the following asymptotic properties.
		
		\noindent
		{\bf 1. Steady-state stability}. The partial sums in \eqref{eq:randomseries} are almost-surely absolutely convergent, namely, we can define the (almost-surely) convergent series:
		\beqa
		\bm{s}^{\mathrm{abs}}(\delta)&\triangleq&\delta\sum_{m=0}^{\infty}(1-\delta)^m \alpha_m |\bm{z}_m|,
		\label{eq:limitdef1}\\
		\bm{s}(\delta)&\triangleq&\delta\sum_{m=0}^{\infty}(1-\delta)^m \alpha_m \bm{z}_m.
		\label{eq:limitdef2}
		\eeqa
		\noindent
		{\bf 2. First moment}. The expectation of $\bm{s}(\delta)$ is:
		\beq
		\E[\bm{s}(\delta)]={\sf m}_z\delta\sum_{m=0}^{\infty}(1-\delta)^m \alpha_m=\alpha\,{\sf m}_z + O(\delta),
		\label{eq:expeclemma}
		\eeq
		where $O(\delta)$ is a quantity such that the ratio $O(\delta)/\delta$ remains bounded as $\delta\rightarrow 0$.
		
		\noindent
		{\bf 3. Weak law of small step-sizes}. 
		The series $\bm{s}(\delta)$ converges to $\alpha\,{\sf m}_z$ in probability as $\delta\rightarrow 0$, namely, for all $\epsilon>0$ we have that:
		\beq
		\lim_{\delta\rightarrow 0}\P\left[|\bm{s}(\delta)-\alpha\,{\sf m}_z|>\epsilon\right]=0.
		\label{eq:weaklawequ}
		\eeq
		\noindent
		{\bf 4. Second moment}. If:
		\beq
		\sigma^2_z \triangleq\VAR[\bm{z}_m]<\infty,
		\label{eq:VARsingle}
		\eeq
		then:
		\beqa
		\VAR[\bm{s}(\delta)]&=&
		\sigma^2_z\delta^2\sum_{m=0}^{\infty}(1-\delta)^{2m} \alpha^2_m\nonumber\\
		&=&\frac{\alpha^2\sigma^2_z}{2}\,\delta+O(\delta^2).
		\label{eq:VARlemma}
		\eeqa
		\noindent
		{\bf 5. Asymptotic normality}. If $\bm{z}_m$ has finite variance $\sigma^2_z$, then the following convergence in distribution holds:
		\beq
		\frac{\bm{s}(\delta)-{\sf m}_z}{\sqrt{\delta}}\stackrel{\delta\rightarrow 0}{\rightsquigarrow} 
		\mathscr{G}\Big(0,\alpha^2\sigma^2_z/2\Big),
		\label{eq:CLTlemma}
		\eeq
		and, hence, $\bm{s}(\delta)$ is asymptotically normal as $\delta\rightarrow 0$.
		
		\begin{figure}[t]
			\centering
			\includegraphics[width=.9\linewidth]{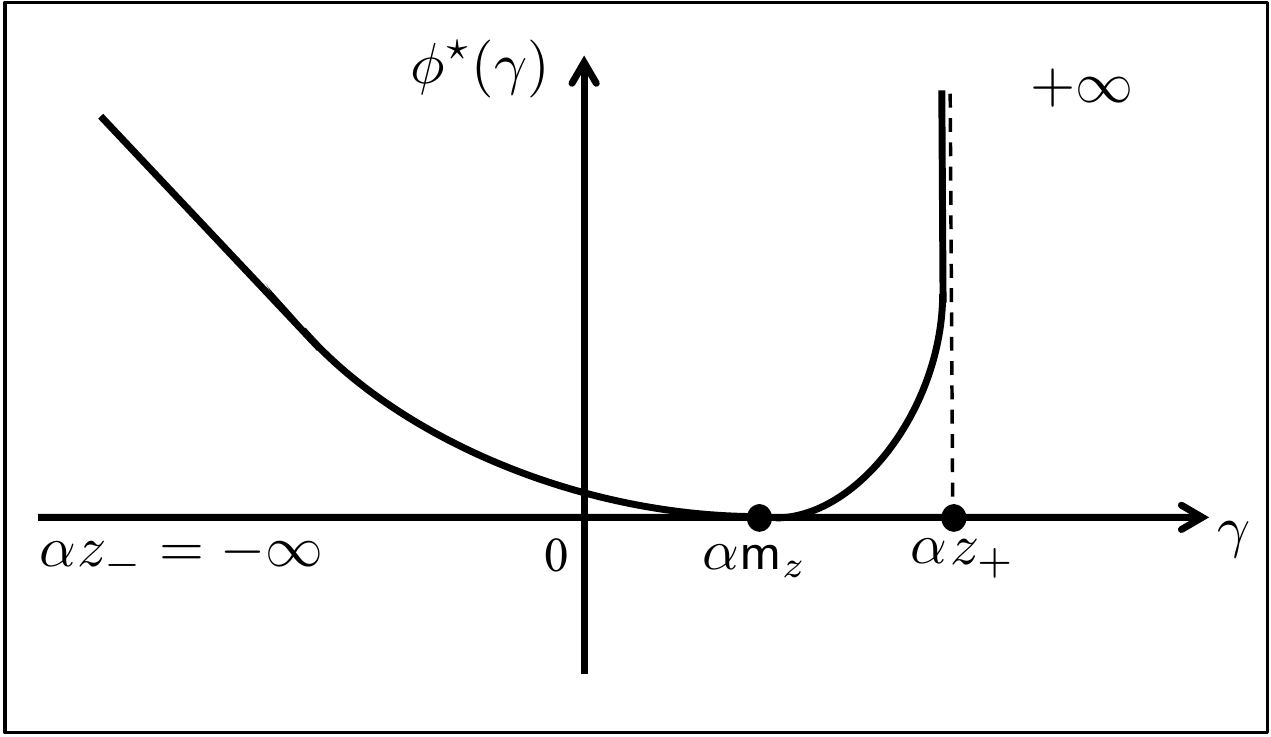}
			\caption{Typical shape of the rate function.}
			\label{fig:ratefun}
		\end{figure}
		\noindent
		{\bf 6. Large deviations}. 
		Assume that $\bm{z}_m$ is non-deterministic and has LMGF finite everywhere:
		\beq
		\Lambda_z(t)=\log\E\left[e^{\bm{z}_m t}\right]<+\infty,~~\forall t\in\mathbb{R}.
		\eeq
		Let $\Lambda_{\alpha z}(t)=\Lambda_z(\alpha t)$ be the LMGF of the scaled variable $\alpha \bm{z}_m$, where $\alpha$ is defined in \eqref{eq:exprate}. 
		Denoting by $\Lambda_{\delta}(t)$ the LMGF of $\bm{s}(\delta)$, we have that:
		\beq
		\lim_{\delta\rightarrow 0} \delta \Lambda_{\delta}(t/\delta)=\phi(t)=\int_{0}^t \frac{\Lambda_{\alpha z}(\tau)}{\tau}d\tau.
		\label{eq:philemma}
		\eeq 
		Then the following Large Deviations Principle (LDP) holds for any measurable set $\mathcal{S}$ (the infimum over an empty set is taken as $+\infty$):
		\beqa
		&&\liminf_{\delta\rightarrow 0} \;\delta\log \P[\bm{s}(\delta)\in \mathcal{S}]
		\geq
		-\!\inf_{\gamma\in\mathcal{S}^{o}}
		\phi^{\star}(\gamma),
		\label{eq:LDPgeneral1}\\
		&&\limsup_{\delta\rightarrow 0} \delta\log \P[\bm{s}(\delta)\in \mathcal{S}]
		\leq 
		-\inf_{\gamma\in\overline{\mathcal{S}}}\phi^{\star}(\gamma),
		\label{eq:LDPgeneral2}
		\eeqa
		where
		\beq
		\phi^{\star}(\gamma)=\sup_{t\in\mathbb{R}} [\gamma t - \phi(t)]
		\label{eq:FenchLeg}
		\eeq
		is the Fenchel-Legendre transform of $\phi(t)$~\cite{DemboZeitouni,DenHollander}. 
		The function $\phi^{\star}(\gamma)$ (which is allowed to be an extended real number) is usually called rate function~\cite{DemboZeitouni,DenHollander} and has the following properties. 
		\begin{itemize}
			\item
			Let $z_{-}$ and $z_{+}$ be the boundaries of the support of $\bm{z}_m$, and let $\mathcal{D}=\{\gamma\in\mathbb{R}: \phi^{\star}(\gamma)<+\infty\}$. Then $\mathcal{D}$ is given by the following open interval: 
			\beq
			\mathcal{D}=(\alpha z_{-}, \alpha z_{+}).
			\eeq 
			\item
			The function $\phi^{\star}(\gamma)$ is smooth and strictly convex on $\mathcal{D}$, and diverges to $+\infty$ at the boundaries of $\mathcal{D}$. In particular, if a boundary is finite, the rate function is equal to $+\infty$ at that boundary. 
			\item
			$\phi^{\star}(\gamma)\geq 0$, with equality if, and only if, $\gamma=\alpha {\sf m}_z$. 
		\end{itemize}
		A typical shape of the rate function is illustrated in Fig.~\ref{fig:ratefun}. Exploiting the aforementioned regularity properties of $\phi^{\star}(\gamma)$, from \eqref{eq:LDPgeneral1}--\eqref{eq:LDPgeneral2} we have in particular that: 
		\beqa
		\lim_{\delta\rightarrow 0}\delta\log\P[\bm{s}(\delta)\geq \gamma]&=&-\phi^{\star}(\gamma),~~~\forall \gamma\geq\alpha {\sf m}_z,\\
		\lim_{\delta\rightarrow 0}\delta\log\P[\bm{s}(\delta)\leq \gamma]&=&-\phi^{\star}(\gamma),~~~\forall \gamma\leq\alpha {\sf m}_z.\label{eq:LDPrelevant}
		\eeqa
	\end{lemma}

	\begin{IEEEproof}
		We prove sequentially the six parts of the lemma.
		
		\vspace*{10pt}
		\noindent
		{\bf Part $\mathbf{1}$.}
		In view of \eqref{eq:absint}, the following series of (absolute) expectations is convergent: 
		\beqa
		\delta \sum_{m=0}^{\infty} (1-\delta)^m  \alpha_m \E\Big[|\bm{z}_m|\Big]
		&=&
		{\sf m}^{\mathrm{abs}}_z\delta \sum_{m=0}^{\infty} (1-\delta)^m  \alpha_m 
		\nonumber\\
		&\leq&
		{\sf m}^{\mathrm{abs}}_z\delta \sum_{m=0}^{\infty}(1-\delta)^m
		\nonumber\\
		&=&{\sf m}^{\mathrm{abs}}_z<+\infty. 
		\label{eq:absconvseries}
		\eeqa
		In view of~\cite{Loeve1951}[Lemma~3.6$'$], convergence of the series of absolute first moments implies that the random series $\bm{s}^{\mathrm{abs}}(\delta)$ is almost-surely finite, which in turn implies that so is $\bm{s}(\delta)$, and part $1$ is proved. 
		
		\vspace*{10pt}
		\noindent
		{\bf Part $\mathbf{2}$.}
		Since the series of (absolute) expectations is convergent, so is the series of expectations:
		\beq
		\sum_{m=0}^{\infty}(1-\delta)^m \alpha_m \E[\bm{z}_m]=
		{\sf m}_z\sum_{m=0}^{\infty}(1-\delta)^m \alpha_m.
		\label{eq:meanexpress}
		\eeq 
		On the other hand, by triangle inequality we have the following upper bound:
		\beq
		|\bm{s}_i(\delta)|
		\leq
		\delta\sum_{m=0}^{i}(1-\delta)^m \alpha_m |\bm{z}_m|\leq\bm{s}^{\mathrm{abs}}(\delta).
		\eeq
		Now we observe that $\bm{s}^{\mathrm{abs}}(\delta)$ is a proper random variable in view of part $1$. 
		Furthermore, it is an integrable random variable from Beppo Levi's monotone convergence theorem~\cite{Durrett}[Th.~1.5.7, p.~27], thanks to the convergence of absolute expectations in \eqref{eq:meanexpress}. 
		
		We conclude that the random sequence $\bm{s}_i(\delta)$ is upper bounded by an integrable random variable. 
		Therefore, the dominated convergence theorem~\cite{Durrett}[Th.~1.5.8, p.~27] implies that the expectation of the almost-sure limit $\bm{s}(\delta)$ is equal to the convergent series of expectations, and the first equality in \eqref{eq:expeclemma} follows. 
		Moreover, we can write:
		\beqa
		\delta\sum_{m=0}^{\infty}(1-\delta)^m \alpha_m&=&
		\delta\sum_{m=0}^{\infty}(1-\delta)^m \left(\alpha_m-\alpha\right)
		\nonumber\\
		&+& \alpha
		\underbrace{\delta\sum_{m=0}^{\infty}(1-\delta)^m}_{=1}.
		\label{eq:singlesum}
		\eeqa
		In view of \eqref{eq:exprate}, the absolute value of the first summation on the RHS in \eqref{eq:singlesum} is dominated by:
		\beq
		\kappa\,\delta\sum_{m=0}^{\infty}\Big(\beta(1-\delta)\Big)^m=\frac{\kappa\,\delta}{1-\beta(1-\delta)}=O(\delta).
		\label{eq:Odelta}
		\eeq
		We conclude from \eqref{eq:meanexpress}, \eqref{eq:singlesum} and \eqref{eq:Odelta} that the second equality in \eqref{eq:expeclemma} holds.
		
		\vspace*{10pt}
		\noindent
		{\bf Part $\mathbf{3}$.}
		Let 
		\beq
		\zeta_m\triangleq \delta(1-\delta)^m\alpha_m,
		\label{eq:zetadef}
		\eeq
		and consider the following centered variables:
		\beq
		\widetilde{\bm{s}}(\delta)=\bm{s}(\delta)-\E[\bm{s}(\delta)],\qquad
		\widetilde{\bm{z}}_m=\bm{z}_m-\E[\bm{z}_m].
		\label{eq:centered1}
		\eeq
		In view of parts $1$ and $2$, the centered partial sums: 
		\beq
		\bm{s}_i(\delta)-\E[\bm{s}_i(\delta)]=
		\sum_{m=0}^i \zeta_m\widetilde{\bm{z}}_m
		\eeq 
		converge in distribution to $\widetilde{\bm{s}}(\delta)$ as $i\rightarrow\infty$. 
		By L\'evy's continuity theorem,  the corresponding characteristic functions must converge~\cite{FellerBookV2}[Th.~2, p.~431].
		Since the $\bm{z}_m$'s are i.i.d. we can write:
		\beq
		\chf(t)\triangleq \E\left[
		e^{j \widetilde{\bm{s}}(\delta) t}
		\right]
		=
		\prod_{m=0}^{\infty} \chfz(\zeta_m t),
		\label{eq:chfcenter1}
		\eeq
		where $j=\sqrt{-1}$. 
		We want to show that $\widetilde{\bm{s}}(\delta)$ converges in probability to $0$ as $\delta\rightarrow 0$. 
		In view of L\'evy's continuity Theorem this is tantamount to showing that $\chf(t)$ converges to $1$ as $\delta\rightarrow 0$. 
		Using \eqref{eq:chfcenter1} we can write:\footnote{The following inequality is known for complex numbers $x_m,y_m$, with $|x_m|\leq 1$ and $|y_m|\leq 1$~\cite{FellerBookV2}: 
			\beq
			\left|\prod_{m=0}^i x_m-\prod_{m=0}^i y_m\right|
			\leq\sum_{m=0}^i |x_m-y_m|,
			\label{eq:prodsumineq}
			\eeq 
		}
		\beq
		\left|
		\chf(t) - 1
		\right|\leq
		\sum_{m=0}^{\infty} |\chfz(\zeta_m t) - 1|.
		\label{eq:prodsumchf}
		\eeq
		Consider, without loss of generality, a positive $t$. 
		Since the random variables $\widetilde{\bm{z}}_m$ have finite expectation, the first derivative of the characteristic function, $\chfz'(t)$, is a continuous function, and by the mean-value theorem we can write (since in particular $\E[\widetilde{\bm{z}}_m]=0$): 
		\beq
		\chfz(\zeta_m t)=1+\zeta_m t\,\chfz'(t_m),\textnormal{ for some $t_m\in(0,\zeta_m t)$}.
		\eeq
		Accordingly we can write:
		\beq
		\left|
		\chfz(\zeta_m t) - 1
		\right|
		\leq
		\zeta_m |t|\,\max_{\tau\in[0,\delta t]}
		|\chfz'(\tau)|,
		\label{eq:meanvalue}
		\eeq
		where the latter inequality follows from the fact that $\zeta_m\leq\delta$, see \eqref{eq:zetadef}.
		Applying \eqref{eq:meanvalue} to \eqref{eq:prodsumchf} we get:
		\beq
		\left|
		\chf(t) - 1
		\right|\leq
		|t|
		\max_{\tau\in[0,\delta t]}
		|\chfz'(\tau)|
		\underbrace{
			\sum_{m=0}^{\infty} 
			\zeta_m}_{\leq 1}.
		\eeq
		On the other hand, since $\chfz'(0)=\E[\widetilde{\bm{z}}_m]=0$, from the continuity of $\chfz'(t)$ it follows that:
		\beq
		\lim_{\delta\rightarrow 0}\max_{\tau\in[0,\delta t]}
		|\chfz'(\tau)|=0,
		\eeq
		which proves that $\bm{s}(\delta)$ converges to $\E[\bm{s}(\delta)]$ in probability as $\delta\rightarrow 0$. The claim in \eqref{eq:weaklawequ} then follows from \eqref{eq:expeclemma}.

		\vspace*{10pt}
		\noindent
		{\bf Part $\mathbf{4}$.}
		Since the variables $\bm{z}_m$ have common finite variance $\sigma^2_z$ and are independent, it is immediate to see that:
		\beq
		\lim_{i\rightarrow\infty} \VAR[\bm{s}_i(\delta)]=
		\sigma^2_z\,\delta^2\sum_{m=0}^{\infty}(1-\delta)^{2m} \alpha^2_m
		<\infty.
		\label{eq:varsum}
		\eeq
		Consider now the squared and centered variables:
		\beqa
		\lefteqn{\left(\bm{s}_i(\delta)-\E\Big[\bm{s}_i(\delta)\Big]\right)^2}\nonumber\\
		&=&
		\delta^2
		\left(
		\sum_{m=0}^{i}(1-\delta)^m \alpha_m \Big(\bm{z}_m-{\sf m}_z\Big)
		\right)^2.
		\eeqa
		In view of parts $1$ and $2$ the quantity on the LHS converges almost surely, as $i\rightarrow\infty$, to:
		\beq
		\left(\bm{s}(\delta)-\E\Big[\bm{s}(\delta)\Big]\right)^2.
		\label{eq:centereds}
		\eeq
		Given the convergence of the variance of the partial sums in \eqref{eq:varsum}, by Fatou's lemma we conclude that~\cite{Durrett}[Th.~1.5.5, p.~26]:
		\beq
		\VAR[\bm{s}(\delta)]
		\leq
		\lim_{i\rightarrow\infty}\VAR[\bm{s}_i(\delta)],
		\label{eq:Fatou}
		\eeq 
		i.e., the limiting variable $\bm{s}(\delta)$ has finite variance. 
		But since the limiting variable $\bm{s}(\delta)$ can be written as:
		\beq
		\bm{s}(\delta)=\bm{s}_i(\delta)+ \delta \sum_{m=i+1}^{\infty}(1-\delta)^m \alpha_m \bm{z}_m,
		\eeq
		with the two quantities on the RHS being statistically independent, the variance of $\bm{s}(\delta)$ cannot be smaller than the variance of $\bm{s}_i(\delta)$ for all $i$, implying that:
		\beq
		\VAR[\bm{s}(\delta)]\geq \lim_{i\rightarrow\infty} \VAR[\bm{s}_i(\delta)].
		\label{eq:lowvar}
		\eeq
		Combining \eqref{eq:Fatou} with \eqref{eq:lowvar} we see that the variance of the almost-sure limit $\bm{s}(\delta)$ is equal to the convergent series of variances, which is the first equality in \eqref{eq:VARlemma}.
		
		In order to prove the second equality in \eqref{eq:VARlemma} we write: 
		\beqa
		\lefteqn{\VAR\left[\delta\sum_{m=0}^{\infty}(1-\delta)^m \alpha_m\bm{z}_m\right]}\nonumber\\
		&=&
		\sigma^2_z\delta^2\sum_{m=0}^{\infty}(1-\delta)^{2m} \alpha^2_m\nonumber\\
		&=&
		\sigma^2_z\delta^2\sum_{m=0}^{\infty}(1-\delta)^{2m} \left(\alpha^2_m-\alpha^2\right)
		\nonumber\\
		&+& \alpha^2\sigma^2_z
		\delta^2\sum_{m=0}^{\infty}(1-\delta)^{2m}.
		\label{eq:singlesum2}
		\eeqa
		Reasoning as done to prove part $2$, we can easily show that the first summation on the RHS in \eqref{eq:singlesum2} is $O(\delta^2)$. The second summation is instead equal to:
		\beq
		\frac{\alpha^2\sigma^2_z\delta^2}{1-(1-\delta)^2}=\frac{\alpha^2\sigma^2_z\delta}{2-\delta},
		\eeq
		and the second equality in \eqref{eq:VARlemma} follows.
		
		\vspace*{10pt}
		\noindent
		{\bf Part $\mathbf{5}$.}
		Let
		\beq
		\sigma^2_{\lim}\triangleq \frac{\alpha^2\sigma^2_z}{2}.
		\label{eq:siglim}
		\eeq
		The claim in \eqref{eq:CLTlemma} is equivalent to prove that the random variable $\frac{\bm{s}(\delta)-{\sf m}_z}{\sqrt{\delta}\sigma_{\lim}}$ converges in distribution to a standard Gaussian. On the other hand, we have that:
		\beq
		\frac{\bm{s}(\delta)-{\sf m}_z}{\sqrt{\delta}\sigma_{\lim}}=
		\frac{\bm{s}(\delta)-\E[\bm{s}(\delta)]}{\sqrt{\delta}\sigma_{\lim}}+
		\frac{\E[\bm{s}(\delta)] - {\sf m}_z}{\sqrt{\delta}\sigma_{\lim}}.
		\label{eq:meanshift}
		\eeq
		Since the second term in \eqref{eq:meanshift} converges to zero in view of \eqref{eq:expeclemma}, from Slutsky's theorem~\cite{shao}[Th.~1.11, p.~60] it suffices to show that the random variable $\frac{\bm{s}(\delta)-\E[\bm{s}(\delta)]}{\sqrt{\delta}\sigma_{\lim}}$ converges in distribution to a standard Gaussian. 
		To this end, we start by introducing, with slight abuse of notation w.r.t. \eqref{eq:zetadef} and \eqref{eq:centered1}, the quantities:
		\beq
		\zeta_m\triangleq\frac{\sqrt{2\delta}(1-\delta)^m\alpha_m}{\alpha},
		\label{eq:zetadef2}
		\eeq
		and:
		\beq
		\widetilde{\bm{s}}(\delta)=\frac{\bm{s}(\delta)-\E[\bm{s}(\delta)]}{\sqrt{\delta}\sigma_{\lim}},\qquad
		\widetilde{\bm{z}}_m=\frac{\bm{z}_m-\E[\bm{z}_m]}{\sigma_z}.
		\label{eq:centered2}
		\eeq
		We notice that $\widetilde{\bm{z}}_m$ has zero mean and unit variance. 
		
		We will now show that $\widetilde{\bm{s}}(\delta)$ converges in distribution to a standard Gaussian.
		In view of L\'evy's continuity theorem, this claim is equivalent to the convergence, as $\delta\rightarrow 0$, of the characteristic function of $\widetilde{\bm{s}}(\delta)$ to the characteristic function $e^{-\frac{t^2}{2}}$. 
		From \eqref{eq:limitdef2}, \eqref{eq:siglim}, \eqref{eq:zetadef2} and \eqref{eq:centered2} we see that:
		\beq
		\widetilde{\bm{s}}(\delta)=\sum_{m=0}^\infty \zeta_m\widetilde{\bm{z}}_m.
		\eeq
		Reasoning as done to compute \eqref{eq:chfcenter1}, the characteristic function of $\widetilde{\bm{s}}(\delta)$ in \eqref{eq:centered2} can be written as: 
		\beq
		\chf(t)
		=
		\prod_{m=0}^{\infty} \chfz(\zeta_m t).
		\label{eq:chfcenter2}
		\eeq 
		Using the triangle inequality for complex numbers we can write:
		\beqa
		\left|\chf(t)-e^{-\frac{t^2}{2}}\right|&\leq&
		\left|\chf(t)-e^{-\frac{\sum_{m=0}^{\infty} \zeta_m^2 t^2}{2}}\right|\nonumber\\
		&+&
		\left|e^{-\frac{\sum_{m=0}^{\infty} \zeta_m^2 t^2}{2}}-e^{-\frac{t^2}{2}}\right|.\nonumber\\
		\label{eq:boundbound}
		\eeqa
		Now, that the second term on the RHS of~(\ref{eq:boundbound}) converges to zero follows from part $4)$, since from \eqref{eq:VARlemma} and the definition of $\zeta_m$ in \eqref{eq:zetadef2} we conclude that:
		\beq
		\lim_{\delta\rightarrow 0}\sum_{m=0}^{\infty} \zeta_m^2=1.
		\eeq
		Let us now focus on the first term on the RHS of~(\ref{eq:boundbound}).
		Since the characteristic functions have magnitude not greater than $1$, in view of~(\ref{eq:prodsumineq}) and \eqref{eq:chfcenter2} we can write:
		\beqa
		\lefteqn{\left|\chf(t)-e^{-\frac{\sum_{m=0}^{\infty}\zeta_m^2 t^2}{2}}\right|}\nonumber\\
		&\leq&
		\sum_{m=0}^{\infty}\left|
		\chfz(\zeta_m t)-e^{-\frac{\zeta_m^2 t^2}{2}}
		\right|\nonumber\\
		&\leq&
		\sum_{m=0}^{\infty}
		\left|
		\chfz(\zeta_m t)
		-
		1
		+
		\frac{\zeta^2_m t^2}{2}
		\right|\nonumber\\
		&+&
		\sum_{m=0}^{\infty}
		\left|
		e^{-\frac{\zeta_m^2 t^2}{2}}
		-
		1
		+
		\frac{\zeta^2_m t^2}{2}
		\right|,
		\label{eq:prodsumbound2}
		\eeqa
		where in the latter step we applied the triangle inequality.
		Now, the last term in \eqref{eq:prodsumbound2} converges to zero since for any positive $s$ we have $|e^{-s}-1+s|\leq s^2/2$, and since it is immediate to show that (see the proof in~\cite{MattaBracaMaranoSayedTIT2016}):
		\beq
		\lim_{\delta\rightarrow 0}\sum_{m=0}^{\infty} \zeta_m^4=0.
		\eeq
		On the other hand, using~\cite{Durrett}[Lemma~3.3.19, p.~134] we can write, for an arbitrarily small $\epsilon>0$:
		\beqa
		\lefteqn{
			\left|
			e^{j\widetilde{\bm{z}}_m \zeta_m t}
			-
			1
			-j\widetilde{\bm{z}}_m \zeta_m t
			+\frac 1 2
			\widetilde{\bm{z}}_m^2 \zeta^2_m t^2
			\right|
		}\nonumber\\
		&\leq&
		\mathbb{I}\Big\{
		|\widetilde{\bm{z}}_m|\zeta_m\leq\epsilon
		\Big\}
		\frac{|\widetilde{\bm{z}}_m \zeta_m t|^3}{6}
		\nonumber\\
		&+&
		\mathbb{I}\Big\{
		|\widetilde{\bm{z}}_m|\zeta_m>\epsilon
		\Big\}(\widetilde{\bm{z}}_m\zeta_m t)^2
		\nonumber\\
		&\leq&
		\epsilon\widetilde{\bm{z}}_m^2\zeta_m^2\,\frac{|t|^3}{6}
		+\widetilde{\bm{z}}_m^2
		\mathbb{I}\Big\{
		|\widetilde{\bm{z}}_m|\zeta_m>\epsilon
		\Big\}\zeta_m^2 t^2
		\nonumber\\
		&\leq&
		\epsilon\widetilde{\bm{z}}_m^2\zeta_m^2\,\frac{|t|^3}{6}
		+\widetilde{\bm{z}}_m^2
		\mathbb{I}\Big\{
		|\widetilde{\bm{z}}_m|>\epsilon\alpha/\sqrt{2\delta}
		\Big\}
		\zeta_m^2 t^2,
		\label{eq:fundamentalbound}
		\eeqa
		where $\mathbb{I}\{\mathcal{E}\}$ is the indicator of event $\mathcal{E}$, and the last inequality follows because $\zeta_m\leq
		\sqrt{2\delta}/\alpha$ --- see \eqref{eq:zetadef2}.
		Let now:
		\beq
		g(\delta)=\E\left[
		\widetilde{\bm{z}}_m^2
		\mathbb{I}\Big\{
		\widetilde{\bm{z}}_m^2>\epsilon \alpha/\sqrt{2\delta}
		\Big\}\right].
		\eeq
		Owing to identical distribution of $\widetilde{\bm{z}}_m$ across index $m$, the function $g(\delta)$ does not depend on $m$. 
		Since $\widetilde{\bm{z}}_m$ has finite variance, we have that $g(\delta)\rightarrow 0$ as $\delta\rightarrow 0$.
		In view of \eqref{eq:fundamentalbound}, recalling that the magnitude of the expectation is upper bounded by the expectation of the magnitude, and that $\widetilde{\bm{z}}_m$ has zero mean and unit variance, we have that:
		\beq
		\left|
		\chfz(\zeta_m t)
		-
		1
		+
		\frac{\zeta^2_m t^2}{2}
		\right|
		\leq\sum_{m=0}^{\infty}\zeta_m^2 
		\left(
		\epsilon\frac{|t|^3}{6}
		+
		t^2 g(\delta)\right),
		\eeq
		and, hence,
		\beq
		\limsup_{\delta\rightarrow 0} \left|
		\chfz(\zeta_m t)
		-
		1
		+
		\frac{\zeta^2_m t^2}{2}
		\right|
		\leq
		\epsilon\,\frac{|t|^3}{6},
		\eeq
		finally implying, due to the arbitrariness of $\epsilon$, that $\chf(t)$ converges to $e^{-t^2/2}$ as $\delta\rightarrow 0$.
		We have therefore shown that $\widetilde{\bm{s}}(\delta)$ in \eqref{eq:centered2} converges to a standard Gaussian as $\delta\rightarrow 0$, and this completes the proof of part $5$.
		
		\vspace*{10pt}
		\noindent
		{\bf Part $\mathbf{6}$.} 
		The convergence in \eqref{eq:philemma} can be proved as done in~\cite[Appendix C]{MattaBracaMaranoSayedTIT2016}. 
		Then the convergence in \eqref{eq:philemma} implies the LDP in \eqref{eq:LDPgeneral1}--\eqref{eq:LDPgeneral2} in view of the G\"{a}rtner-Ellis theorem~\cite{DemboZeitouni}[Th.~2.3.6, p.~44],~\cite{DenHollander}[Th.~V.6, p.~54].
		
		Next we focus on the regularity properties of the Fenchel-Legendre transform $\phi^{\star}(\gamma)$. 
		Following the development used in~\cite[Appendix C]{MattaBracaMaranoSayedTIT2016}, we can prove that $\mathcal{D}^{o}$ is an interval, that $\phi^{\star}(\gamma)$ is smooth and strictly convex for $\gamma\in\mathcal{D}^{o}$, and that $\phi^{\star}(\gamma)\geq 0$ with equality if, and only if, $\gamma=\alpha {\sf m}_z$.
		
		Thus, it remains to characterize the boundaries of $\mathcal{D}^{o}$ and the behavior of the rate function at these boundaries. 
		To this end, it is sufficient to prove the claim with $\alpha=1$ and for the right boundary, since the proof for other values of $\alpha$ and for the left boundary is simply obtained using the scaling and reflection properties of the LMGF~\cite{DemboZeitouni,DenHollander}.

		Now, since it has been shown in~\cite[Appendix C]{MattaBracaMaranoSayedTIT2016} that the right boundary of $\mathcal{D}^{o}$ is equal to $\lim_{t\rightarrow\infty}\Lambda_z(t)/t$, we must now prove that this limit equals $z_{+}$ (recall that we are working with $\alpha=1$). We start by noticing that, letting $z_{-}<\overline{z}<z_{+}$, the LMGF $\Lambda_{z}(t)$ can be written as:
		\beq
		\Lambda_{z}(t)=\log\Big(
		\E\left[
		\mathbb{I}\{\bm{z}_m\leq \overline{z}\}e^{\bm{z}_m t}
		\right]+
		\E\left[
		\mathbb{I}\{\bm{z}_m> \overline{z}\}e^{\bm{z}_m t}
		\right]
		\Big).
		\label{eq:LMGFindicators}
		\eeq
		From \eqref{eq:LMGFindicators} we get, for all $t>0$:
		\beq
		\frac{\Lambda_z(t)}{t}
		\geq
		\frac{\log\Big(
			e^{\overline{z} t}\,
			\E\left[
			\mathbb{I}\{\bm{z}_m> \overline{z}\}
			\right]
			\Big)}{t}
		=
		\overline{z} + \frac{\log q}{t},
		\label{eq:Lambdazlowbound}
		\eeq
		where we set $q=\P[\bm{z}_m> \overline{z}]$. We remark that $0<q<1$ since $\overline{z}$ is internal to the support of $\bm{z}_m$. From \eqref{eq:Lambdazlowbound} we get:
		\beq
		\liminf_{t\rightarrow\infty}\frac{\Lambda_z(t)}{t}\geq \overline{z}.
		\label{eq:liminfLambdaz}
		\eeq
		If $z_{+}=+\infty$ the result is proved due to arbitrariness of $\overline{z}$. 
		If $z_{+}<+\infty$, we can choose $\overline{z}=z_{+}-\epsilon$, and conclude that the limit inferior in \eqref{eq:liminfLambdaz} is equal to $z_{+}$. The fact that the corresponding limit superior is equal to $z_{+}$ follows by observing that, in view of \eqref{eq:LMGFindicators}, for all $t>0$ the quantity  $\Lambda_z(t)/t$ is upper bounded by $z_{+}$.

		Finally, we characterize the behavior of the rate function at the boundaries of $\mathcal{D}^{o}$. 
		We focus again on the right boundary $z_{+}$. 
		When $z_{+}=+\infty$, it suffices to notice that the rate function $\phi^{\star}(\gamma)$ is strictly convex in $\mathcal{D}^{o}$ and is strictly increasing for $\gamma>{\sf m}_z$ (see Fig.~\ref{fig:ratefun}) to conclude that the rate function diverges to $+\infty$ as $\gamma\rightarrow z_{+}$.
		
		We move on to examine the case $z_{+}<+\infty$. Exploiting \eqref{eq:LMGFindicators} we can write, for all $t>0$:
		\beqa
		\Lambda_{z}(t)
		&\leq&
		\log
		\left(
		(1-q) e^{\overline{z} t}+q e^{z_{+} t}
		\right)
		\nonumber\\
		&=&
		z_{+}t+\log
		\left(
		(1-q) e^{-(z_{+} - \overline{z}) t}+q
		\right).
		\label{eq:binobound}
		\eeqa
		Since $z_{+}>\overline{z}$, for any $\epsilon>0$ there exists $t_{\epsilon}>0$ such that:
		\beq
		(1-q) e^{-(z_{+} - \overline{z}) t}\leq \epsilon q, ~~\textnormal{for all }t\geq t_{\epsilon},
		\label{eq:tepsilon}
		\eeq
		implying, in view of \eqref{eq:binobound}:
		\beq
		\Lambda_{z}(t)\leq z_{+} t +\log((1+\epsilon)q), ~~\textnormal{for all }t\geq t_{\epsilon}.
		\label{eq:Lambdazuseful}
		\eeq
		Using \eqref{eq:Lambdazuseful} in \eqref{eq:philemma} we can thus write:
		\beqa
		\phi(t)&=&\int_{0}^t \frac{\Lambda_{z}(\tau)}{\tau}d\tau
		=\int_{0}^{t_{\epsilon}} \frac{\Lambda_{z}(\tau)}{\tau}d\tau
		+\int_{t_{\epsilon}}^{t} \frac{\Lambda_{z}(\tau)}{\tau}d\tau
		\nonumber\\
		&\leq&
		\phi(t_{\epsilon})
		+z_{+}(t-t_{\epsilon})
		+\int_{t_{\epsilon}}^{t} \frac{\log\left((1+\epsilon)q\right)}{\tau}d\tau\nonumber\\
		&=&
		\phi(t_{\epsilon})
		+z_{+}(t-t_{\epsilon})
		+\log\left((1+\epsilon)q\right)\,\log\frac{t}{t_{\epsilon}}.
		\label{eq:chaineqlemmapart6}
		\eeqa
		Plugging the latter inequality in \eqref{eq:FenchLeg} we get:
		\beqa
		\phi^{\star}(z_{+})
		&\geq&
		\sup_{t\geq t_{\epsilon}}[z_{+}t - \phi(t)]
		\geq
		-\phi(t_{\epsilon})+z_{+}t_{\epsilon}\nonumber\\
		&+& \log\frac{1}{(1+\epsilon) q}\,\sup_{t\geq t_{\epsilon}} \log\frac{t}{t_{\epsilon}}=+\infty,
		\label{eq:divergboundary}
		\eeqa
		where we have chosen $\epsilon$ so small to ensure that $(1+\epsilon) q<1$.
		Finally, in view of \eqref{eq:FenchLeg} we can write, for a generic $t\in\mathbb{R}$:
		\beq
		\lim_{\gamma\rightarrow z_{+}} \phi^{\star}(\gamma)
		\geq
		\lim_{\gamma\rightarrow z_{+}}[\gamma t - \phi(t)]=
		[z_{+} t - \phi(t)],
		\eeq
		and from \eqref{eq:divergboundary} we conclude that $\phi^{\star}(\gamma)\rightarrow+\infty$ as $\gamma\rightarrow z_{+}$.
	\end{IEEEproof}

	\section{}
	\label{app:Th1}
	
	\begin{IEEEproof}[Proof of Theorem~\ref{theor:steady}]
		We are interested in characterizing, for each agent $k$, the {\em joint} behavior of the random variables $\widehat{\bm{\lambda}}^{(\delta)}_{k,i}(\theta)$ for all values of $\theta\neq\theta_0$. 
		To this end, it is useful to consider the $(H-1)\times 1$ vector $\widehat{\bm{\lambda}}^{(\delta)}_{k,i}$ similarly defined as the vector in \eqref{eq:logbelvec}.
		We also introduce, for a fixed time epoch $i$, the $N\times (H-1)$ data matrix $\bm{X}_i$, whose entries, for $\ell=1,2,\ldots,N$ and $\theta\neq\theta_0$, are:
		\beq
		[\bm{X}_i]_{\ell \theta}=\bm{x}_{\ell,i}(\theta).
		\eeq
		In light of \eqref{eq:lambdarec} we can write:
		\beq
		\widehat{\bm{\lambda}}^{(\delta)}_{k,i}=f_{k,i}^{(\delta)}(\bm{X}_1,\bm{X}_2,\ldots,\bm{X}_i),
		\label{eq:lambdafun}
		\eeq
		to highlight that the random vector $\widehat{\bm{\lambda}}_{k,i}^{(\delta)}$ is a certain function $f_{k,i}^{(\delta)}$ of the data matrices $\bm{X}_1, \bm{X}_2,\ldots, \bm{X}_i$. Since the data are i.i.d. over time, reversing the order of the data matrices in \eqref{eq:lambdafun} does not change the distribution of the resulting random vector, i.e.:
		\beq
		\widetilde{\bm{\lambda}}_{k,i}^{(\delta)}=f_{k,i}^{(\delta)}(\bm{X}_i,\bm{X}_{i-1},\ldots,\bm{X}_1)\stackrel{\mathrm{d}}{=}\widehat{\bm{\lambda}}_{k,i}^{(\delta)},
		\label{eq:equadist}
		\eeq
		where $\stackrel{\mathrm{d}}{=}$ denotes equality in distribution. 
		Considering this reversed order of the data matrices in \eqref{eq:lambdarec} and exchanging the order of summation we obtain:
		\beq
		\widetilde{\bm{\lambda}}^{(\delta)}_{k,i}(\theta)=
		\sum_{\ell=1}^N \delta \sum_{m=0}^{i-1} (1-\delta)^m [A^{m+1}]_{\ell k}\, \bm{x}_{\ell,m+1}(\theta).
		\label{eq:lambdarecrev}
		\eeq
		From part $1)$ of Lemma~\ref{lem:mainlemma} in the Appendix, each of the $N$ inner partial sums (scaled by $\delta$) converges {\em almost surely}.
		In fact, the random variables $\bm{x}_{\ell,m+1}(\theta)$ have finite first moment in view of Assumption~\ref{assum:integrable}, and the weights $[A^{m+1}]_{\ell k}$ fulfill condition \eqref{eq:exprate} in view of Property~\ref{prop:Perron}.
		It makes thus sense to define a proper random variable as the (almost-surely convergent) value of the random series in \eqref{eq:lambdarecrev}, which corresponds to \eqref{eq:convseries}. This in turn implies the following almost-sure convergence, as $i\rightarrow\infty$, of the {\em vector} with reversed ordering, $\widetilde{\bm{\lambda}}^{(\delta)}_{k,i}$, to the limiting random vector $\widetilde{\bm{\lambda}}^{(\delta)}_{k}$. 
		In view of \eqref{eq:equadist}, this almost-sure convergence implies the convergence {\em in distribution} of the original (i.e., with correct ordering of the data matrices $\bm{X}_i$) vector $\widehat{\bm{\lambda}}_{k,i}$, finally yielding the claim of the theorem.
	\end{IEEEproof}
	
	\section{}
	\label{app:Th2}
	\begin{IEEEproof}[Proof of Theorem~\ref{theor:weaklaw}]
		We start by proving \eqref{eq:wlawintermediate}. Examining \eqref{eq:convseries} we see that each one of the $N$ inner series matches the conditions in Lemma~\ref{lem:mainlemma}, part $3$, implying that the $\ell$-th inner series converges in probability, as $\delta\rightarrow 0$, to the expected value $\pi_{\ell}\E[\bm{x}_{\ell,m+1}(\theta)]=\pi_{\ell} d_{\ell}(\theta)$. 
		As a result, $\widetilde{\bm{\lambda}}^{(\delta)}_k(\theta)$ converges in probability to $\mnet(\theta)$, which implies, for any $\epsilon>0$:
		\beq
		\lim_{\delta\rightarrow 0}
		\P\left[
		\widetilde{\bm{\lambda}}^{(\delta)}_k(\theta)<\mnet(\theta)-\epsilon
		\right]=0.
		\eeq
		Since under Assumption~\ref{assum:globo} the quantity $\mnet(\theta)$ is strictly positive, we conclude that:
		\beq
		\lim_{\delta\rightarrow 0}
		\P\left[
		\widetilde{\bm{\lambda}}^{(\delta)}_k(\theta)\leq 0
		\right]=0,
		\eeq
		which, by application of the union bound, in light of \eqref{eq:errprob} gives:
		\beqa
		p^{(\delta)}_k&=&\P\left[
		\exists \theta\neq\theta_0: \widetilde{\bm{\lambda}}^{(\delta)}_k(\theta)\leq 0
		\right]
		\nonumber\\
		&\leq&
		\sum_{\theta\neq\theta_0} \P\left[
		\widetilde{\bm{\lambda}}^{(\delta)}_k(\theta)\leq 0
		\right]\stackrel{\delta\rightarrow 0}{\longrightarrow} 0,
		\eeqa
		and the claim of the theorem is proved.
	\end{IEEEproof}
	
	\section{}
	\label{app:Th3}
	\begin{IEEEproof}[Proof of Theorem~\ref{theor:CLT}]
		In the following we will refer to the elements $\theta_1,\theta_2,\ldots,\theta_{H-1}$ in the set $\Theta\setminus\theta_0$ --- see \eqref{eq:wrongset}. 
		Consider a zero-mean Gaussian random vector: 
		\beq
		\bm{g}=[\bm{g}(\theta_1),\bm{g}(\theta_2),\ldots,\bm{g}(\theta_{H-1})]\T,
		\eeq
		with covariance matrix equal to $\Cnet/2$. We recall that the $(\theta,\theta')$-th entry of $\Cnet$ is the covariance $\cnet(\theta,\theta')$ defined in \eqref{eq:cnet}. What we want to show is that the random vector:
		\beq
		\frac{\widetilde{\bm{\lambda}}^{(\delta)}_k - \mnet}{\sqrt{\delta}}
		\eeq
		converges in distribution to $\bm{g}$.
		
		When dealing with convergence in distribution of random vectors, the standard path is to reduce the vector problem to a scalar problem through the following argument. In view of L\'evy's continuity theorem for random vectors, convergence in distribution takes place if, and only if, convergence of the pertinent (multivariate) characteristic functions takes place \cite{shao}. This implies that\footnote{This corollary of L\'evy's continuity theorem is also known as Cram\'er-Wold device or theorem \cite{shao}[Th.~1.9, p.~56].} our claim will be proved if we show that, for any sequence of real numbers $t(\theta_1), t(\theta_2),\ldots,t(\theta_{H-1})$:
		\beq
		\sum_{\theta\neq\theta_0} t(\theta) \, \frac{\widetilde{\bm{\lambda}}^{(\delta)}_k(\theta) - \mnet(\theta)}{\sqrt{\delta}}
		\stackrel{\delta\rightarrow 0}{\rightsquigarrow}\sum_{\theta\neq\theta_0} t(\theta) \bm{g}(\theta).
		\label{eq:Gaussum}
		\eeq
		Obviously, the linear combination on the RHS in \eqref{eq:Gaussum} is a Gaussian random variable with zero mean and with variance:
		\beq
		\VAR\left[\sum_{\theta\neq\theta_0} t(\theta) \bm{g}(\theta)\right]=
		\sum_{\theta\neq\theta_0}\sum_{\theta'\neq\theta_0} t(\theta) t(\theta') \frac{\cnet(\theta,\theta')}{2}.
		\label{eq:desiredGvar}
		\eeq
		Let us now examine the LHS in \eqref{eq:Gaussum}. Using \eqref{eq:lambdarecrev} we get:
		\beqa
		\lefteqn{\sum_{\theta\neq\theta_0} t(\theta) \widetilde{\bm{\lambda}}^{(\delta)}_k(\theta)}
		\nonumber\\
		&=&
		\sum_{\ell=1}^N 
		\delta\sum_{m=0}^{\infty}
		(1-\delta)^m 
		[A^{m+1}]_{\ell k}
		\sum_{\theta\neq\theta_0}
		t(\theta)
		\bm{x}_{\ell,m+1}(\theta),\nonumber\\
		\eeqa
		whereas using \eqref{eq:delldef} we have:
		\beq
		\sum_{\theta\neq\theta_0} t(\theta) \mnet(\theta)=
		\sum_{\ell=1}^N\pi_\ell
		\sum_{\theta\neq\theta_0} t(\theta) d_{\ell}(\theta).
		\eeq
		Let us now set:
		\beqa
		\bm{z}^{(\ell)}_m&\triangleq&\sum_{\theta\neq\theta_0}
		t(\theta)
		\bm{x}_{\ell,m+1}(\theta), 
		\label{eq:zelldef1}\\
		\alpha^{(\ell)}_m&\triangleq&[A^{m+1}]_{\ell k},
		\label{eq:aelldef1}\\
		\bm{s}^{(\ell)}(\delta)&\triangleq&
		\delta\sum_{m=0}^{\infty}
		(1-\delta)^m \alpha^{(\ell)}_m\bm{z}^{(\ell)}_m.
		\label{eq:selldef1}
		\eeqa
		We observe that:
		\beqa
		\E\left[\bm{z}_m^{(\ell)}\right]&=&\sum_{\theta\neq\theta_0} t(\theta) d_{\ell}(\theta),
		\label{eq:meanzellj}\\
		\VAR\left[\bm{z}_m^{(\ell)}\right]&=&
		\sum_{\theta\neq\theta_0}\sum_{\theta'\neq\theta_0} t(\theta)t(\theta') \rho_{\ell}(\theta,\theta').
		\label{eq:varzellj}
		\eeqa
		Exploiting Eqs. \eqref{eq:zelldef1}--\eqref{eq:meanzellj}, the LHS in \eqref{eq:Gaussum} can be cast in the form:
		\beq
		\sum_{\ell=1}^N \frac{\bm{s}^{(\ell)}(\delta) - \E\left[\bm{z}^{(\ell)}_m\right]}{\sqrt{\delta}}.
		\eeq
		We see from Eqs. \eqref{eq:zelldef1}--\eqref{eq:selldef1} that the random variables $\bm{s}^{(\ell)}(\delta)$ match the structure of the random series used  in Lemma~\ref{lem:mainlemma}. 
		We now verify that $\bm{s}^{(\ell)}(\delta)$ fulfills the conditions of part $5$ in Lemma~\ref{lem:mainlemma}, for every $\ell=1,2,\ldots,N$. 
		First we note that $\bm{z}^{(\ell)}_m$ has finite variance since it is a linear combination of random variables that have finite variance. 
		Second we see that condition \eqref{eq:exprate} is verified in view of Property~\ref{prop:Perron}. 
		We conclude then from part $5$ of Lemma~\ref{lem:mainlemma} that the following convergence in distribution holds:
		\beq
		\frac{\bm{s}^{(\ell)}(\delta) - \E\left[\bm{z}^{(\ell)}_m\right]}{\sqrt{\delta}}
		\stackrel{\delta\rightarrow 0}{\rightsquigarrow}\mathscr{G}\left(
		0,\frac{\pi^2_{\ell}}{2}
		\VAR\left[
		\bm{z}^{(\ell)}_m
		\right]
		\right).
		\label{eq:individCLT}
		\eeq
		Since the data are independent across agents, we have that the random variables $\bm{s}^{(\ell)}(\delta)$ are independent across index $\ell$. For this reason, and in view of \eqref{eq:individCLT}, we conclude that the LHS in \eqref{eq:Gaussum} is asymptotically normal, with zero mean and with variance given by:
		\beqa
		\lefteqn{
			\frac{\pi^2_{\ell}}{2}\sum_{\ell =1}^N \VAR\left[
			\bm{z}^{(\ell)}_m
			\right]}\nonumber\\
		&=&
		\sum_{\theta\neq\theta_0}\sum_{\theta'\neq\theta_0}t(\theta)t(\theta')
		\sum_{\ell=1}^N\frac{\pi^2_{\ell}}{2}\rho_{\ell}(\theta,\theta')
		\nonumber\\
		&=&
		\sum_{\theta\neq\theta_0}\sum_{\theta'\neq\theta_0}t(\theta)t(\theta')\frac{\cnet(\theta,\theta')}{2},
		\label{eq:finalvari}
		\eeqa
		where we have used \eqref{eq:varzellj}. 
		Since the RHS in \eqref{eq:finalvari} coincides with the variance in \eqref{eq:desiredGvar}, the proof is complete.
	\end{IEEEproof}
	
	\section{}
	\label{app:Th4}
	\begin{IEEEproof}[Proof of Theorem~\ref{theor:LD}]
		In light of \eqref{eq:errprob}, the error probability of {\em not} choosing $\theta_0$ can be bounded as follows (with the lower bound holding for every $\theta\neq\theta_0$):
		\beq
		\P\left[
		\widetilde{\bm{\lambda}}^{(\delta)}_{k,i}(\theta) \leq 0
		\right]
		\leq 
		p^{(\delta)}_{k,i} 
		\leq
		\sum_{\theta\neq\theta_0} \P\left[
		\widetilde{\bm{\lambda}}^{(\delta)}_{k,i}(\theta)\leq 0
		\right],
		\label{eq:errprobound1}
		\eeq
		where the upper bound is the union bound.
		At the steady state, Eq. \eqref{eq:errprobound1} implies:
		\beq
		\P\left[
		\widetilde{\bm{\lambda}}^{(\delta)}_k(\theta)\leq 0
		\right]
		\leq 
		p^{(\delta)}_k
		\leq
		\sum_{\theta\neq\theta_0} \P\left[
		\widetilde{\bm{\lambda}}^{(\delta)}_k(\theta)\leq 0
		\right].
		\label{eq:errprobound2}
		\eeq
		One key point to prove the claim of the theorem is the exponential characterization of the probability $\P\left[\widetilde{\bm{\lambda}}^{(\delta)}_k(\theta)\leq 0\right]$. 
		Preliminarily, let us set:
		\beqa
		\bm{z}^{(\ell)}_m&\triangleq&\bm{x}_{\ell,m+1}(\theta), 
		\label{eq:zelldef2}\\
		\alpha^{(\ell)}_m&\triangleq&[A^{m+1}]_{\ell k},
		\label{eq:aelldef2}\\
		\bm{s}^{(\ell)}(\delta)&\triangleq&
		\delta\sum_{m=0}^{\infty}
		(1-\delta)^m \alpha^{(\ell)}_m\bm{z}^{(\ell)}_m,
		\label{eq:selldef2}
		\eeqa
		which yields:
		\beq
		\widetilde{\bm{\lambda}}^{(\delta)}_k(\theta)=\sum_{\ell=1}^N \bm{s}^{(\ell)}(\delta).
		\eeq 
		Recall that the log-likelihood ratios $\bm{x}_{\ell,m}(\theta)$ are assumed to be independent across agents (i.e., across $\ell$). Thus $\bm{s}^{(\ell)}(\delta)$ are also independent random variables.
		Now, part $6$ of Lemma~\ref{lem:mainlemma} would provide the required exponential characterization for the individual variable $\bm{s}^{(\ell)}(\delta)$.  We need instead the characterization for $\widetilde{\bm{\lambda}}^{(\delta)}_k(\theta)$, which is the sum of the (independent) variables $\bm{s}^{(\ell)}(\delta)$. Let us elaborate on this aspect. 
		The starting point to prove part $6$ in Lemma~\ref{lem:mainlemma} is the convergence in \eqref{eq:philemma}. 
		Exploiting additivity of the LMGF for independent variables, we conclude that the LMGF of $\widetilde{\bm{\lambda}}^{(\delta)}_k$, scaled by $\delta$ and evaluated at $t/\delta$, converges to the sum:
		\beq
		\sum_{\ell=1}^N \int_{0}^{t} \frac{\Lambda_{\ell}(\pi_{\ell}\tau;\theta)}{\tau}d\tau=
		\int_{0}^{t}\frac{\omeganet(\tau;\theta)}{\tau}d\tau\dfz \phi(t;\theta),
		\label{eq:tothesum}
		\eeq 
		where: $i)$ we used the fact that the LMGF of $\bm{z}^{(\ell)}_m$ is $\Lambda_{\ell}(t;\theta)$; $ii)$ the intermediate equality comes from \eqref{eq:LMGFav} (having exchanged the integral with the sum); and $iii)$ the last equality comes from \eqref{eq:phittheta}.
		Moreover, the properties of the rate function in part $6$ of Lemma~\ref{lem:mainlemma}  depend only on the fact that $\Lambda_{\alpha z}(t)$ is a logarithmic moment generating function that is finite for all $t\in\mathbb{R}$. 
		Since $\omeganet(\tau;\theta)$ is the LMGF of the average variable $\xnet(\theta)$ (and is finite for all $t\in\mathbb{R}$ by assumption), all the remaining results in part $6$ of Lemma~\ref{lem:mainlemma} hold true, provided that the properties pertaining to $\alpha \bm{z}_m$ are now referred to $\xnet(\theta)$.
		
		We conclude that it is legitimate to use the exponential characterization provided in Lemma~\ref{lem:mainlemma}. In particular, since we have $\gamma=0<\mnet(\theta)$, the pertinent relation is given by \eqref{eq:LDPrelevant} with the choice $\gamma=0$, yielding:
		\beq
		\lim_{\delta\rightarrow 0}\delta
		\log\P\left[
		\widetilde{\bm{\lambda}}^{(\delta)}_k(\theta)\leq 0
		\right]=
		-\Phi(\theta),
		\label{eq:LDexp}
		\eeq 
		where the exponent $\Phi(\theta)$ is accordingly computed as the value of the rate function at $\gamma=0$, namely, 
		\beq
		\Phi(\theta)=\sup_{t\in\mathbb{R}}[-\phi(t;\theta)]=-\inf_{t\in\mathbb{R}}\phi(t;\theta).
		\label{eq:Phitheta}
		\eeq
		Using the lower bound in \eqref{eq:errprobound2}, we can readily conclude from \eqref{eq:LDexp} and from the definitions appearing in \eqref{eq:theorLDclaim} and \eqref{eq:Phitheta} that:
		\beq
		\liminf_{\delta\rightarrow 0}\delta \log p^{(\delta)}_k\geq \max_{\theta\neq\theta_0} \Big(-\Phi(\theta)\Big)=
		-\min_{\theta\neq\theta_0}\Phi(\theta)=-\Phi.
		\label{eq:liminf}
		\eeq
		Let us now focus on the upper bound in \eqref{eq:errprobound2}. 
		By definition, for all $\theta\neq\theta_0$ we have that $\Phi\leq\Phi(\theta)$. 
		Accordingly, the convergence in \eqref{eq:LDPrelevant} implies that, given an arbitrary $\epsilon>0$, for sufficiently small $\delta$ we can write:
		\beq
		\P\left[
		\widetilde{\bm{\lambda}}^{(\delta)}_k(\theta)\leq 0
		\right]
		\leq
		e^{-(1/\delta)(\Phi-\epsilon)}.
		\label{eq:simpleUB}
		\eeq
		Exploiting \eqref{eq:simpleUB}, the upper bound in \eqref{eq:errprobound2} yields:
		\beq
		\delta \log p_k^{(\delta)}\leq \delta\log(H-1) -\Phi + \epsilon,
		\eeq
		where we recall that $H$ is the number of hypotheses or admissible models. Due to the arbitrariness of $\epsilon$, we have:
		\beq
		\limsup_{\delta\rightarrow 0} \delta \log p_k^{(\delta)}\leq - \Phi.
		\label{eq:limsup}
		\eeq
		Bridging \eqref{eq:liminf} and \eqref{eq:limsup} implies the desired claim.
	\end{IEEEproof}

	\section{}
	\label{app:theor5proof}
\begin{figure}[t]
		\centering
		\includegraphics[width=.9\linewidth]{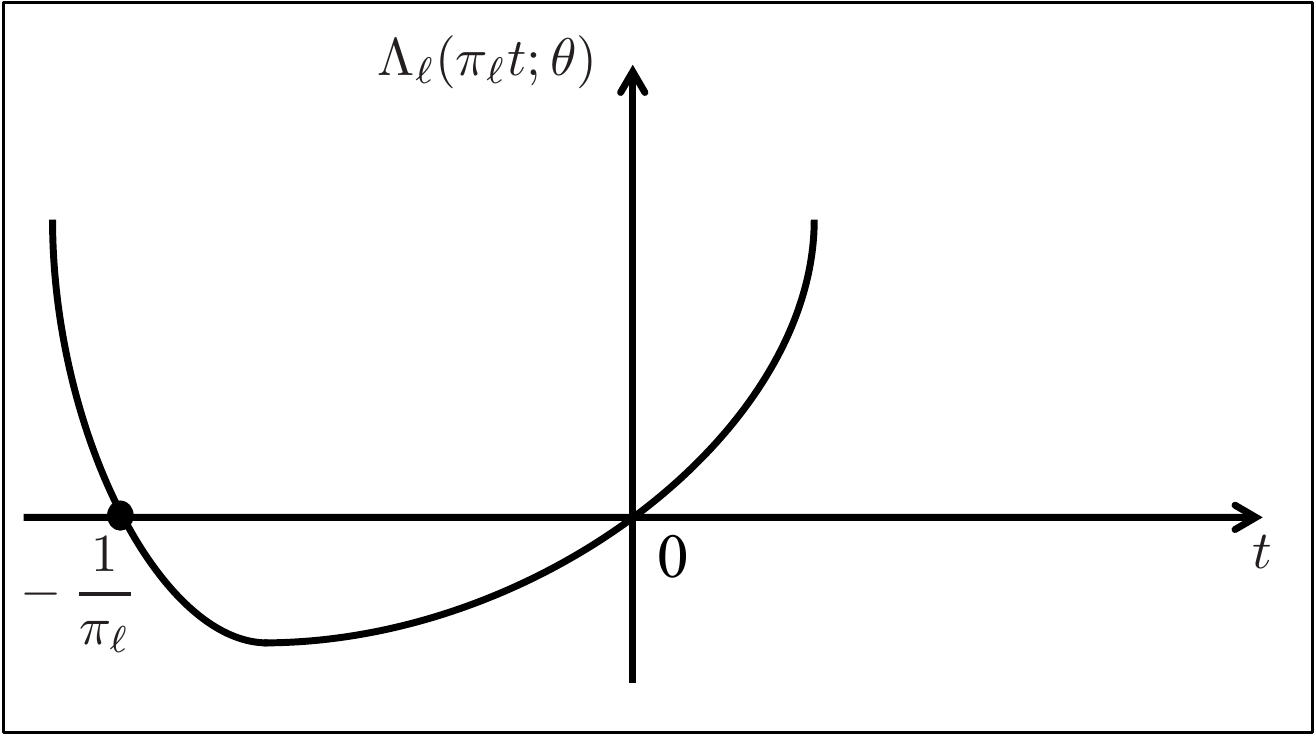}
		\caption{Typical shape of the LMGF of the $\ell$-th likelihood.}
		\label{fig:fconvexphi}
	\end{figure}

	We start by proving an auxiliary lemma.
	\begin{lemma}[Useful properties of the LMGF $\Lambda_{\ell}(t;\theta)$]
		\label{lem:uspropLMGFapp}
		The lemma is proved under the same assumptions used in Theorem~\ref{theor:LD}. 
		Let 
		\beq
		\Lambda_{\ell}(t;\theta)=\log\E\left[e^{t\,\bm{x}_{\ell,i}(\theta)}\right]=\log\E\left[e^{t \log\frac{L_{\ell}(\bm{\xi}_{\ell,i}|\theta_0)}{L_{\ell}(\bm{\xi}_{\ell,i}|\theta)}}\right]
		\label{eq:verybasicdefLambdaell}
		\eeq
		be the LMGF of the log-likelihood at the $\ell$-th agent, let
		\beq
		\Lambda_{\mathrm{ave}}(t;\theta)=\log\E\left[e^{t\,\bm{x}_{\mathrm{ave},i}(\theta)}\right]=\sum_{\ell=1}^N \Lambda_{\ell}(\pi_{\ell}t;\theta)
		\label{eq:Lambdaverepeatapp}
		\eeq
		be the LMGF of the network average of log-likelihoods, $\bm{x}_{\mathrm{ave},i}(\theta)=\sum_{\ell=1}^N \pi_{\ell} \bm{x}_{\ell,i}(\theta)$, and let:
		\beq
		\phi(t;\theta)=\int_{0}^t\frac{\Lambda_{\mathrm{ave}}(\tau;\theta)}{\tau}d\tau.
		\label{eq:phitandthetalemmapp}
		\eeq
		Then, we have the following properties:
		\begin{itemize}
			\item[P1)]
			The error exponent $\Phi(\theta)$ is given by:
			\beq
			\Phi(\theta)=-\inf_{t\in\mathbb{R}}\phi(t;\theta)=-\phi(t^{\star}_{\theta};\theta),
			\label{eq:errexpodeflemmapp}
			\eeq
			where $t^{\star}_{\theta}<0$ is the unique solution to:
			\beq
			\frac{\Lambda_{\mathrm{ave}}(t^{\star}_{\theta};\theta)}{t^{\star}_{\theta}}=0.
			\label{eq:errexpodeflemmapp2}
			\eeq
			\item[P2)] 
			For all $t\in\mathbb{R}$ we have:
			\beq
			\Lambda_{\ell}(t;\theta)\geq d_{\ell}(\theta) t,
			\eeq
			implying in particular that:
			\beq
			\Phi(\theta)\leq |t^{\star}_{\theta}| \mnet(\theta).
			\eeq
			\item[P3)]
			Let $\pi_{\min}$ and $\pi_{\max}$ be the minimum and maximum entry of the Perron eigenvector, respectively.
			Then we have:
			\beq
			\frac{1}{\pi_{\max}}\leq |t^{\star}_{\theta}|\leq \frac{1}{\pi_{\min}}.
			\label{eq:tstarbounds}
			\eeq
		\end{itemize}
	\end{lemma}
	\begin{IEEEproof}
		From the convexity properties of $\phi(t;\theta)$ (see~\cite{MattaBracaMaranoSayedTIT2016,MattaBracaMaranoSayedTSIPN2016} for a detailed summary) we know that the infimum of $\phi(t;\theta)$ in \eqref{eq:errexpodeflemmapp} is in fact a unique minimum located at the solution $t^{\star}_{\theta}$ to the stationary equation:
		\beq
		\phi'(t^{\star}_{\theta};\theta)=0,
		\label{eq:stationequatappendix}
		\eeq
		where $'$ denotes derivative w.r.t. $t$. 
		Therefore, Eq. \eqref{eq:errexpodeflemmapp2} follows from \eqref{eq:phitandthetalemmapp}.
	On the other hand, in view of the convexity properties of $\phi(t;\theta)$, the function $\phi'(t;\theta)$ is strictly increasing in $t$, and since $\phi'(0;\theta)=\Lambda'_{\mathrm{ave}}(0;\theta)=\mnet(\theta)>0$ (we use the fact that the first derivative of the LMGF evaluated in $0$ is equal to the mean of the relative random variable), from \eqref{eq:stationequatappendix} we conclude that the value $t^{\star}_{\theta}$ that minimizes the function $\phi(t)$ in \eqref{eq:errexpodeflemmapp} cannot but be negative, and the proof of property P1) is complete~\cite{MattaBracaMaranoSayedTIT2016,MattaBracaMaranoSayedTSIPN2016}.
			
		Regarding property P2), from the convexity of the local LMGF $\Lambda_{\ell}(t;\theta)$ we can write, for all $t\in\mathbb{R}$:
		\beq
		\Lambda_{\ell}(t;\theta)\geq t \Lambda'_{\ell}(0;\theta)=t d_{\ell} (\theta).
		\label{eq:Lambdaprime}
		\eeq
		Exploiting \eqref{eq:Lambdaverepeatapp}, \eqref{eq:phitandthetalemmapp}, \eqref{eq:errexpodeflemmapp}, and \eqref{eq:Lambdaprime}, we obtain:
		\beqa
		\Phi(\theta)&=&-\phi(t^{\star}_{\theta};\theta)=
		-\int_0^{t^{\star}_{\theta}} \frac{\Lambda_{\mathrm{ave}}(\tau;\theta)}{\tau}d\tau
		\nonumber\\
		&=&
		\sum_{\ell=1}^N
		\int_{t^{\star}_{\theta}}^0 \frac{\Lambda_{\ell}(\pi_{\ell}\tau;\theta)}{\tau}d\tau\nonumber\\
		&\leq&
		|t^{\star}_{\theta}| \sum_{\ell=1}^N \pi_{\ell} d_{\ell}(\theta)=
		|t^{\star}_{\theta}|\mnet(\theta),
		\eeqa
		and property P2) is proved. 
		
		Finally we prove property P3). 
		Making explicit the definition of $\Lambda_{\mathrm{ave}}(t;\theta)$, Eq. \eqref{eq:errexpodeflemmapp2} can be written as:
		\beq
		\frac{\sum_{\ell=1}^N \Lambda_{\ell}(\pi_{\ell}t^{\star}_{\theta};\theta)}{t^{\star}_{\theta}}=0.
		\label{eq:phiprimeqapp}
		\eeq
		In view of \eqref{eq:verybasicdefLambdaell}, with expectation computed under the model $L_{\ell}(\xi|\theta_0)$, we have that $\Lambda_{\ell}(-1;\theta)=0$. 
		Accordingly, when $\pi_{\ell}=1/N$ for all $\ell$, Eq. \eqref{eq:tstarbounds} is obvious. 
		Let us focus on the case where the Perron eigenvector is not uniform. From the strict convexity of $\Lambda_{\ell}(t;\theta)$, we know that:
		\beqa
		\Lambda_{\ell}(\pi_{\ell} t;\theta)&>&0~~\textnormal{for }t<-\frac{1}{\pi_{\ell}},\nonumber\\
		\Lambda_{\ell}(\pi_{\ell} t;\theta)&<&0~~\textnormal{for }-\frac{1}{\pi_{\ell}}<t<0,
		\label{eq:solutsigncond}
		\eeqa
		see Fig.~\ref{fig:fconvexphi}. 
		Since the equality in \eqref{eq:phiprimeqapp} requires that $\Lambda_{\ell}(\pi_{\ell}t^{\star}_{\theta};\theta)$ takes on at least one positive and one negative value, Eq. \eqref{eq:solutsigncond} implies property P3). 
	\end{IEEEproof}
	
	\begin{IEEEproof}[Proof of Theorem~\ref{theor:instanterrprobfin}]
		From \eqref{eq:equadist} we know that $\widehat{\bm{\lambda}}_{k,i}^{(\delta)}(\theta)$ and $\widetilde{\bm{\lambda}}_{k,i}^{(\delta)}(\theta)$ share the same distribution. Thus, from \eqref{eq:withtransient} and \eqref{eq:lambdarec} we have that (we recall that $\stackrel{\mathrm{d}}{=}$ denotes equality in distribution):
		\begin{align}
		\lefteqn{\bm{\lambda}_{k,i}^{(\delta)}(\theta)}
		\nonumber\\&\stackrel{\mathrm{d}}{=}
		\widetilde{\bm{\lambda}}_{k,i}^{(\delta)}(\theta)+(1-\delta)^i\sum_{\ell=1}^N[A^i]_{\ell k}\lambda_{\ell,0}(\theta)\nonumber\\
		&\geq
		\widetilde{\bm{\lambda}}_{k,i}^{(\delta)}(\theta)+(1-\delta)^i\sum_{\ell=1}^N \pi_{\ell}\lambda_{\ell,0}(\theta)\nonumber\\
		&-
		\kappa (1-\delta)^i\beta^i\sum_{\ell=1}^N |\lambda_{\ell,0}(\theta)|\nonumber\\
		&=\widetilde{\bm{\lambda}}_{k,i}^{(\delta)}(\theta)+(1-\delta)^i\sum_{\ell=1}^N \pi_{\ell}\lambda_{\ell,0}(\theta)
		-\frac{{\sf K}_2(\theta)}{|t^{\star}_{\theta}|}(1-\delta)^i\beta^i,
		\label{eq:veryfirstboundappinst}
		\end{align}
		where the inequality follows from Property~\ref{prop:Perron}, and in the last equality we used \eqref{eq:C2theta}. 
		In view of \eqref{eq:veryfirstboundappinst}, and since $t^{\star}_{\theta}<0$, we can write:
		\beqa
		\lefteqn{
			\P[
			\bm{\lambda}_{k,i}^{(\delta)}(\theta)\leq 0
			]}\nonumber\\
		&\leq&
		\P\left[
		\widetilde{\bm{\lambda}}_{k,i}^{(\delta)}(\theta)\leq 
		-(1-\delta)^i \lambda_{\mathrm{ave},0}(\theta)
		+\frac{{\sf K}_2(\theta)}{|t^{\star}_{\theta}|}(1-\delta)^i \beta^i
		\right]\nonumber\\
		&\stackrel{\textnormal{(a)}}{=}&
		\P\left[
		\frac{t^{\star}_{\theta}}{\delta}\widetilde{\bm{\lambda}}_{k,i}^{(\delta)}(\theta)\!\geq\!
		\frac{|t^{\star}_{\theta}|}{\delta}(1-\delta)^i \lambda_{\mathrm{ave},0}(\theta)
		\!-\!\frac{{\sf K}_2(\theta)}{\delta}(1-\delta)^i \beta^i
		\right]
		\nonumber\\
		&\stackrel{\textnormal{(b)}}{\leq}&
		\frac{
			\E\left[
			\exp\left\{\frac{t^{\star}_{\theta}}{\delta}\,\widetilde{\bm{\lambda}}_{k,i}^{(\delta)}(\theta)
			\right\}
			\right]}
		{\exp\left\{
			\frac{|t^{\star}_{\theta}|}{\delta}\,(1-\delta)^i \lambda_{\mathrm{ave},0}(\theta)-\frac{{\sf K}_2(\theta)}{\delta}(1-\delta)^i \beta^i
			\right\}
		}
		\nonumber\\
		&\stackrel{\textnormal{(c)}}{=}&
		e^{\frac{1}{\delta}
			\left[
			\delta\Lambda^{(\delta)}_{k,i}\left(\frac{t^{\star}_{\theta}}{\delta};\theta\right)
			-(1-\delta)^i |t^{\star}_{\theta}|\lambda_{\mathrm{ave},0}
			+{\sf K}_2(\theta)(1-\delta)^i \beta^i
			\right]
		},
		\label{eq:variousineqchain}
		\eeqa
		where (a) follows from multiplying by $t^{\star}_{\theta}/\delta$ both sides of the inequality in the probability brackets and taking into account the fact that $t^{\star}_{\theta}<0$; (b) follows from applying Chernoff's bound; and in (c) we applied Property P3) and introduced the LMGF of $\widetilde{\bm{\lambda}}^{(\delta)}_{k,i}(\theta)$, which can be explicitely defined as:
		\begin{align}
		\Lambda^{(\delta)}_{k,i}(t;\theta)&=\log\E\left[
		e^{t \widetilde{\bm{\lambda}}^{(\delta)}_{k,i}(\theta)}\right]\nonumber\\
		&=\sum_{\ell=1}^N\sum_{m=0}^{i-1} \Lambda_{\ell}\left(\delta(1-\delta)^m [A^{m+1}]_{\ell k} t \,;\, \theta \right),
		\label{eq:LMGFexplic1}
		\end{align}
		with $\Lambda_{\ell}(t;\theta)$ being the LMGF of the log-likelihood ratio $\bm{x}_{\ell,m+1}(\theta)$. 
		Now, letting
		\beq
		c_i\dfz(1-\delta)^{i-1},
		\eeq
		and applying Eqs.~(85) and~(86) from~\cite{MattaBracaMaranoSayedTSIPN2016} to the inner summation in \eqref{eq:LMGFexplic1}, we have the following representation:
		\beqa
		\lefteqn{\Lambda^{(\delta)}_{k,i}\left(\frac{t^{\star}_{\theta}}{\delta};\theta\right)}\nonumber\\
		&=&
		\frac 1\delta\left[
		\sum_{\ell=1}^N
		\int_{c_i \pi_{\ell} t^{\star}_{\theta}}^{\pi_{\ell}t^{\star}_{\theta}}\frac{\Lambda_{\ell}(\tau;\theta)}{\tau}d\tau + \mathcal{O}(\delta)\right]\nonumber\\
		&\stackrel{\textnormal{(a)}}{=}&
		\frac 1\delta\left[
		\phi(t^{\star}_{\theta};\theta)-
		\sum_{\ell=1}^N
		\int_0^{c_i \pi_{\ell} t^{\star}_{\theta}}\frac{\Lambda_{\ell}(\tau;\theta)}{\tau}d\tau + \mathcal{O}(\delta)\right]\nonumber\\
		&\stackrel{\textnormal{(b)}}{=}&
		\frac 1\delta\left[
		-\Phi(\theta)+ \int_{-c_i\pi_{\ell} |t^{\star}_{\theta}|}^0 \frac{\Lambda_{\ell}(\tau;\theta)}{\tau}d\tau + \mathcal{O}(\delta)\right]\nonumber\\
		&\stackrel{\textnormal{(c)}}{\leq}&
		\frac 1\delta\left[
		-\Phi(\theta)+ c_i|t^{\star}_{\theta}|\sum_{\ell=1}^N\pi_{\ell}d_{\ell}(\theta)
		+ \mathcal{O}(\delta)\right],\label{eq:boundonLMGFfinalapp}
		\eeqa
		where (a) follows from \eqref{eq:phitandthetalemmapp}, while (b) and (c) from properties P1) and P2) in Lemma~\ref{lem:uspropLMGFapp}, respectively.
		Using now \eqref{eq:boundonLMGFfinalapp} in \eqref{eq:variousineqchain} and using the definition of ${\sf K}_1(\theta)$ in \eqref{eq:C1theta} we get the upper bound in \eqref{eq:insterrprobmainbound}.
	\end{IEEEproof}
	
	\section{}
	\label{app:corolla}
	\begin{IEEEproof}[Proof of Corollary~\ref{cor:corollaryinsterr}]
		We now determine the adaptation time as the critical instant after which we stay close to the exponent $\Phi$, in the precise sense specified by \eqref{eq:whatwemeanbyadaptime}.
		Let us consider first the case where $\lambda_{\mathrm{ave}}(\theta)\geq \mnet(\theta)$ for all $\theta\neq\theta_0$. 
		In this case, we have ${\sf K}_1(\theta)\leq 0$ for all $\theta$ and, hence, in view of \eqref{eq:insterrprobmainbound}, condition \eqref{eq:whatwemeanbyadaptime} will be met if we ensure that:
		\beq
		i > \frac{1}{\log\beta^{-1}}\log\frac{{\sf K}_2}{\epsilon\,\Phi}\Rightarrow
		{\sf K}_2\,\beta^{i}<\epsilon\Phi,
		\eeq
		which shows that the choice for ${\sf T}_{\sf ASL}$ in \eqref{eq:adaptime1} guarantees \eqref{eq:whatwemeanbyadaptime} for all $i>{\sf T}_{\sf ASL}$.
		
		We continue by examining the unfavorable case where $\lambda_{\mathrm{ave}}(\theta)< \mnet(\theta)$ for at least one value $\theta\neq\theta_0$. 
		In this case we have ${\sf K}_1=\max_{\theta\neq\theta_0} {\sf K}_1(\theta)>0$, and we can write:
		\beq
		i>\frac{1}{\log(1-\delta)^{-1}}
		\log\frac{
			{\sf K}_1
		}{\epsilon\,\Phi}
		\Rightarrow
		(1-\delta)^i {\sf K}_1<\epsilon\Phi.
		\label{eq:fracofPhi}
		\eeq
		Then, if we set the adaptation time ${\sf T}_{\sf ASL}$ according to the law in \eqref{eq:fracofPhi}, the quantity $\beta^i$ appearing in \eqref{eq:insterrprobmainbound} would decay to zero as $\approx\beta^{1/\delta}$, and, hence, would be incorporated into the higher-order term $\mathcal{O}(\delta)$, and the claim of the corollary is proved.
	\end{IEEEproof}
	
\end{appendices}

%







\end{document}